\documentclass[11pt]{article}
\pdfoutput=1 
\usepackage{amssymb}
\usepackage{amsmath}
\usepackage{amstext}
\usepackage{graphicx,epsfig}
\usepackage{epsfig}
\usepackage{verbatim} 
\usepackage{fancybox}
\usepackage{color}
\usepackage{ulem}
\usepackage{enumitem}
\usepackage{subfigure}
\usepackage{bbm}
\usepackage{parskip}
\usepackage{dsfont}
\usepackage[numbers,sort&compress]{natbib}
\usepackage{ytableau}
\usepackage{framed}

\usepackage{tikz}
\usetikzlibrary{decorations}
\pgfdeclaredecoration{complete sines}{initial}
{
    \state{initial}[
        width=+0pt,
        next state=upsine,
        persistent precomputation={\pgfmathsetmacro\matchinglength{
            \pgfdecoratedinputsegmentlength / int(\pgfdecoratedinputsegmentlength/\pgfdecorationsegmentlength)}
            \setlength{\pgfdecorationsegmentlength}{\matchinglength pt}
        }] {}
    \state{upsine}[width=\pgfdecorationsegmentlength,next state=downsine]{
        \pgfpathsine{\pgfpoint{0.25\pgfdecorationsegmentlength}{0.5\pgfdecorationsegmentamplitude}}
        \pgfpathcosine{\pgfpoint{0.25\pgfdecorationsegmentlength}{-0.5\pgfdecorationsegmentamplitude}}
    }
    \state{downsine}[width=\pgfdecorationsegmentlength,next state=upsine]{
        \pgfpathsine{\pgfpoint{0.25\pgfdecorationsegmentlength}{-0.5\pgfdecorationsegmentamplitude}}
        \pgfpathcosine{\pgfpoint{0.25\pgfdecorationsegmentlength}{0.5\pgfdecorationsegmentamplitude}}
}
    \state{final}{}
}

\newcommand{\vev}[1]{\bigl\langle#1\bigr\rangle}

\newcommand{\Comment}[1]{{}}
\definecolor{darkblue}{rgb}{0.15,0.35,0.55}
\definecolor{comment}{rgb}{1,0.4,0.4}
\definecolor{reddish}{rgb}{0.65, 0.2, 0.2}
\usepackage[linktocpage=true]{hyperref}
\hypersetup{
colorlinks=true,
citecolor=darkblue,
linkcolor=reddish,
urlcolor=darkblue,
pdfauthor={},
pdftitle={},
pdfsubject={}
}

\newcommand{\be}{\begin{equation}}
\newcommand{\ee}{\end{equation}}
\newcommand{\bea}{\begin{eqnarray}}
\newcommand{\eea}{\end{eqnarray}}
\newcommand{\beas}{\begin{eqnarray*}}
\newcommand{\eeas}{\end{eqnarray*}}

\def\({\left(}
\def\){\right)}

\newcommand{\rd}{{\rm d}}
\newcommand{\vp}{\varphi}

\def\gsim{ \lower .75ex \hbox{$\sim$} \llap{\raise .27ex \hbox{$>$}} }
\def\lsim{ \lower .75ex \hbox{$\sim$} \llap{\raise .27ex \hbox{$<$}} }

\def\xyma{\xymatrix@M.7em}
\def\xymas{\xymatrix@M.1em}
\newcommand{\ba}{\begin{eqnarray}}
\newcommand{\ea}{\end{eqnarray}}

\definecolor{darkred}{rgb}{0.7,0.3,0.3}
\definecolor{darkgreen}{rgb}{0.2,0.7,0.3}
\definecolor{greyish}{rgb}{.90,.90,.90}
\definecolor{greyish2}{rgb}{.96,.96,.96}
\definecolor{darkblue2}{rgb}{0.3,0.4,0.9}
\usepackage{pifont}
\newcommand{\cmark}{\ding{51}}%
\newcommand{\xmark}{\ding{55}}%
\usepackage{xcolor,colortbl}
\newcommand{\shade}{\cellcolor{greyish}}
\usepackage{tcolorbox}

\title{}
\author{}

\numberwithin{equation}{section}

\begin{document}
%
\setcounter{page}{1}
\renewcommand{\thefootnote}{\fnsymbol{footnote}}
~
\vspace{.80truecm}
\begin{center}
{\LARGE \bf{Inflationary soft theorems revisited:}}\\ \vspace{.2cm}
{\LARGE \bf{A generalized consistency relation}}
\end{center} 

\vspace{1truecm}
\thispagestyle{empty}
\centerline{{\large Lam Hui,\footnote{\href{mailto:lh399@columbia.edu}{\texttt{lh399@columbia.edu}}} Austin Joyce,\footnote{\href{mailto:austin.joyce@columbia.edu}{\texttt{austin.joyce@columbia.edu}}} and Sam S. C. Wong\footnote{\href{mailto:sw3266@columbia.edu}{\texttt{sw3266@columbia.edu}}}}}
\vspace{.5cm}
 
 \centerline{{\it Center for Theoretical Physics, Department of Physics,}}
 \centerline{{\it Columbia University, New York, NY 10027}} 
 \vspace{.25cm}

 \vspace{.8cm}
\begin{abstract}
\noindent
We reconsider the derivation of soft theorems associated with
nonlinearly-realized symmetries in cosmology. Utilizing the
path integral, we derive a generalized consistency relation that
relates a squeezed $(N+1)$-point correlation function to an $N$-point
function, where the relevant soft mode is at {\it early} rather than late
time. This generalized (early-late-time) version has wider applicability than the
standard consistency relation where all modes are evaluated at late
times. We elucidate the conditions under which the latter follows from
the former. A key ingredient is the physical mode condition: that the
nonlinear part of the symmetry transformation must match the time
dependence of the dominant, long wavelength physical mode. 
This is closely related to, but distinct from, the adiabatic mode condition. 
Our derivation sheds light on a number of otherwise puzzling features
of the standard consistency relation: (1) the underlying
nonlinearly-realized symmetries (such as dilation and special
conformal transformation SCT) originate as residual gauge redundancies,
yet the consistency relation has physical content---for instance, it
can be violated; (2) the standard consistency
relation is known to fail in ultra-slow-roll inflation, but 
since dilation and SCT remain good symmetries, there should be a
replacement for the standard relation; (3) in large scale structure applications, it is
known that the standard consistency relation breaks down if the long wavelength
power spectrum is too blue. The early-late-time consistency relation
helps address these puzzles. We introduce a toy model where
explicit checks of this generalized consistency relation are simple to
carry out. Our methodology can be adapted to
cases where violations of the standard consistency relation involve
additional light degrees of freedom beyond the inflaton.
\end{abstract}

\newpage

\setcounter{tocdepth}{2}
\tableofcontents

\newpage
\renewcommand*{\thefootnote}{\arabic{footnote}}
\setcounter{footnote}{0}

\section{Introduction}

Consistency relations are soft theorems in cosmology that relate a
squeezed $(N+1)$-point correlation function to an $N$-point function,
of the following schematic form:
\begin{eqnarray}
\label{CRschematic}
\lim_{\vec q \rightarrow 0} {1\over P_\zeta (q)} \langle
  \zeta_{\vec k_1} \zeta_{\vec k_2} ... \zeta_{\vec k_N} \zeta_{\vec q} \rangle' \sim
  \langle \zeta_{\vec k_1} \zeta_{\vec k_2} ... \zeta_{\vec k_N}
  \rangle' \, ,
\end{eqnarray}
where $\zeta$ is the curvature perturbation,\footnote{
In other words, the spatial part of
the unitary gauge metric goes as $\rd s^2 \propto (e^{2\zeta} \delta_{ij}
+ e^{\gamma_{ij}}) \rd x^i \rd x^j$, where $\zeta$ is the scalar perturbation
and $\gamma_{ij}$ is the tensor perturbation which is trace-less and transverse.} $P_\zeta$ is its power
spectrum (two-point function), ${\vec k_1}, \cdots, {\vec k_N}$ label the
$N$ hard momenta, ${\vec q}$ labels a soft momentum and
$\langle \cdots\rangle'$ refers to a correlation function with the
overall delta function removed. Note that the right hand side should
contain operators ({\it i.e.}, derivatives) acting on the $N$-point function,
which are not shown explicitly.\footnote{Also, the hard modes can generally be replaced by other observables at high momenta.}
The consistency relations can be thought of as 
analogous to soft theorems familiar in high energy physics, with
$\zeta_{\vec q}$ playing the role of the soft Nambu--Goldstone boson,
or soft pion. The consistency relations are associated with
nonlinearly realized symmetries, the prime example of which is
spatial dilation under which $\zeta$ transforms as
$\zeta \mapsto \zeta + \delta\zeta$ where
\begin{eqnarray}
\delta\zeta =
  \lambda (1 + \vec x \cdot \vec \nabla \zeta) \, ,
\end{eqnarray}
with $\lambda$ being a small parameter. 
Maldacena first pointed out a dilation consistency relation exists
in the context of single field inflation 
\cite{Maldacena:2002vr,Creminelli:2004yq,Cheung:2007sv}.
It was later pointed out that an analogous consistency relation
exists for spatial special conformal transformation (SCT) under which:
\begin{eqnarray}
\delta\zeta = 2 \vec b \cdot \vec x + 
(2 (\vec b \cdot \vec x) \, \vec x - \vec x {}^2 \, \vec b) \cdot \vec \nabla
  \zeta \, ,
\end{eqnarray}
where $\vec b$ is a small constant vector
\cite{Creminelli:2012ed,Hinterbichler:2012nm,Assassi:2012zq,
  Kehagias:2012pd,Goldberger:2013rsa}. 
The above symmetries give scalar consistency relations. There are also tensor
consistency relations where the soft mode is a tensor mode
\cite{Maldacena:2002vr,Creminelli:2012ed}. 
More generally, there is an infinite tower of consistency relations,
each associated with a symmetry transformation of the following schematic form:
\begin{eqnarray}
\delta \zeta \sim x^n + x^{n+1} \partial \zeta\, , \quad\quad\quad\quad 
\delta \gamma \sim x^n + x^{n+1} \partial \gamma \, ,
\end{eqnarray}
where the index structure is suppressed, $\gamma$ is the tensor
perturbation, $n = 0, 1, 2, \ldots$, 
and the corresponding consistency relation involves in
general both a soft $\zeta$ and a soft $\gamma$
\cite{Hinterbichler:2013dpa}. The symmetry at each $n$ constrains the $q^n$ behavior
of the left hand side of eq.~\eqref{CRschematic}.  
Purely scalar consistency relations like eq.~\eqref{CRschematic} exist
for $n=0$ (dilation) and $n=1$ (special conformal transformation),
which we will largely focus on.
These relations have also 
been generalized to multiple soft
legs~\cite{Joyce:2014aqa,Mirbabayi:2014zpa} and to 
non-Bunch--Davies initial
states~\cite{Flauger:2013hra,Berezhiani:2014kga,Collins:2014fwa,Shukla:2016bnu}. The other type of consistency relation in the collapsed limit is the Suyama-Yamaguchi inequality from unitarity \cite{Suyama:2007bg}.
Parallel to these developments in cosmology, there has been a recent resurgence of interest in the soft behavior of 
flat space scattering amplitudes, both in gauge theories and effective field theory (see, {\it e.g.},~\cite{Strominger:2017zoo,Cheung:2014dqa}). A deep connection between these soft theorems and asymptotic symmetries of asymptotically-flat spaces has been suggested \cite{Strominger:2013jfa,He:2014laa,Campiglia:2016jdj,Campiglia:2016hvg,Conde:2016csj,Campiglia:2016efb,Conde:2016rom,Hamada:2018vrw}. In a similar way, cosmological soft theorems are related to the 
asymptotic symmetries of cosmological
spacetimes~\cite{Mirbabayi:2016xvc,Ferreira:2016hee,Hinterbichler:2016pzn}.

The consistency relations---being symmetry statements---are
exact and hold even if the high momentum fluctuations are nonlinear or non-perturbative.
This makes them especially interesting in the late universe, that is, in
large scale structure studies, where modes measured with the highest
precision are in the nonlinear regime.
In this context, Kehagias and Riotto as well as 
Peloso and Pietroni (KRPP) \cite{Kehagias:2013yd,Peloso:2013zw} 
pointed out that consistency relations exist even in the Newtonian
(sub-Hubble) regime relevant for large scale structure.
It can be shown that, of the whole tower of symmetries mentioned
above, two remain non-trivial in the Newtonian limit, which in
Newtonian gauge reduce to 
a shift symmetry of the gravitational
potential and the KRPP translation symmetry
\cite{Creminelli:2013mca,Horn:2014rta}.\footnote{We should mention that in the large scale structure context, there is
also a different kind of consistency relation that is approximate in
nature and not based on symmetries in the sense discussed here
\cite{Valageas:2013zda,Kehagias:2013paa}. They are less robust, for instance they are violated if
the hard modes are galaxy count fluctuations, while the consistency relations
we focus on hold even for realistic high-momentum galaxy observables
(see \cite{Horn:2014rta} for a discussion).}

By now, there are several different derivations of the consistency
relations, using the background wave method
\cite{Maldacena:2002vr,Creminelli:2004yq}, 
the Ward identity~\cite{Hinterbichler:2013dpa},
the effective action~\cite{Goldberger:2013rsa},
the wave functional~\cite{Pimentel:2013gza,Kundu:2015xta}
and the Slavnov--Taylor identity~\cite{Berezhiani:2013ewa,Binosi:2015obq}.\footnote{There are also indirect derivations using holography such as \cite{Schalm:2012pi,McFadden:2014nta,Isono:2016yyj}.}
Even if an important result deserves multiple derivations, the reader
might reasonably ask: why would yet another derivation be useful at
this point?

Our motivation is the existence of several long-standing puzzles connected with
consistency relations. We hope to clarify and illuminate certain implicit assumptions made in previous derivations.
Our goal is more than conceptual clarification: we wish to find
a generalization of consistency relations that is more robust and
would open the way to writing down exact relations in more general contexts.
The puzzles we wish to address are as follows:
\begin{itemize}

\item The nonlinearly realized symmetries used in consistency
  relations originate as gauge symmetries. For instance, the
  transformation $\delta\zeta = \lambda (1 + \vec x \cdot \vec
  \nabla)\zeta$ follows from performing the diffeomorphism $\vec x
  \mapsto (1 - \lambda) \vec x$ and working out how the metric
  transforms. (See Section~\ref{sec:generalderivation} below.) The relevant diffeomorphisms are ones
  that do not vanish at spatial infinity (``large'' gauge
  transformations). Even so, it is perhaps surprising that a physical
  statement follows from what is a gauge redundancy. We
  know the consistency relations are physical ({\it i.e.}, not trivial
  identities) because they can be
  broken; explicit examples exist, such as
  when additional light fields are present during inflation
  \cite{Lyth:2001nq,Kofman:2003nx,Dvali:2003em,Chen:2009we,Chen:2009zp,Noumi:2012vr,Arkani-Hamed:2015bza,Lee:2016vti,Arkani-Hamed:2018kmz}.\footnote{The consistency relations can be rewritten in a way that
    make them look very simple (in particular the analogue of the right
    hand side of eq. (\ref{CRschematic}) vanishes) \cite{Tanaka:2011aj,Pajer:2013ana,Horn:2015dra}, but it does
    not mean they are physically empty statements, as evidenced by the
    fact that they can be broken.}
  There is a general recognition that this has something to do with the
  adiabatic mode condition---that the ``large'' diffeomorphism
  generates a mode that is smoothly connected to the long wavelength
  limit of a physical mode. We wish to sharpen this intuition,
  particularly in light of another example, namely ultra-slow-roll
  inflation \cite{Tsamis:2003px,Kinney:2005vj}, where an adiabatic mode exists, yet the consistency
  relations are violated \cite{Namjoo:2012aa,Martin:2012pe}. Related to this issue is that several
  existing derivations of the consistency relations assume the initial
  wave function(al) is invariant under the ``large'' diffeomorphisms. We
  will show why this is not true, despite insisting that the
  hamiltonian and momentum constraints annihilate the wave function.

\item Paradoxically, exact statements---like the consistency relations---are
  most interesting when they fail; their possible breakdown 
  means testing them experimentally is meaningful and would teach
  us about the violations of certain basic assumptions
  underlying these statements. Two examples of breakdown are
  (1) scenarios where there are light fields present in addition to
  the inflaton~\cite{Lyth:2001nq,Kofman:2003nx,Dvali:2003em,Chen:2009we,Chen:2009zp,Noumi:2012vr,Arkani-Hamed:2015bza,Lee:2016vti,Arkani-Hamed:2018kmz}, and
  (2) ultra-slow-roll inflation: a single-field inflation model where
  the potential is exactly flat and the inflaton has a non-zero
  kinetic energy that is rapidly diminished by expansion 
  \cite{Tsamis:2003px,Kinney:2005vj, Namjoo:2012aa,Martin:2012pe}.
  In these cases, the relevant (spatial) diffeomorphisms remain good
  symmetries, and one is led to ask:
  is there some generalized form of the
  consistency relations that replace the standard ones?
  For scenarios involving extra fields, it is not surprising that the
  symmetry transformation of those fields should be relevant and the
  standard consistency relations should get corrected.\footnote{For scenarios such as the curvaton model, one might be
    tempted to say as long as one is interested in observables at late
  times, after the curvaton has long decayed, that there's no need to
  keep track of the curvaton fluctuations. Our more robust consistency
relation, {\it i.e.}, the early-late time one, shows that this is not the
case. We will have more to say about this later.} Consistency relation in the presence of extra light field have been studied before \cite{Assassi:2012zq,Gong:2017wgx}. However they reduce to the known single field consistency relation in the single field case and are not satisfied in ultra slow  roll.
  For ultra-slow-roll inflation, the situation is more puzzling
  because there is no extra field that requires keeping track of. Recently, it was
  pointed out that the shift symmetry of the inflaton, combined with
  temporal and spatial diffeomorphisms, can be used to deduce a
  modified consistency relation 
  \cite{Mooij:2015yka,Pajer:2017hmb,Finelli:2017fml,Finelli:2018upr,Bravo:2017wyw}. There is still the question of what
  happens to the original symmetries (spatial dilation and SCT)---they remain good symmetries on their
  own, without mixing with the inflaton's shift symmetry---what are
  their consequences? We will show that indeed they, on their own, imply
  consistency relations, albeit of an unequal-time form, with the soft
  mode in the far past. This is contrasted with the standard
  consistency relations, eq.~\eqref{CRschematic}, where all modes,
  including the soft one, are evaluated at the same late time.
  The ultra-slow-roll inflation model is admittedly a rather special
  example.\footnote{The fact that $\zeta$ grows for the super-Hubble
    modes in the ultra-slow-roll model means its late-time
    (post-inflation) predictions depend on the details of how
    inflation ends. This is different from the {\it normal} slow-roll
    models where $\zeta$ is conserved outside Hubble, and its conservation is
    unaffected by reheating. The growing of $\zeta$ and the exponential deceleration of $\dot{\bar{\phi}}$ also signals the breakdown of $\zeta$-gauge at some point. } We view it as a
  useful warm-up exercise---to understand the violation of the standard
  consistency relations in a case that involves no extra fields---before tackling the more phenomenologically interesting examples of
  having additional light degrees of freedom.

\item As mentioned above, the application of consistency relations to
  large scale structure is particularly interesting because of the
  non-perturbative nature of both. It is known from an explicit
  perturbative check that the consistency relations would be violated
  if the mass power spectrum $P_\delta (q)$ ($\delta$ is the
  fractional mass density fluctuation) has a slope that is too blue in
  the $q \rightarrow 0$ limit. More precisely, if $P_\delta (q) \sim
  q^m$ for small $q$'s, $m$ must be less than $3$ for the KRPP consistency
  relation to hold, and less than $4$ for a lower order consistency
  relation associated with the shift symmetry of the gravitational
  potential \cite{Horn:2014rta}.
  Why this should be so is not clear from existing derivations of the
  consistency relations. We will show that the generalized version,
  where the soft mode is in the far past, has no such requirement on
  $m$. Note that the KRPP consistency relation in large scale structure \cite{Kehagias:2013yd,Peloso:2013zw} 
  is already in an unequal-time form, and is in some sense a
  forerunner of our early-late-time version.
The relation between the two will be discussed below.
\end{itemize}

Clearly one of our central tasks is to elucidate under what
conditions the early-late-time version of the consistency relations
imply the standard, purely late-time version.
We will introduce the idea of the {\it physical mode condition}, and
distinguish it from the {\it adiabatic mode condition}. The physical mode condition plays a
crucial role in connecting the two versions of the consistency
relations.

The early-late-time version of the identity can be explicitly checked in {\it normal
slow-roll} inflation ({\it i.e.}, the inflaton is on an trajectory where
Hubble friction balances the slight tilt of the potential). For
technical reasons that will be explained below, it is difficult to do
the same for {\it ultra-slow-roll} inflation. We introduce a toy model
that exhibits the same spatial dilation and SCT as {\it global} symmetries, where a perturbative check
is straightforward to carry out. The model has no gravity, yet
contains a lot of the features describing inflationary perturbations,
making it a useful playground to investigate the consistency relation in its
many manifestations. The model can be adjusted to resemble
ultra-slow-roll, normal slow-roll or even to describe fluctuations in Minkowski
space (reminiscent of the ghost condensate example studied by \cite{Goldberger:2013rsa}).

The paper is organized as follows.
In Section~\ref{sec:generalderivation} we revisit the derivation of the standard
cosmological consistency relation, and derive, using the path
integral, a generalized version, relating correlation functions with
an initial time soft insertion to 
the symmetry-transformed late-time correlators.\footnote{We often switch between the singular and plural form for
  ``consistency relation". The singular form refers to the one that
  holds for all relevant symmetries (such as~\eqref{eq:mainID}), whereas the plural
  form counts one for each symmetry separately (such as dilation
  consistency relation, SCT consistency relation and so on).}
We discuss the circumstances under which this identity can be
promoted to a late-time 
identity and introduce the idea of physical mode condition in Section \ref{sec:physicalmodes}.\footnote{We follow convention and sometimes refer to the consistency
  relation as an ``identity", but it should be emphasized the
  consistency relation has non-trivial physical content, as explained
  above and further in Section \ref{sec:physicalmodes}.
}
We then discuss the application of this identity to inflation in Section~\ref{sec:inflation}.
In Section~\ref{sec:toymodel1} we introduce a toy model and discuss
its symmetries for various choices of time dependence of the
couplings, and analyze the corresponding consistency relations.
We collect some useful information tangential to the main 
development in a number of appendices. 
In Appendix~\ref{app:4point} we discuss the normalization of the path integral and its effects on the final identity we derive. In Appendix~\ref{app:wavefgaugetrans} we show explicitly in electromagnetism and pure gravity that the wavefunctional can and does transform under large gauge transformations, despite being invariant under small gauge transformations. In Appendix~\ref{sec:pertproof} we give a perturbative argument for the necessity of the physical mode condition in order to be able to write an identity for  late-time correlation functions. In Appendix~\ref{app:toymodel} we collect many computations of correlation functions in the toy model we introduce in Section~\ref{sec:toymodel1}. Finally, in Appendix~\ref{app:classicalcheck} we discuss the classical analogue of our results, focusing on their relevance for correlation functions of large scale structure observables.

\section{Derivation of a generalized consistency relation}
\label{sec:generalderivation}

Our goal in this section is to first derive an early-late-time
consistency relation in Section~\ref{sec:pathintderiv} and then discuss in
Section~\ref{sec:physicalmodes} under
what condition this early-late-time version leads to the purely
late-time (standard) version of the consistency relation.
In this section, to emphasize the general nature of the derivation,
we use $\varphi$ to refer to the fluctuating field of interest
{\it i.e.}, it would be $\zeta$ in many applications, but it could
also be some other variable. We use the symbol $\phi$ to denote
the corresponding field operator.

In this section, we assume there is some general nonlinearly-realized symmetry
and work out its consequences. For the application to the inflationary
consistency relations, we are interested in ``large'' gauge
transformations (ones that do not vanish at spatial infinity).
The connection between a large gauge
transformation and a nonlinearly-realized symmetry will be discussed in the next section---in
particular, we will show that the wavefunction transforms
non-trivially under a ``large'' gauge transformation, much like a
nonlinearly-realized global symmetry (see also Appendix~\ref{app:wavefgaugetrans}).

\subsection{A general path integral Ward identity}
\label{sec:pathintderiv}

Schr\"odinger field theory provides a convenient language in which to
think about Ward identities for correlation functions. A similar
approach was employed
in~\cite{Goldberger:2013rsa,Berezhiani:2013ewa,Berezhiani:2014kga,Avery:2015rga}. In
this approach, to a given quantum state, $\lvert\psi\rangle$, we assign
a wavefunctional, $\Psi[\vp_a] =\langle \vp_a\rvert \psi\rangle$,
which is the projection of the state $\lvert\psi\rangle$ onto the
basis of field eigenstates, which satisfy $\phi(\vec
x,\tau_a)\lvert\vp_a\rangle = \vp_a(\vec
x)\lvert\vp_a\rangle$, where $\phi(\vec x, \tau_a)$ is the field
operator.\footnote{Although the $\lvert\vp_a\rangle$ are
  Heisenberg picture quantum states, they carry implicit time
  dependence to ensure that they are eigenstates of the $\phi(\vec
  x,\tau)$ operator at time $\tau=\tau_a$.} Here $\vp_a(\vec x)$ is the
spatial profile of the (real) scalar field $\phi$, at time $\tau_a$.

In terms of this wavefunctional, equal-time correlation functions 
(at some time $\tau_f$) are defined in the usual way
\be
\langle \vp_f(\vec x_1)\cdots\vp_f(\vec x_N)\rangle = {\int{\cal
    D}\vp_f \,\vp_f(\vec x_1)\cdots\vp_f(\vec
  x_N)\left\lvert\Psi[\vp_f]\right\rvert^2} \, ,
\label{eq:schrodingercorrs}
\ee
so that $\left\lvert\Psi[\vp_f]\right\rvert^2$ provides a probability
distribution for spatial $\phi$ profiles at time $\tau_f$. This assumes
the wavefunctional is properly normalized: ${\int{\cal D}\vp_f \left\lvert\Psi[\vp_f]\right\rvert^2} = 1$.
Written as such, the above expectation value represents exactly
the quantum expectation value:
\be
\langle \vp_f(\vec x_1)\cdots\vp_f(\vec x_N)\rangle = \langle \psi | \phi (\vec x_1, \tau_f) ... \phi (\vec x_N,
\tau_f) | \psi \rangle \, .
\ee
The late-time wavefunctional, $\Psi[\vp_f]$, can itself be constructed
as a path integral from some initial vacuum state, $\lvert 0_{\rm
  in}\rangle$, by inserting a complete set of field eigenstates
\be
\Psi[\vp_f] = \int{\cal D}\vp_i\,
\langle\vp_f\rvert\vp_i\rangle\langle\vp_i\rvert 0_{\rm in}\rangle = 
\int {\cal D}\vp_i
\raisebox{.09cm}{$\underset{\substack{{\scriptscriptstyle \phi(\tau_f)=\varphi_f} \\ {\scriptscriptstyle\phi(\tau_i) =\varphi_i}}}{\displaystyle\int}$}   \hspace{-.4cm}
{\cal D}\phi \,e^{iS[\phi]}\Psi_0[\vp_i],
\label{eq:wavefunctional}
\ee
where we have represented the inner product $\langle\vp_f\rvert\vp_i\rangle$ as a restricted path integral that
sums over all possible field configurations for $\phi$, subject to the
boundary conditions $\phi(\vec x, \tau_i) = \vp_i(\vec x)$ and $\phi(\vec
x, \tau_f) = \vp_f(\vec x)$.\footnote{In this context, $\phi$ is of course merely a variable of integration
  and not an operator.}
We have also represented the  vacuum wavefunctional as $\Psi_0[\vp_i]\equiv \langle \varphi_i | 0_{\rm in} \rangle$.

By standard arguments \cite{Weinberg:1995mt}, $\Psi_0[\varphi_i]$, for
$\tau_i$ in the far past, can be equated with the free vacuum wavefunctional,
which is gaussian:
\be
\Psi_0[\vp_i]  \propto
\exp\left(-\frac{1}{2}\int\frac{\rd^3k}{(2\pi)^3}
\, {\cal E}_i (k) \,
\vp_i(\vec k)\vp_i(-\vec k)\right) \, ,
\label{eq:gaussianwavef}
\ee
where ${\cal E}_i$ is some kernel and we assumed rotational symmetry. The power spectrum $P_{\phi_i} (k)$ defined by
$\langle \varphi_i (\vec k) \varphi_i (\vec k') \rangle =
\langle 0_{\rm in} |  \phi_i (\vec k) \phi_i (\vec k') | 0_{\rm in} \rangle = (2\pi)^3
\delta_D (\vec k + \vec k') P_{\phi_i} (k)$, where $\phi_i (\vec k)
\equiv \phi (\vec k, \tau_i)$, is related to this
kernel by 
\be
P_{\phi_i} (k) = {1 \over 2 {\,\rm Re\,}{\cal E}_i (k)} \, .
\label{Pvarphii}
\ee
In the context of inflation, $|0_{\rm in} \rangle$ is the Bunch--Davies
vacuum (Appendix \ref{app:vacuumfunctional}). Other choices of the initial state are possible---the initial wavefunctional would then take a non-gaussian form (except
for special cases), and our derivation would have to be modified accordingly.
We have not been specific about the overall normalization of the
wavefunctional
because it does not affect the results; we divide out by the
normalization in expectation values (see Appendix \ref{app:4point}).

Consider the expectation value of a general operator ${\cal
  O}_\vp(\vec k_1,\cdots, \vec k_N)$, built from the fundamental
$\phi$ field ({\it e.g.}, a product of $\phi$'s at different momenta):
\begin{equation}
\langle{\cal O}_{\vp_f}(\vec k_1,\cdots, \vec k_N)\rangle = \int{\cal D}\vp_f\, {\cal O}_{\vp_f}(\vec k_1,\cdots, \vec k_N)\,\left\lvert\Psi[\vp_f]\right\rvert^2.
\label{eq:genfunctional}
\end{equation}
Next, consider an abstract symmetry which acts on the fields $\phi$, and which may consist of both a nonlinear and a linear piece. We can write such a symmetry schematically as
\begin{equation}
\delta \vp_f = \delta_{\rm NL}\vp_f+\delta_{\rm L}\vp_f,
\end{equation}
where $\delta_{\rm NL}\vp_f$ is the nonlinear transformation of $\vp_f$ and $\delta_{\rm L}\vp_f$ is the part of the transformation linear in the field.\footnote{Strictly speaking, the transformation $\delta_{\rm NL}\vp_f$ is {\it sub-linear}, in that it is independent of the field $\vp$. In keeping with standard terminology, we call such transformations nonlinear, as they are not proportional to the field itself.} We now perform the change of variable $\vp_f\mapsto\vp_f+\delta\vp_f$ in the integral~\eqref{eq:genfunctional}. This does not affect the late-time correlation function, because $\vp_f$ is merely a dummy variable of integration. This implies that
\begin{equation}
\langle{\cal O}_{\vp_f}(\vec k_1,\cdots, \vec k_N)\rangle= \int{\cal D}(\vp_f+\delta \vp_f)\, \left({\cal O}_{\vp_f}(\vec k_1,\cdots, \vec k_N)+\delta{\cal O}_{\vp_f}(\vec k_1,\cdots, \vec k_N) \right)\,\left\lvert\Psi[\vp_f+\delta\vp_f]\right\rvert^2,
\label{eq:generalwardID}
\end{equation}
where $\delta{\cal O}_{\vp_f}$ is the transformation rule for the
operator ${\cal O}_{\vp_f}$ inherited from its dependence on
$\vp_f$. We assume that the path integral measure is invariant under
this change of variable (${\cal D}\delta\vp_f = 0$).\footnote{In general,
the measure does change by a field-independent normalization (see \cite{Kaya:2018zpa} regarding the issue of measure and FP determinant).
This is relatively harmless as long as one enforces the proper normalization of
the wavefunctional. We discuss in Appendix \ref{app:4point} how to account for this.}

Next, we expand out~\eqref{eq:generalwardID} in powers of the field variation, $\delta\vp_f$, treating it as infinitesimal. The $0^{\rm th}$-order piece in $\delta\vp_f$ is precisely the right-hand side of~\eqref{eq:genfunctional} so we find that the pieces first order in $\delta\vp_f$  must sum up to zero, leading to the identity
\begin{equation}
0 = \int{\cal D}\vp_f\delta{\cal O}_{\vp_f}(\vec k_1,\cdots, \vec k_N)
\left|\Psi[\vp_f]\right|^2+
\left[ \int{\cal D}\vp_f{\cal O}_{\vp_f}(\vec k_1,\cdots, \vec
  k_N) \left( \Psi^*[\vp_f]\delta \Psi[\vp_f]+{\rm c.c.} \right) \right] \, ,
\label{eq:wardidunsimplfied}
\end{equation}
where ${\rm c.c.}$ denotes complex conjugate.

To proceed, we need an expression for $\delta\Psi$. We do not want to make any assumptions about how the late-time wavefunctional transforms, but instead infer its transformation rules from those of the initial vacuum, which we assume is gaussian.\footnote{Another possibility is to make some assumption about the transformation properties of the late-time wavefunctional directly. This approach was utilized in~\cite{Pimentel:2013gza,Kundu:2015xta} to derive Maldacena's consistency relation by assuming that the late-time wavefunctional is invariant under the symmetry transformations we consider (except for a shift of the 1-point function). However, the very fact that the dilation consistency relation is violated in certain models (for instance ultra slow-roll) implies that the late-time wavefunctional does not transform in the prescribed way in these cases. We therefore instead choose to use the known gaussian form of the vacuum wavefunctional to infer how the late-time wavefunctional should transform by time-evolving the transformed initial state using the path integral.}
Recall that the wavefunctional can be written as~\eqref{eq:wavefunctional}, which implies
\begin{equation}
\Psi[\vp_f+\delta \vp_f] = \int {\cal D}\vp_i
\raisebox{.09cm}{$\underset{\substack{{\scriptscriptstyle
        \phi(\tau_f)=\varphi_f + \delta\vp_f} \\ {\!\!\!\!\!\!\!\!\!\!\!\!\!\!\scriptscriptstyle\phi(\tau_i) =\varphi_i}}}{\displaystyle\int}$}   \hspace{-.4cm}
{\cal D}\phi \,e^{iS[\phi]}\Psi_0[\vp_i],
\end{equation}
where only the final boundary condition on the path integral is
shifted by the symmetry transformation. We perform a similar transformation
on $\vp_i$ (a dummy variable of integration) as we did on $\vp_f$ by
sending $\vp_i \mapsto \vp_i+\delta \vp_i $ (again assuming the path
integral measure is invariant) to obtain the expression
\begin{equation}
\label{PsiVarphifDelta}
\Psi[\vp_f+\delta \vp_f] = \int {\cal D}\vp_i 
\raisebox{.09cm}{$\underset{\substack{{\scriptscriptstyle
        \phi(\tau_f)=\varphi_f + \delta\vp_f} \\ {\!\scriptscriptstyle\phi(\tau_i)
        =\varphi_i + \delta\vp_i}}}{\displaystyle\int}$}   \hspace{-.4cm}
{\cal D}\phi \,e^{iS[\phi]}
\Psi_0[\vp_i+\delta\vp_i].
\end{equation}
The manipulations so far work for {\it any} transformation $\delta\varphi_f$
and $\delta\varphi_i$; the fact that the transformation is a {\it symmetry}
is important for the next step: we use the fact that for a symmetry
\be
\label{symmetrystatement1}
\raisebox{.09cm}{$\underset{\substack{{\scriptscriptstyle
        \phi(\tau_f)=\varphi_f + \delta\vp_f} \\
      {\!\scriptscriptstyle\phi(\tau_i) =\varphi_i + \delta \vp_i}}}{\displaystyle\int}$}   \hspace{-.4cm}
{\cal D}\phi \,e^{iS[\phi]} =
\raisebox{.09cm}{$\underset{\substack{{\scriptscriptstyle \phi(\tau_f)=\varphi_f} \\ {\scriptscriptstyle\phi(\tau_i) =\varphi_i}}}{\displaystyle\int}$}   \hspace{-.4cm}
{\cal D}\phi \,e^{iS[\phi]}  \, .
\ee
That this is true can be seen by changing the variable of
integration as $\phi \mapsto \phi + \delta\phi$, 
and using the symmetry statement $S[\phi+\delta\phi] = S[\phi]$
(assuming again the invariance of the measure).\footnote{In general, the symmetry transformation could depend on
  time {\it i.e.}, $\phi(\tau,\vec x) \mapsto \phi(\tau,\vec x) + \delta
  \phi(\tau,\vec x)$, in which
  case $\delta\vp_f$ and $\delta \vp_i$ are the the corresponding transformations at the
  final time $\tau=\tau_f$ and initial time $\tau=\tau_i$, respectively. }
A succinct way to make the same symmetry statement is:
\be
\label{symmetrystatement2}
\langle \vp_f + \delta\vp_f | \vp_i + \delta\vp_i \rangle = \langle
\vp_f | \vp_i \rangle \, .
\ee
Combining eqs.~\eqref{PsiVarphifDelta} and~\eqref{symmetrystatement1}
gives the first order variation:
\begin{equation}
\label{deltaPsif}
\delta\Psi[\vp_f] = \int{\cal D}\vp_i \langle\vp_f\rvert\vp_i\rangle \delta\Psi_0[\vp_i] .
\end{equation}

To proceed further, we use the explicit form of the initial vacuum
wavefunctional,~\eqref{eq:gaussianwavef}. This implies
\begin{equation}
\label{deltaPsi0}
 \delta\Psi_0[\vp_i]  =
 -\frac{1}{2}\int\frac{\rd^3p}{(2\pi)^3}{\cal E}_i (p) \left[
   \vp_i(\vec{p}) \, \delta
   \vp_i(-\vec{p}) + \delta \vp_i(\vec{p}) \, \vp_i(-\vec{p})\right]
 \Psi_0[\vp_i].
\end{equation}
Recall that $\delta\vp_i = \delta_{\rm NL} \vp_i + \delta_{\rm L}
\vp_i$. We are interested in cases where the nonlinear part of the
transformation does not vanish at spatial infinity {\it i.e.}, in Fourier
space, $\delta_{\rm NL} \vp_i$ takes the form
\be
\label{Dqdef}
\delta_{\rm NL} \vp_i (\vec q) = \left[ (2\pi)^3
\delta (\vec q) \right]{D}_{-\vec q} (\tau_i) \, ,
\ee
where ${D}_{-\vec q} (\tau_i)$ is an operator that in general depends
on the momentum $\vec q$, and possibly on time as well. 
In general, we expect $\delta_{\rm NL} \vp_i (\vec q)$ to contain
derivatives of the delta function; the above expression implicitly
assumes an integration by parts has been performed.
For instance, under SCT $\delta_{\rm NL}\vp_i (\vec x) = 2 \vec b \cdot \vec
x$, implying $\delta_{\rm NL} \vp_i (\vec q) = - 2 i \vec b \cdot \vec
\nabla_{q} [(2\pi)^3 \delta (\vec q)]$ and following
eq.~\eqref{Dqdef}, we write this as 
$[(2\pi)^3 \delta (\vec q)] (2i \vec b \cdot \vec\nabla_q)$
{\it i.e.}, $D_{-\vec q} (\tau_i) = 2i \vec b \cdot \vec\nabla_q$ for SCT.
The case of dilation, where $\delta_{\rm NL} \vp_i (\vec x) = \lambda$, involves
no integration by parts, and so $D_{-\vec q} (\tau_i) = \lambda$. 
With this definition, eq.~\eqref{deltaPsi0} can be rewritten as:
\be
\label{deltaPsi0b}
 \delta\Psi_0[\vp_i]  =
- D_{\vec q} (\tau_i) \left[ {\cal E}_i (q) \vp_i ({\vec q})\right] \Big|_{\vec
  q = 0} \Psi_0[\vp_i]
 -\int\frac{\rd^3p}{(2\pi)^3}{\cal E}_i (p) \left[
   \vp_i(\vec{p}) \,\delta_{\rm L}
   \vp_i(-\vec{p})\right]
\Psi_0[\vp_i] \, ,
\ee
where we have assumed ${\cal E}_i (p)$ does not depend on the
direction of $\vec p$. 
Substituting this into~\eqref{deltaPsif} and~\eqref{eq:wardidunsimplfied}, we arrive at:
\begin{eqnarray}
\label{eq:wald00}
&& -\int{\cal D}\vp_f{\cal O}_{\vp_f}(\vec k_1,\cdots, \vec
  k_N)\Psi^*[\vp_f]\left(\int{\cal D}\vp_i
  \langle\vp_f\rvert\vp_i\rangle\left(\int\frac{\rd^3p}{(2\pi)^3}
{\cal E}_i (p) \vp_i(\vec p)
\delta_{\rm L}\vp_i(-\vec p)\right)\Psi_0[\vp_i]\right)
\nonumber \\ 
&& - \, {D}_{\vec q} (\tau_i)\left({\cal E}_i(q) \int{\cal D}\vp_f{\cal
  O}_{\vp_f}(\vec k_1,\cdots, \vec k_N)\Psi^*[\vp_f]\int{\cal D}\vp_i
  \langle\vp_f\rvert\vp_i\rangle\vp_i(\vec{q})
  \Psi_0[\vp_i]\right)\bigg\rvert_{\vec q =0}  + ``{\rm c. c.}"
\nonumber \\
&& + \, \langle\delta{\cal O}_{\vp_f}(\vec k_1,\cdots, \vec
     k_N)\rangle = 0 \, ,
\end{eqnarray}
where we put the complex conjugate of the left hand side in quotes
$``{\rm c. c.},"$ meaning all terms should be conjugated except ${\cal
  O}_{\vp_f}$. 
This expression can be simplified somewhat by removing complete sets of states to
write things back in terms of Heisenberg picture operators:\footnote{To derive the the second term on the left,
we assume $D_{\vec q} = D_{-\vec q}^*$, motivated by
the fact that $D_{\vec q}$ generally consists of derivatives with
respect to $\vec q$ and each derivative carries an $i$. 
We also assume $\langle 0_{\rm in} | {\cal O}_{\phi_f} (\vec k_1,
\cdots, \vec k_N) \phi_i (\vec q) | 0_{\rm in} \rangle = \langle 0_{\rm in} | {\cal O}_{\phi_f} (-\vec k_1,
\cdots, -\vec k_N) \phi_i (-\vec q) | 0_{\rm in} \rangle$ which follows
from symmetry under parity, and that ${\cal O}$ and $\phi$ are
hermitian operators in real space, such that
$\langle 0_{\rm in} | {\cal O}_{\phi_f} (\vec k_1,
\cdots, \vec k_N) \phi_i (\vec q) | 0_{\rm in} \rangle^* = 
\langle 0_{\rm in} | \phi_i (-\vec q) {\cal O}_{\phi_f} (-\vec k_N,
\cdots, -\vec k_1) | 0_{\rm in} \rangle$. Note the ${\rm c. c.}$ in
this term is actual complex conjugation ({\it i.e.}, not in quotes as in
$``{\rm c. c.}"$).
}
\begin{align}
\label{eq:almostfinalward}
\tiny
& {\cal V} + {D}_{\vec q}(\tau_i)\left({\cal E}_i (q) \langle 0_{\rm in} | {\cal
  O}_{\phi_f}(\vec k_1,\cdots,\vec k_N)\phi_i(\vec{q}) | 0_{\rm in}
   \rangle + {\rm c. c.} \right)\bigg\rvert_{\vec q =0}
= \langle 0_{\rm in} | \delta{\cal O}_{\phi_f}(\vec k_1,\cdots,\vec k_N)
  | 0_{\rm in} \rangle \, ,
\end{align}
where we have written the field insertions as $\phi_a(\vec k) \equiv
\phi(\tau_a,\vec k)$ to emphasize the distinction from the $\vp$'s, which
are functions, whereas the $\phi$'s are operators.
The term ${\cal V}$ is defined by
\begin{eqnarray}
\label{Vdef}
\tiny
&& {\cal V} \equiv \int\frac{\rd^3p}{(2\pi)^3} {\cal E}_i (p) 
\Big[ \langle 0_{\rm in} | {\cal O}_{\phi_f}(\vec k_1,\cdots,\vec k_N) \phi_i(\vec p) \delta_{\rm L} \phi_i(-\vec p) |
  0_{\rm in} \rangle 
\nonumber \\
&&
\quad \quad \quad \quad - \langle 0_{\rm in} | {\cal O}_{\phi_f}(\vec
   k_1,\cdots,\vec k_N) | 0_{\rm in} \rangle 
\langle 0_{\rm in} | \phi_i(\vec p) \delta_{\rm L} \phi_i(-\vec p) |
  0_{\rm in} \rangle \Big] + \, ``{\rm c. c.}" \, .
\end{eqnarray}
Equation~\eqref{eq:almostfinalward}, with ${\cal V}$ defined above,
does not strictly follow from eq.~\ref{eq:wald00}.
We have introduced a correction
to account for the fact that the integration measure could change
by an overall normalization under the transformation in question---this is the origin of the $\langle 0_{\rm in} | {\cal O}_{\phi_f} |
0_{\rm in} \rangle \langle 0_{\rm in} | \phi_i \delta_{\rm L} \phi_i
| 0_{\rm in} \rangle$ term in ${\cal V}$
(see Appendix \ref{app:4point} for details).

There are several terms in eq.~\eqref{eq:almostfinalward}: on the right hand
side is an $N$-point function, and on the left is an $(N+2)$-point
function (the term ${\cal V}$) followed by an
$(N+1)$-point function. 
There are several further subtleties we must address before reducing
this Ward identity to its final form. 
First, it can be shown that the $(N+2)$-point term vanishes in the far
past limit for $\tau_i$ ({\it i.e.}, $\tau_i \rightarrow -\infty$). 
Essentially, the reason is
that once the term is rewritten in interaction-picture, it can be seen
that each $\phi_i (\vec p)$ carries an oscillating factor of $e^{\pm i p
  \tau_i}$ 
which in the far past becomes infinitely oscillating, leading
to the vanishing of the integral over $\vec p$.\footnote{There is,
  potentially, a non-vanishing contribution from disconnected pieces
  in the limit where some subset of the hard momenta add up to zero
  ({\it i.e.}, an internal collapsed limit). The possibility of having
  an internal collapsed limit can only be probed by studying a
  squeezed $4$-point function or higher i.e. $N+1 \ge 4$ (in this paper, all
  explicit perturbative checks are done at the level of a squeezed $3$-point
  function).
  Thus, one should keep in mind
  there is a technical assumption in the consistency relation we 
  derive by dropping ${\cal V}$: that we do not
  have a squeezed $N+1$-point function where a subset of the hard
  modes have their momenta summed to zero. 
  See Appendix~\ref{app:4point} for a more complete discussion.} There is the
possibility that the oscillating factors in $\phi_i (\vec p)$ and
$\delta_{\rm L} \phi_i (-\vec p)$ (the latter is generally proportional to
$\phi_i (-\vec p)$) cancel, but that possibility is precluded by the
subtraction of the particular combination of disconnected pieces in
${\cal V}$.
A more detailed discussion can be found in Appendix \ref{app:4point}.
It is worth pointing out that, in contrast, the $(N+1)$-point term does not vanish
because it arises from the nonlinear part of the transformation which
carries a delta function at zero momentum. In other words, the $(N+1)$-point term should be
understood as the result of the following limiting procedure: $\vec q
\rightarrow 0$, $\tau_i \rightarrow -\infty$ and $q \tau_i \rightarrow
0$. We emphasize that if $\tau_i$ is not taken to be in the far past,
this $(N+2)$-point term is really present and should be accounted for.
This is not easy to see in an explicit (perturbative) quantum mechanical computation,
where typically an $i\epsilon$ prescription together with $\tau_i
\rightarrow -\infty$ is used to project onto the desired
vacuum. In a classical computation, however, it is possible to
perturbatively verify the analogue
of eq.~\eqref{eq:almostfinalward}, and one can see explicitly the
presence of the ${\cal V}$ term,
unless a large separation limit between $\tau_f$ and $\tau_i$ is taken.
This is discussed in Appendix \ref{app:classicalcheck}.

\begin{figure}
   \centering
   \includegraphics[scale=0.6]{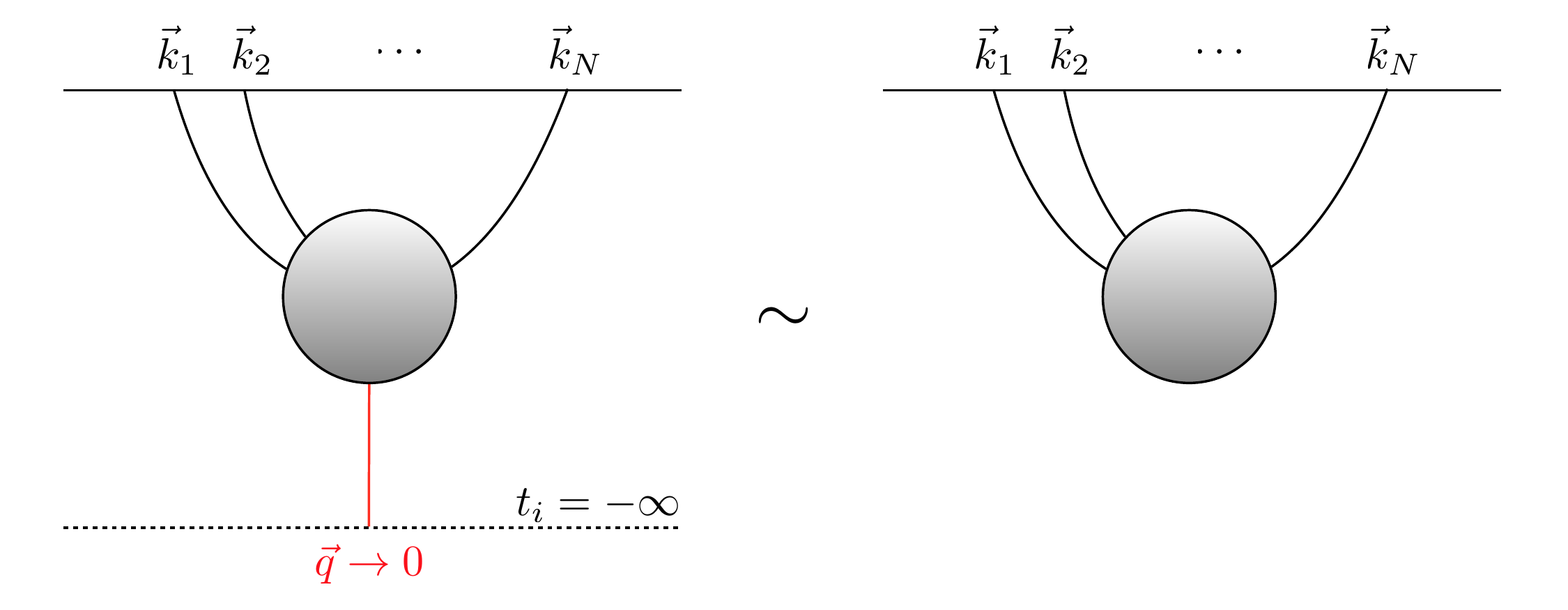}
   \caption{\small The new identity relates $(N+1)$ point functions, with the soft mode inserted in the far past, to $N$ point functions. }
   \label{fig:1}
\end{figure}

Second, the remaining terms, the $(N+1)$-point function on the left and
$N$-point function on the right, can be re-written as connected
correlation functions, provided $\delta {\cal O}_{\phi_f}$ on the
right is replaced by $\delta_{\rm L} {\cal O}_{\phi_f}$, that is, the
linear part of the transformation. That this can be done in general
was shown in Appendix C of \cite{Hinterbichler:2013dpa}.
Thus, we have:
\begin{eqnarray}
\label{eq:almostfinalward2}
{D}_{\vec q}(\tau_i)\left({\cal E}_i (q) \langle 0_{\rm in} | {\cal
  O}_{\phi_f}(\vec k_1...\vec k_N)\phi_i(\vec{q}) | 0_{\rm in}
   \rangle_c  + {\,\rm c. c.} \right)\bigg\rvert_{\vec q =0}
= \langle 0_{\rm in} | \delta_{\rm L} {\cal O}_{\phi_f}(\vec k_1...\vec k_N)
  | 0_{\rm in} \rangle_c \, ,
\end{eqnarray}
where the subscript $c$ denotes the connected part of the correlation function.
For practical applications, it is useful to remove the overall
momentum conserving delta functions. Removing delta functions
for a particular consistency relation generally depends on the
consistency relation at a lower level---for instance, removing
delta functions for the SCT consistency relation (where the linear
part of the transformation scales roughly as $x^2$) relies on the
dilation consistency relation (where the nonlinear part of the
transformation scales as $x$) and rotational invariance; the same holds for symmetry
transformations that go as $x^n$ \cite{Hinterbichler:2013dpa}. 
Assuming this structure, we arrive at the following early-late-time
consistency relation:
\vspace{.15cm}
\begin{tcolorbox}[colframe=white,arc=0pt,colback=greyish2]
\vspace{-.42cm}
\begin{eqnarray}
\label{eq:mainID}
{D}_{\vec q} (\tau_i)\left({\cal E}_i (q) \langle 0_{\rm in} | {\cal
  O}_{\phi_f}(\vec k_1...\vec k_N)\phi_i(\vec{q}) | 0_{\rm in}
   \rangle_c {}' + {\,\rm c. c.} \right)\bigg\rvert_{\vec q =0} 
= \langle 0_{\rm in} | \delta_{\rm L} {\cal O}_{\phi_f}(\vec k_1...\vec k_N)
  | 0_{\rm in} \rangle_c {}'\, ,
\end{eqnarray}
\end{tcolorbox}
\noindent where the prime ${}'$ denotes a correlation function with the delta
function removed. Figure~\ref{fig:1} gives a pictorial representation of this relation between an $(N+1)$-point function and an $N$-point function. On the left hand side of this relation, the operator
$D_{\vec q}$ (defined in~\eqref{Dqdef}, related to the nonlinear part
of the transformation)
acts on the soft momentum $\vec q$, with the
understanding that $\vec k_N$ also has $\vec q$ dependence
because $\vec k_N = - \vec q - \vec k_1 - \cdots -\vec k_N$. 
On the right hand side of the consistency relation,
the $\delta_{\rm L}$ operator (the linear part of the transformation)
should be understood as acting on
$N-1$ momenta {\it i.e.},  $\vec k_N$ is not independently varied and is
instead treated as $\vec k_N = - \vec k_1 - \vec k_2- \cdots - \vec
k_{N-1}$, while $\vec k_1, \cdots, \vec k_{N-1}$ are independently
varied. 
More explicitly, for the dilation symmetry we have:
\be
D_{\vec q} (\tau_i) = \lambda, \qquad \qquad 
 \delta_{\rm L} =
\sum_{a=1}^{N-1} -\lambda (3 + \vec k_a \cdot \vec \nabla_{k_a}) \, ,
\ee
where $\lambda$ is a transformation parameter that can be removed from both
sides of the consistency relation.
For special conformal transformations:
\be
D_{\vec q} (\tau_i) = - 2 i \vec b \cdot \vec \nabla_{q}  , \qquad
\delta_{\rm L} = \sum_{a=1}^{N-1} 
i \left( 6 \vec b \cdot \vec \nabla_{k_a} - (\vec b \cdot \vec k_a)
  \nabla^2_{k_a} + 2 (\vec k_a \cdot \vec \nabla_{k_a}) (\vec b \cdot
  \vec \nabla_{k_a}) \right) \, ,
\ee
where $\vec b$ can be removed from both sides as well.

We have thus derived a soft theorem for correlation functions where
the soft mode is inserted at the initial time while the hard modes are
at some later time.
Note that the assumptions we have made in deriving this generalized
consistency relation are
minimal, it essentially follows from invariance of the action under a
(nonlinearly realized) symmetry transformation and the fact that the
path integral is insensitive to changes of integration
variables. Consequently, we expect this Ward identity to hold in a wider
range of situations than the standard consistency relation.
For applications to inflation, the symmetry transformations originate as
large diffeomorphisms. The adiabatic mode condition---that the large
diffeomorphism generates a fluctuation that is the long wavelength
limit of a physical mode---must be obeyed for
a large diffeomorphism to be treated as a bona fide nonlinearly
realized symmetry, a point we will discuss in detail in Section~\ref{sec:inflation}.

\subsection{The physical mode condition and the late-time Ward identity}
\label{sec:physicalmodes}

Our main task in this section is to explicate under what condition
the early-late-time consistency relation \eqref{eq:mainID} implies
the standard consistency relation where all modes including the soft
one are at late times.

We are interested in the fluctuation $\phi$ which generally
obeys a second order equation of motion so that there are typically two solutions. Suppose the dominant mode (usually called the growing mode)
has the time dependence $g(\tau)$ in the $\vec q \rightarrow 0$ limit. A more
precise quantum statement is (condition 1):
\be
\label{condition1}
\lim_{\vec q \rightarrow 0} \,\, \phi_i (\vec q) |0_{\rm in} \rangle = \phi_f (\vec q) |0_{\rm in}
\rangle {g(\tau_i) \over g(\tau_f)} \, ,
\ee
or, applying it to the correlation function of interest:
\be
\label{condition1b}
\langle 0_{\rm in} | {\cal O}_{\phi_f} (\vec k_1 ... \vec k_N)
\phi_i (\vec q) |0_{\rm in} \rangle_c {}' = 
\langle 0_{\rm in} | {\cal O}_{\phi_f} (\vec k_1 ... \vec k_N)
\phi_f (\vec q) |0_{\rm in} \rangle_c {}' \, {g(\tau_i) \over g(\tau_f)} 
+ \Delta(q)\, ,
\ee
where $\Delta(q)$ denotes a correction that is suppressed relative
to the first term on the right by some positive powers of $q$.
Note that $\Delta(q)$ is in general complex, as is the left hand side of
this equation,
but the equal (final-time) correlation on the right is real (at least
in the case where $O_{\phi_f}$ consists of the product of a number of
$\phi$'s).
In other words, the real part of $\langle {\cal O}_{\phi_f} \phi_i
\rangle_c {}'$ is dominated by $\langle {\cal O}_{\phi_f} \phi_f
\rangle_c {}' g(\tau_i)/g(\tau_f)$.

The quantity ${\cal E}_i (q)$ that shows up in the consistency
relation~\eqref{eq:mainID} originates as a kernel that defines the
gaussian initial wavefunctional \eqref{eq:gaussianwavef}.
It is in general complex. Let us write it as
\be
{\cal E}_i (q) = {1 \over 2 P_{\phi_i} (q)} + i {\,\rm Im\,} {\cal
  E}_i (q) \, ,
  \label{eq:wavefcoeffcomplex}
\ee
where we have used \eqref{Pvarphii}.
We are interested in
this quantity in the limit $q \rightarrow 0$, 
$\tau_i \rightarrow -\infty$ and $q \tau_i \rightarrow 0$.
In the context of inflation (either normal or ultra slow-roll), in
this limit, the imaginary part dominates: $|{\rm Im\,} {\cal E}_i (q)| \gg |{\rm Re\,}{\cal E}_i
(q) |$. The large imaginary part is related to the fact the wavefunction
approaches a squeezed state in the long wavelength limit
(see~\cite{Guth:1985ya} for a discussion). 
On the other hand, in a flat-space toy example introduced below, 
${\cal E}_i (q)$ is purely real. It is useful to  keep in mind these
different possibilities as we proceed. 

A corollary of condition 1 is that, in the soft limit, the late time power spectrum
and the early time power spectrum are related:
\begin{eqnarray}
\label{condition1c}
\lim_{\vec q \rightarrow 0} \,\, P_{\phi_i} (q) = P_{\phi_f} (q)
  \left( {g(\tau_i) \over g(\tau_f)} \right)^2 \, .
\end{eqnarray}
Alternatively this can be phrased as a small $q$ statement,
analogous to \eqref{condition1b}:
\begin{eqnarray}
\label{condition1d}
P_{\phi_i} (q) = P_{\phi_f} (q)
  \left( {g(\tau_i) \over g(\tau_f)} \right)^2+ \tilde\Delta(q) \, ,
\end{eqnarray}
where $\tilde\Delta(q)$ is suppressed compared to the first term on the
right by positive powers of $q$. Here $\tilde\Delta(q)$ is real.

Next we introduce the physical mode condition (condition 2):
\be
D_{\vec q} (\tau_i) = D_{\vec q} (\tau_f) {g (\tau_i) \over g
  (\tau_f)} \, .
\ee
Recall that $D_{\vec q} (\tau)$ has to do with the nonlinear part of
the transformation~\eqref{Dqdef}. The condition is that the nonlinear
part of the transformation depends on time in the same way
as the dominant mode exhibited in~\eqref{condition1}.
Thus the physical mode condition can be stated as:
\vspace{.15cm}
\begin{tcolorbox}[colframe=white,arc=0pt,colback=greyish2]
{\bf Physical mode condition:} {\it that the non-linear part of the
  transformation has the same time dependence as the
zero momentum dominant (growing) mode of the field.}
\end{tcolorbox}
This should be distinguished from the adiabatic mode condition,
which stipulates that the large diffeomorphism generates a mode
(the nonlinear part of its transformation) that is the long wavelength
limit of a physical mode. Under the adiabatic mode condition, there's no
requirement or guarantee that the relevant physical mode is the
dominant ({\it i.e.}, growing) mode. 
We will have more to say about the distinction between the
two conditions in the next section.

Putting together conditions 1 and 2, the left hand side of
the early-late-time consistency relation \eqref{eq:mainID} can be
rewritten as:
\begin{eqnarray}
\tiny
&& {D}_{\vec q} (\tau_i)\left({\cal E}_i (q) \langle 0_{\rm in} | {\cal
  O}_{\phi_f}(\vec k_1...\vec k_N)\phi_i(\vec{q}) | 0_{\rm in}
   \rangle_c {}' + {\,\rm c. c.} \right)
\nonumber \\
&& = D_{\vec q} (\tau_f) {g(\tau_i) \over g(\tau_f)} 
\Big[ \left( {1\over
   2 P_{\phi_f} (q)}  \left({g(\tau_f) \over g(\tau_i)}\right)^2 +
   \tilde\Delta(q) + i
   {\,\rm Im\,} {\cal E}_i (q) \right)
\nonumber \\ && \quad \quad \quad \quad \quad \quad \quad 
\left(
\langle 0_{\rm in} | {\cal
  O}_{\phi_f}(\vec k_1...\vec k_N)\phi_f(\vec{q}) | 0_{\rm in}
   \rangle_c {}' {g(\tau_i) \over g(\tau_f)} + \Delta(q) \right)
+
   {\rm c. c.} \Big].
\end{eqnarray}
A third condition for the standard consistency relation to hold is
that (condition 3)
\begin{eqnarray}
{\cal E}_i (q) \langle 0_{\rm in} | {\cal O}_{\phi_f} (\vec k_1 ... \vec
  k_N) \phi_i (\vec q) | 0_{\rm in} \rangle_c {}'
= \left( {\rm Re\, } {\cal E}_i (q) \right) \left( {\, \rm Re\,} \langle 0_{\rm in} | {\cal O}_{\phi_f} (\vec k_1 ... \vec
  k_N) \phi_i (\vec q) | 0_{\rm in} \rangle_c {}' \right) + \bar \Delta(q)
\end{eqnarray}
with $\bar\Delta(q)$ a small correction that is suppressed by powers
of $q$. It is not sufficient that $\bar\Delta(q)$ becomes negligible as
$q$ goes to zero. If $D_{\vec q}$ carries $n$ derivatives with respect
to $q$ (for instance $n=0$ for dilation and $n=1$ for SCT), we must
have $\bar\Delta(q)$ going as a steeper power of $q$ than $q^n$. In other
words, we need
\begin{eqnarray}
\lim_{\vec q \rightarrow 0} D_{\vec q} (\tau_i) \bar\Delta(q) = 0 \, .
\end{eqnarray}

\begin{figure}
   \centering
   \includegraphics[scale=0.6]{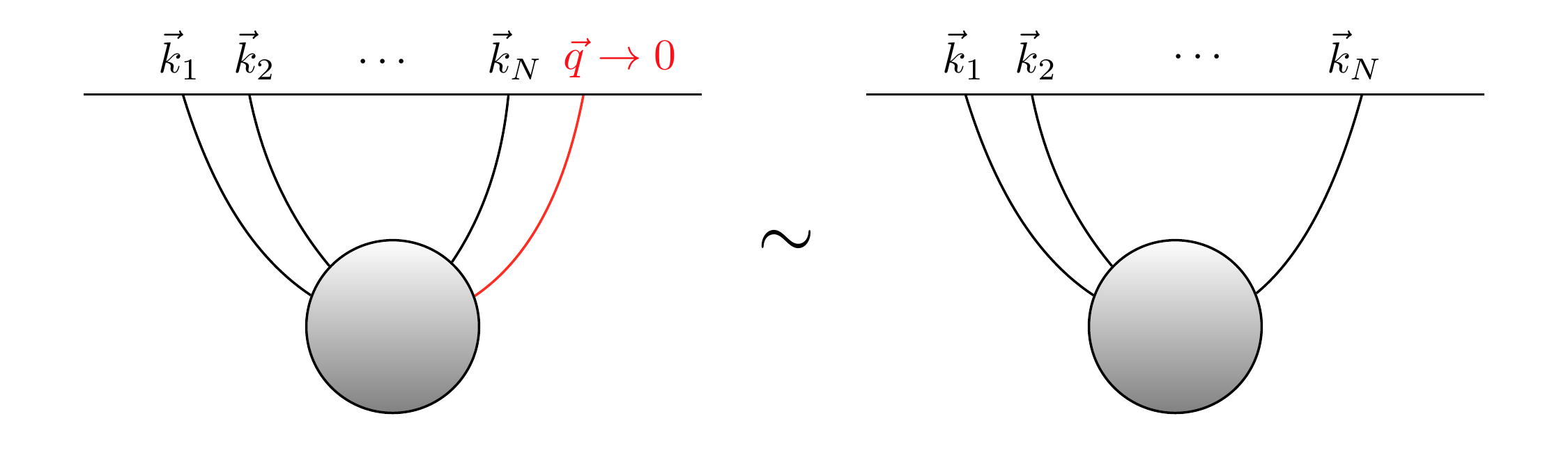}
   \caption{\small The usual soft theorems relates a squeezed $(N+1)$-point equal time correlator to an $N$-point correlator.}
   \label{fig:2}
\end{figure}

With all three conditions satisfied, we arrive at the standard
consistency relation where all modes are at late times:
\be
\label{finaltimeWard}
\left.D_{\vec q} (\tau_f)
\left( {1\over
   P_{\phi_f} (q)}  
\langle 0_{\rm in} | {\cal
  O}_{\phi_f}(\vec k_1...\vec k_N)\phi_f(\vec{q}) | 0_{\rm in}
   \rangle_c {}' \right)\right\rvert_{\vec q = 0} = \langle 0_{\rm in} | \delta_{\rm L} {\cal O}_{\phi_f}(\vec k_1...\vec k_N)
  | 0_{\rm in} \rangle_c {}' \, .
\ee

We give a schematic representation of this relation in Figure~\ref{fig:2}.
It is worth stressing that the conditions we have laid out for
going from the early-late-time consistency relation \eqref{eq:mainID}
to the standard, late-time one \eqref{finaltimeWard} are merely
sufficient conditions.
We do not know if they are necessary. 
The physical mode condition makes intuitive sense; condition 3 on the
other hand seems a bit obscure.
It is satisfied by normal slow-roll inflation. This is non-trivial:
in the small $q$ limit, ${\cal E}_i (q)$ is in fact dominated by its
imaginary part, and there's the danger that the contribution
from the product ${\,\rm Im\,} {\cal E}_i (q)  \, {\,\rm Im
  \,} \Delta(q) $ might be larger than, or comparable to, the combination that shows up in the
  standard consistency relation. This does not happen in normal
  slow-roll, but we will see that it happens in a simplified model of
  ultra-slow-roll. 
  In Appendix~\ref{sec:pertproof}, we give a perturbative derivation of the late-time soft theorems which uses the physical mode condition in an essential way, but where condition 3 is automatically satisfied, indicating that the physical mode condition is the more physically meaningful assumption.
  We will also see that in class of toy
  models we consider, there are examples where condition 3 holds, with $\bar\Delta(q)$
  vanishes fast enough (with $q$) for the standard dilation
  consistency relation to hold, but not fast enough for the standard
  SCT consistency relation.
  
We therefore see that the early-late time consistency relation requires only the adiabatic mode condition in order to be valid, whereas the late-time consistency relation further requires the physical mode condition. This makes it clear that these are two distinct requirements.

\section{Application to inflation}
\label{sec:inflation}
The derivation of Section~\ref{sec:generalderivation} gives us a new perspective on soft theorems, and we would now like to apply these general lessons to the inflationary universe. In particular we will focus on single-field inflation where we elucidate the connection between the physical mode condition we have introduced and the construction of adiabatic modes. We then verify that the improved consistency relation that we have derived~\eqref{eq:mainID} holds in attractor models of inflation. Next, we turn to considering the interplay between the physical mode condition and ultra-slow-roll inflation, which clarifies the violation of Maldacena's consistency relation in this case, and we argue that our improved identity should hold even in this case.

\subsection{Adiabatic modes and symmetries}
\label{sec:infsymms}
Our derivation of the soft theorem~\eqref{eq:mainID} utilized a change of variables in the path integral corresponding to a symmetry of the classical action. One might rightly wonder if our derivation applies to {\it gauge symmetries}, which are the relevant transformations in the inflationary context. Here we briefly review how residual large gauge transformations act effectively as global symmetries of the gauge-fixed action. It is then clear that the derivation presented in Section~\ref{sec:generalderivation} applies. A crucial ingredient in translating between large gauge transformations and global symmetries is the so-called adiabatic mode condition---we will explain how this condition is related to, but logically distinct from, the physical mode condition introduced in the previous section.

For concreteness, we focus on single-field inflationary scenarios where the matter driving the cosmological background can be described by a scalar field of the $P(X,\phi)$ type,
\be
S = \int\rd^4x\sqrt{-g}\left(\frac{M_{\rm Pl}^2R}{2} + \Lambda^4P(X,\phi)\right),
\ee
where $X\equiv -\frac{1}{2\Lambda^4}(\partial\phi)^2$ is the canonical kinetic term. We assume that the functional form of $P$ is such that a classical background solution exists where the scalar field is homogeneous in space and the metric is of FLRW form
\be
\rd s^2 = -\rd t^2 +a(t)^2 \rd\vec x^2,
\ee
with $a(t)$ the cosmological scale factor. The derivative of the scale factor is related to the  Hubble parameter, $H = \dot a/a$.

In order to analyze fluctuations around the homogeneous background, we parameterize the metric using ADM variables
\be
\rd s^2 = -N^2 \rd t^2 +h_{ij}(\rd x^i+N^i\rd t)(\rd x^j+N^j\rd t).
\ee
and perturb the scalar field as $\phi = \bar\phi(t)+\vp$.
Under a diffeomorphism, with parameter $\xi^\mu = (\xi,\xi^i)$, the fields transform as
\begin{subequations}
\begin{align}
\delta h_{ij} &= \nabla_i\xi_j+\nabla_j\xi_i+\xi\dot h_{ij}+N_i\partial_j\xi+N_j\partial_i\xi, \\
\delta N^i &= \xi^j\partial_jN^i-\partial_j\xi^i N^j+\frac{\rd}{\rd t}\left(\xi N^i\right)+\dot\xi^i-\left(N^2 h^{ij}+N^iN^j\right)\partial_j\xi, \\
\delta N &= \xi^i\partial_iN+\frac{\rd}{\rd t}\left(\xi N\right)-NN^i\partial_i\xi ,\\
\delta\phi &= \xi^\mu\partial_\mu\phi.
\end{align}
\label{eq:difftransforms}
\end{subequations}
It is convenient to fix this gauge freedom by going to so-called $\zeta$-gauge, which is defined by having an unperturbed scalar field profile, $\vp = 0$, and a spatial metric of the form
\be
h_{ij} =a^2 e^{2\zeta}(e^\gamma)_{ij}~~~~~{\rm with}~~~\gamma^i_i = \partial_i\gamma^i_j = 0.
\ee
In this formulation, the lapse, $N$, and shift, $N_i$, are Lagrange multipliers which enforce constraints. To obtain an action for only the propagating degrees of freedom, we solve these constraints perturbatively and substitute back into the action~\cite{Maldacena:2002vr}.

The equation of motion for the shift vector is the so-called momentum constraint, to first order it takes the form
\be
\partial_i\left(HN^{(1)}-\dot\zeta\right) - \frac{1}{4a^2}\nabla^2N_i^T = 0,
\label{eq:momconst}
\ee
where $N^{(1)}$ is the linear lapse perturbation $N = 1+N^{(1)}+\cdots$, and we have split
the shift into a transverse and a longitudinal piece: $N_i = N_i^T +\partial_i\psi$, with $\partial^i N_i^T = 0$. Variation with respect to the lapse enforces the hamiltonian constraint\footnote{The sound speed of fluctuations is defined in terms of the function $P$ as
\be
c_s^2 = \frac{P_{,X}}{P_{,X}+2P_{,XX}}.
\ee
}
\be
-a^{-2}\partial_i\left(\partial^i\zeta+H\partial^i\psi\right)+\frac{H^2\epsilon}{c_s^2}N^{(1)}-3H\left(HN^{(1)}-\dot\zeta\right) = 0.
\label{eq:hamconst}
\ee
Here $\epsilon$ is the first slow-roll parameter: $\epsilon\equiv - \dot H/H^2$.
Solving the equations~\eqref{eq:momconst} and~\eqref{eq:hamconst} for $N, N_i$ in terms of $\zeta$ and substituting back into the action we obtain an action purely for the physical degrees of freedom $\zeta$ and $\gamma_{ij}$.

We can now ask whether this action possesses any symmetries. In fact, it has an infinite number, which are remnants of the gauge invariances of the original action~\cite{Hinterbichler:2012nm}. Though $\zeta$-gauge completely fixes all diffeomorphisms that go to zero at spatial infinity (small gauge transformations), there is an infinite number of residual gauge transformations which preserve $\zeta$-gauge, but which change the asymptotic form of the metric (large gauge transformations). As we will see, these transformations act nontrivially on states in the theory, and therefore really deserve to be thought of as global symmetries.\footnote{In a sense, the fields $\zeta$ and $\gamma_{ij}$ can be thought of as goldstone modes for the spontaneous breaking of these symmetries by the FLRW background. This perspective is closely related to the study of asymptotic symmetries in flat space~\cite{Bondi:1962px,Sachs:1962wk,Strominger:2013jfa} and in de Sitter~\cite{Anninos:2010zf,Ferreira:2016hee,Hinterbichler:2016pzn}.} The majority of these transformations mix the scalar and tensor perturbations (and therefore give rise to Ward identities relating scalar and tensor correlation functions), but a subset act only in the scalar sector~\cite{Hinterbichler:2013dpa}. We will focus on these.

Explicitly, we look for residual diffeomorphisms that keep us in $\zeta$-gauge. Gauge transformations induce particular profiles for the fields, and we need to ensure that these are consistent with the constraint equations that we have used. At lowest order in the gauge parameter, under a general diffeomorphism the fields experience the following nonlinear transformations
\begin{align}
\delta_{\rm NL}\vp  &= \dot{\bar\phi}\xi\\
\label{eq:nonlinearNtrans}
\delta_{\rm NL} N^{(1)} &= \dot\xi\\
\delta_{\rm NL} N^i &= \dot\xi^i-a^{-2}\partial^i\xi\\
\delta_{\rm NL}\zeta &= \frac{1}{3}\partial_i\xi^i+H\xi\\
\delta_{\rm NL}\gamma_{ij} &= \partial_i\xi_j+\partial_j\xi_i-\frac{2}{3}\partial_k\xi^k\delta_{ij}.
\label{eq:gammatrans}
\end{align}

We now restrict the possible $\xi^\mu$ by demanding that they keep us in $\zeta$-gauge. In this gauge the tensor perturbation is transverse---for this to continue to be true, we must have
\be
\partial^i\delta\gamma_{ij} = \nabla^2\xi_j+\frac{1}{3}\partial_j\partial^i\xi_i = 0.
\label{eq:divtensor}
\ee
This is nothing but the divergence of the conformal Killing equation for ${\mathbb R}^3$, so any diffeomorphism which is a conformal Killing vector will preserve traceless-ness of $\gamma_{ij}$ and will lead to a shift of $\zeta$ by~\cite{Hinterbichler:2012nm}
\be
\delta\zeta = \frac{1}{3}\partial_i\xi^i+\xi^i\partial_i\zeta.
\ee
As an example, a time-dependent dilation is such a symmetry
\be
\delta\zeta = \lambda(t) \left(1+\vec x\cdot\vec\nabla\zeta\right),
\label{eq:timedepdilation}
\ee
and so we might be tempted to posit that it is a symmetry of the cubic $\zeta$-action. This is however too quick. This statement is correct before we integrate out the lapse and shift, where $N$ and $N_i$ transform as in~\eqref{eq:difftransforms}, but once we integrate them out, the situation is more constrained. We are not allowed to do diffeomorphisms which generate arbitrary lapse and shift profiles, but only ones which generate profiles for $\{\zeta, N^{(1)}, N^i\}$ which are compatible with the constraint equations~\eqref{eq:momconst}~\&~\eqref{eq:hamconst}.

This constraint is subtle: when we solve~\eqref{eq:momconst}~\&~\eqref{eq:hamconst} for the lapse and shift, we invert the spatial laplacian. This makes the requirement that diffeomorphisms preserve the constraints nontrivial. In order for a diffeomorphism to be a symmetry of the $\zeta$ action once we have integrated out the lapse and shift, it cannot solve the constraints in the trivial way of being annihilated by spatial derivatives. In a sense diffeomorphisms generate infinite-wavelength field profiles, and it is necessary for these profiles to be the $\vec q\to 0$ limit of a profile which itself solves the constraints at finite momentum in order to be off-shell symmetries of the action~\cite{Weinberg:2003sw}. This guarantees that the profile generated by a gauge transformation is smoothly connected to a physical solution to the equation of motion. This requirement is referred to as the {\it adiabatic mode condition}.

We are therefore looking for solutions to~\eqref{eq:divtensor} which solve the constraint equations~\eqref{eq:momconst}~\&~\eqref{eq:hamconst} at finite momentum. This can be done systematically~\cite{Hinterbichler:2013dpa}, the result is that diffeomorphisms of the form~\cite{Mirbabayi:2014zpa,Pajer:2017hmb}
\be
\xi^i(t,\vec x) = \left(1 +\int^t\rd t_1\,a^{-3}\int^{t_1}\rd t_2\, a\,\nabla^2\right)\bar\xi^i(\vec x),
\label{eq:adaiabaticmodetimedep}
\ee
where $\bar\xi^i$ is a time-independent solution to~\eqref{eq:divtensor}, generate symmetries of the action with the lapse and shift integrated out. 

Note that the time dependence of the transformations~\eqref{eq:adaiabaticmodetimedep} is {\it fixed} by the adiabatic mode condition. In particular, adiabatic modes can only ever generate soft-$\zeta$ profiles which are {\it constant} in time. This is most straightforward to see from the divergence of~\eqref{eq:momconst}; after inverting the laplacian, this equation fixes the lapse in terms of $\zeta$:
\be
N^{(1)} = \frac{\dot\zeta}{H}.
\ee
To stay in $\zeta$-gauge, residual diffeomorphisms cannot have a time component, so from~\eqref{eq:nonlinearNtrans} we see that the lapse does not transform under the nonlinear part of residual gauge transformations, this implies that
\be
\partial_t\left(\delta_{\rm NL}\zeta\right) = 0.
\ee
This is a surprisingly strong constraint. It tells us that the only long-wavelength solutions to the equation of motion for $\zeta$ that can possibly be adiabatic modes must go to a constant. Note that the adiabatic mode condition guarantees that the field profile generated by a large gauge transformation will solve the $\zeta$ equation of motion, but it does {\it not} guarantee that this solution is the growing mode solution ({\it i.e.}, the solution which dominates at late times). The coincidence of the profile generated by a nonlinear symmetry and the growing mode solution is the {\it physical} mode condition. 
We will see that in ultra-slow-roll inflation, the growing mode evolves in time at long wavelength, so it cannot be an adiabatic mode, it therefore violates the physical mode condition, and the late-time dilation identity does not exist.

There are two choices of $\xi^i$ for which~\eqref{eq:gammatrans} vanishes, these are therefore symmetries purely of the scalar sector. Since eq.~\eqref{eq:gammatrans} is the conformal Killing equation of ${\mathbb R}^3$ it is not surprising that these take the form of a dilation and a special conformal transformation:
\begin{align}
\bar \xi^i_D &= \lambda x^i\\
\bar \xi^i_K & = 2b^jx_j x^i - b^i\vec x^2 - 2b^i\int^t\rd\tilde t_1\,a^{-3}\int^{t_1}\rd t_2\, a,
\end{align}
where the adiabatic mode condition requires that we supplement a special conformal transformation by a spatial translation with a particular time dependence. Typically we consider situations which are spatially translation invariant, and so we drop the additional time translation piece. Under these symmetries, $\zeta$ transforms infinitesimally as (here we drop the infinitesimal parameters of the transformations)
\begin{align}
\label{eq:infdil}
\delta_D\zeta &= 1+\vec x\cdot\nabla\zeta\\
\delta_{K^i}\zeta&=2x^i +\left(2x^i \vec{x}\cdot\vec\nabla -x^2 \nabla^i \right)\zeta.
\label{eq:infsct}
\end{align}

A more pragmatic viewpoint on the adiabatic mode condition is that it fixes the time-dependence of residual gauge transformations to ensure that they are global symmetries of the $\zeta,\gamma$ action after the lapse and shift have been integrated out. To illustrate this point, consider the dilation symmetry~\eqref{eq:timedepdilation}. This transformation is only a symmetry of the $\zeta$ action for $\lambda = {\rm constant}$. To see this, notice that a necessary requirement is that the nonlinear transformation $\delta_{\rm NL}\zeta = \lambda(t)$ be a symmetry of the quadratic $\zeta$-action
\be
S_2 = \int\rd t\,\rd^3x\,a^3\epsilon\left(\dot\zeta^2 - \frac{1}{a^{2}}(\nabla\zeta)^2\right).
\ee
It is clear, however, that this nonlinear transformation is only a symmetry when $\lambda = {\rm const.}$, which coincides with the adiabatic mode condition.

We have identified a set of global symmetries which act on the scalar sector of the action for cosmological perturbations, regardless of the background evolution. We now want to explore the Ward identities that follow from these symmetries and to discuss the consequences of the physical mode condition we have introduced above.

A technical point that deserves comment is that in Section~\ref{sec:generalderivation}, one of the key ingredients is that the initial state wavefunction transforms under the relevant symmetries. A reasonable question is whether this is true in the current case as well; particularly because it is often assumed that the wavefunctional does not transform under diffeomorphisms, and the symmetries of interest originate as diffeomorphisms. However, the wavefunction {\it can} transform under large gauge transformations. In Appendix~\ref{app:wavefgaugetrans}, we explain how this happens in the context of electrodynamics and gravity.

\subsection{Unequal time soft theorem for inflation}
Now that we have identified the relevant symmetries of the scalar sector in $\zeta$-gauge, we can apply the formalism of Section~\ref{sec:generalderivation} to obtain inflationary soft theorems. These statements should be widely applicable, holding even in cases where the standard consistency relations are violated.

\subsubsection{Dilation}
We first consider the dilation symmetry~\eqref{eq:infdil}. In this case $\delta_{\rm NL}\zeta(\vec x,\tau) = 1$ so that we have $\delta_{\rm NL}\zeta_{\vec q} = (2\pi)^3\delta(\vec q)$. Converting the linear transformation $\delta_{\rm L}\zeta = \vec x\cdot\vec\nabla\zeta$ to Fourier space, we obtain\footnote{In this case, removing the momentum-conserving delta function shifts the action of the dilation operator on the hard momentum modes~\cite{Maldacena:2011nz,Hinterbichler:2013dpa}.}
\be
\lim_{\vec{q}\to0}\, {\cal E}_i(q)\langle\zeta^f_{\vec k_1}\cdots\zeta^f_{\vec k_N}\zeta^i_{\vec q}\rangle'+{\rm c.c.} = -\left(3(N-1) + \sum_{a=1}^N \vec k_a \cdot \vec\nabla_{k_a}\right)\langle\zeta^f_{\vec k_1}\cdots\zeta^f_{\vec k_N}\rangle'\,.
\label{eq:dilmainIDinf}
\ee
This identity bears a resemblance to Maldacena's consistency relation, except the soft mode is inserted at the initial time, and instead of the power spectrum appearing on the left-hand side we have the initial-time wavefunctional coefficient~\eqref{eq:wavefcoeffcomplex}, which in general is complex.

\subsubsection{Special conformal transformation}
There is also an identity following from nonlinearly-realized special conformal transformations~\eqref{eq:infsct}. In this case, the Fourier transform of the nonlinear transformation is $\delta_{\rm NL} \vp_i (\vec q) = - 2 i \vec b \cdot \vec
\nabla_{q} [(2\pi)^3 \delta (\vec q)]$ which implies that the differential operator
$D_{-\vec q} (t_i) = 2i \vec b \cdot \vec\nabla_q$.
Putting this together with the Fourier transform of the special conformal operator leads to the identity
\be
\partial_{\vec q}\left({\cal E}_i(q)\langle\zeta^f_{\vec k_1}\cdots\zeta^f_{\vec k_N}\zeta^i_{\vec q}\rangle'+{\rm c.c.}\right)\bigg\rvert_{\vec q=0}  = \frac{1}{2}\sum_{a=1}^N\left(-6\nabla_{k_a}^i+ k_a^i\nabla_{k_a}^2-2\vec k_a\cdot\vec\nabla_{k_a}\nabla_{k_a}^i\right)\langle\zeta^f_{\vec k_1}\cdots\zeta^f_{\vec k_N}\rangle'\,.
\label{eq:SCTmainIDinf}
\ee
This expression is also slightly different from the usual SCT soft theorem.

\subsection{Slow roll and other attractor models}
\label{sec:SRcheck}
We first consider the soft theorems~\eqref{eq:dilmainIDinf} and~\eqref{eq:SCTmainIDinf} for conventional models of inflation. In these models, the scale factor grows nearly exponentially in cosmic time, $a(t) = e^{Ht}$. In terms of conformal time ($a(\tau)\rd\tau = \rd t$) this corresponds to
\be
a(\tau) = \frac{1}{H(-\tau)}.
\ee
where the time variation of the Hubble parameter
is very small, $\epsilon \ll 1$. In addition to the $\epsilon$ parameter, there is an entire hierarchy of slow-roll parameters, the next-leading parameter is $\eta \equiv \frac{\dot\epsilon}{H\epsilon}$, here we assume that all the slow-roll parameters are small.

In this Section, we would like to show explicitly that the soft theorem~\eqref{eq:dilmainIDinf} is satisfied in attractor models of inflation, at least in the case of the relation between the 3-point function and the power spectrum.
One approach would be to compute directly the unequal time 3-point correlation function in a concrete model and check it, but we can take a somewhat simpler route.

We can write the part of the soft theorem involving the unequal time correlator as,
\begin{equation}
 \lim_{\vec q \to 0}{\cal E}_i(q) \vev{\zeta^f_{\vec k_1}\zeta^f_{\vec k_2} \zeta^i_{\vec q} }' +{\rm c.c.}  = \lim_{\vec q \to 0}\left( 2\mbox{Re}\,{\cal E}_i(q) \,\mbox{Re}\, \vev{\zeta^f_{\vec k_1}\zeta^f_{\vec k_2} \zeta^i_{\vec q} }'  - 2\mbox{Im}\,{\cal E}_i(q)\, \mbox{Im}\, \vev{\zeta^f_{\vec k_1}\zeta^f_{\vec k_2} \zeta^i_{\vec q} }' \right).
\end{equation}
The strategy is to show that, perturbatively, at leading order in $q$,
\begin{equation}
\lim_{\vec q \to 0}2\mbox{Re}\,{\cal E}_i(q)\, \mbox{Re}\, \vev{\zeta^f_{\vec k_1}\zeta^f_{\vec k_2} \zeta^i_{\vec q} }'  = \lim_{q \to 0}  \frac{1}{P_{\zeta}(q) } \vev{\zeta^f_{\vec k_1}\zeta^f_{\vec k_2} \zeta^f_{\vec q} }' +{\cal O}(q^2),
\label{eq:ReReuneq}
\end{equation}
and that the imaginary parts, $\mbox{Im}\,{\cal E}_i(q)\, \mbox{Im}\, \vev{\zeta^f_{\vec k_1}\zeta^f_{\vec k_2} \zeta^i_{\vec q} }$, in the slow-roll case are of higher order in $q$, and therefore do not contribute to the soft limit. We then infer that the unequal time identity is satisfied because the well-known final time identity $ \lim_{\vec q \to 0}  \frac{1}{P_{\zeta}(q) } \vev{\zeta^f_{\vec k_1}\zeta^f_{\vec k_2} \zeta^f_{\vec q} }' = -(3+\vec k_1\cdot\partial_{\vec k_1})\vev{\zeta^f_{\vec k_1}\zeta^f_{-\vec k_1}}' $ is satisfied.\footnote{We would like to emphasize that this does {\it not} imply that the validity of the early-time identity relies in any way on the validity of the late-time identity. We are merely using known facts about computations of the late-time squeezed limit to infer information about the early-time squeezed limit without computing it explicitly.}

We now show that~\eqref{eq:ReReuneq} is true at tree level. The leading order unequal time three point function is computed using the formula
\begin{align} \label{eqn:3ptperteval}
\nonumber
   \vev{\zeta^f_{\vec k_1}\zeta^f_{\vec k_2} \zeta^i_{\vec q} }' &= \vev{0 \big\lvert i \int _{\tau_i}^{\tau}\rd\tau' \left[ H_{\rm int}^{(3)}(\tau') ,\zeta^I_{\vec k_1}(\tau) \zeta^I_{\vec k_2}(\tau)  \right] \zeta^I_{\vec q}(\tau_i) \big\rvert 0}  \\
    &= i u^*_{k_1}(\tau) u^*_{k_2}(\tau)u^*_{q}(\tau_i )   \int_{\tau_i}^\tau \rd\tau' H_{\rm int}^{(3)}\left[ u_{k_1}(\tau') u_{ k_2}(\tau')u_{q}(\tau' ) \right] \\
     & \quad  -i u_{k_1}(\tau) u_{ k_2}(\tau)u^*_{ q}(\tau_i )   \int_{\tau_i}^\tau \rd\tau' H_{\rm int}^{(3)}\left[ u^*_{ k_1}(\tau) u^*_{ k_2}(\tau)u_{q}(\tau_i ) \right]  \nonumber  + {\rm perms}.,
\end{align}
where $\zeta^I_{\vec k}(\tau) $ are interaction picture fields, $H_{\rm int}^{(3)}$ is the cubic interaction hamiltonian, and the mode functions, $u_k(\tau)$ are given by (to leading order in slow-roll)
\begin{equation}
  u_{q}(\tau) = \frac{ H}{ 2 \sqrt{\epsilon q^3 } } (1+i q \tau) e^{-i q \tau} + {\cal O} (\epsilon^{1/2}, \epsilon^{-1/2}\eta) \overset{q\to 0}{ =} \frac{ H}{ 2 \sqrt{\epsilon q^3 } } \bigg(1+ \frac{1}{2} q^2 \tau^2 - \frac{i}{3} q^3 \tau^3 \bigg)+ {\cal O} (\epsilon^{1/2}, \epsilon^{-1/2}\eta) ;
\end{equation}
in the second equality we have taken the $\vec q\to 0$ limit.

We therefore see that the three-point function in the late time limit, at leading order in $\epsilon$, satisfies
\begin{equation}
 \lim_{\vec q \to 0}\vev{\zeta^f_{\vec k_1}\zeta^f_{\vec k_2} \zeta^i_{\vec q} }'  =   \lim_{\vec q\to 0}\vev{\zeta^f_{\vec k_1}\zeta^f_{\vec k_2} \zeta^f_{\vec q} }'  + \dots + i \bigg( b(k_1;\tau_i) q^0 +\dots \bigg),
\end{equation}
where $b$ is some function whose form we will not need. We therefore see that
 the real part starts at order $q^{-3}$ and the imaginary part starts at order $q^0$. The form of the integrals in~\eqref{eqn:3ptperteval} in the squeezed limit is exactly the same as those that appear in the computation of $\vev{\zeta^f_{\vec k_1}\zeta^f_{\vec k_2} \zeta^f_{\vec q} }' $. Using the known computations of the squeezed limit of the late time 3-point function,
\begin{equation}
 \lim_{\vec q\to 0} \vev{\zeta^f_{\vec k_1}\zeta^f_{\vec k_2} \zeta^f_{\vec q} }'  = (2\epsilon -\eta)P_{\zeta}(k_1)  \frac{H^2}{4\epsilon q^3}  + \dots ,
\end{equation}
we obtain the leading behavior of $\vev{\zeta^f_{\vec k_1}\zeta^f_{\vec k_2} \zeta^i_{\vec q} }'$ . The other ingredient we need is the Bunch--Davies wavefunction~\eqref{eq:generalwavefunctionform}. Using this, we see that in the soft limit
\begin{equation}
\lim_{\vec q \to 0}{\cal E}_i(q) =  \frac{2\epsilon }{H^2}q^3 +\dots +i \left( - \frac{2\epsilon }{H^2 \tau_i}q^2 + \dots \right).
\end{equation} 
Putting these two ingredients together, we can compute the squeezed limit of the unequal time 3-point function
\begin{equation}
 \lim_{q \to 0} \mbox{Re} \,{\cal E}_i(q) \, \mbox{Re}\, \vev{\zeta^f_{\vec k_1}\zeta^f_{\vec k_2} \zeta^i_{\vec q} }'  =\lim_{q\to 0} \frac{1}{2} \frac{1}{P_{\zeta}(q) } \vev{\zeta^f_{\vec k_1}\zeta^f_{\vec k_2} \zeta^f_{\vec q} }' = \frac{1}{2}(2\epsilon -\eta)P_{\zeta}(k_1) ,
\end{equation}
and since $\lim_{\vec q\to 0}\mbox{Im}\,{\cal E}_i(q)\, \mbox{Im}\, \vev{\zeta^f_{\vec k_1}\zeta^f_{\vec k_2} \zeta^i_{\vec q} }'  ={\cal O}(q^2)$, we can conclude that 
\begin{equation}
  \lim_{\vec q \to 0}{\cal E}_i(q) \vev{\zeta^f_{\vec k_1}\zeta^f_{\vec k_2} \zeta^i_{\vec q} }' +{\rm c.c.} = (2\epsilon -\eta)P_{\zeta}(k_1).
\end{equation}
This implies that the unequal-time dilation soft theorem is satisfied for the 3-point function. Repeating the same argument, we also find that the SCT soft theorem~\eqref{eq:SCTmainIDinf} is satisfied, because the imaginary parts do not contribute until ${\cal O}(q^2)$, so that Condition 3 from Section~\ref{sec:physicalmodes} is satisfied.  It is also worth emphasizing that the early-late time identity does not follow from the known late-time soft theorems, so the fact that it is satisfied in slow-roll inflation is a non-trivial check.

For attractor models of inflation, the long-wavelength limit of the growing mode solution is $\zeta_{\vec q\to 0} \sim {\rm const}.$, which has the same time dependence as the nonlinear part of the dilation symmetry. We therefore see that dilations satisfy the physical mode condition in these cases.
This is consistent with the fact that it is possible to write a late-time identity corresponding to the soft theorem~\eqref{eq:dilmainIDinf}. In this particular case, the physical mode condition and the adiabatic mode condition coincide, but in the next Section we will see an example where this is not the case.

\subsection{Ultra-slow roll}
\label{sec:usruneqtime}
Ultra-slow-roll inflation~\cite{Tsamis:2003px,Kinney:2005vj} is a somewhat peculiar model of inflation, where the normal slow-roll hierarchy breaks down. These models are of particular interest because Maldacena's consistency relation can be violated~\cite{Namjoo:2012aa,Martin:2012pe}. We will see that the essential reason for this violation is that the dilation symmetry---though it continues to be an adiabatic mode---ceases to be a physical mode in the ultra-slow-roll background. Though ultra-slow roll may not be the most physically-relevant scenario which violates the consistency relation, we view it as a valuable conceptual test case, where we can explore violations of the inflationary soft theorems without the added complications of additional degrees of freedom.

The simplest ultra-slow-roll model can be defined in the context of canonical scalar field models, where the action takes the form
\be
S = \int\rd^4x\sqrt{-g}\left(\frac{M_{\rm Pl}^2R}{2} -\frac{1}{2}(\partial\phi)-V(\phi)\right),
\label{eq:simpleUSR}
\ee
The equation of motion for the homogeneous background of the scalar field is given by
\be
\ddot\phi + 3H\dot\phi + V_{,\phi} = 0.
\ee
In normal slow-roll inflation, the $\ddot\phi$ term is negligible compared to the others, and the slow-roll attractor solution fixes $\dot\phi$ in terms of the slope of the potential $\dot\phi \simeq V_{,\phi}/(3H)$. The situation in ultra-slow roll is different. In this case, the potential is taken to be exactly flat, $V(\phi)  = V_0$, so that the equation of motion takes the form\footnote{In order for inflation to end, this cannot be exactly true, so realistic models require the potential to only be sufficiently flat over some finite range in field space. For our purposes, it suffices to consider this idealized situation.}
\be
\ddot\phi+ 3H\dot\phi = 0.
\ee
This equation can be integrated to find that the field velocity redshifts as $\dot\phi \propto a^{-3}$. The field is therefore slowing down very rapidly, hence the name ultra-slow roll.\footnote{In some sense this name is misleading because the field is actually rolling more quickly than it would be on the slow-roll attractor solution, where $\dot\phi \propto V_{,\phi}$, which vanishes on an exactly flat potential.} After a very short time, the kinetic energy density, which scales as $\dot\phi^2\sim a^{-6}$, is completely negligible compared to the potential, so that the Friedmann equation, $3M_{\rm Pl}^2 H^2 \simeq V_0$, implies that $a(t) = e^{Ht}$, which is an inflationary background.

During this de Sitter phase, the field slowing down leads to a rapidly changing equation-of-state: $\epsilon = \epsilon_0a^{-6}$,
where $\epsilon_0$ is the initial value of $\epsilon$ at the onset of the ultra-slow-roll phase. This rapid evolution of the equation of state is important for the dynamics of perturbations. It is interesting to compute the next slow-roll parameter, $\eta  = -6+2\epsilon$,
which is large in this case, in contrast to normal slow roll. There is therefore a phase where the field is slowing down, but the background evolution is nevertheless quasi-de Sitter. It is possible to have arbitrarily many $e$-folds of this phase for sufficiently flat potentials and sufficiently large initial kinetic energy~\cite{Dimopoulos:2017ged}.

We now turn to fluctuations around the ultra-slow-roll solution. We have
\be
a^2\epsilon = \epsilon_0(H\tau)^{4},
\ee
so that the linearized equation of motion for the scalar fluctuations in Fourier space is
\be
\zeta_k''+\frac{4}{\tau}\zeta_k'+k^2\zeta_k = 0.
\label{eq:usrlineareom}
\ee
In the long-wavelength limit, the two solutions to this differential equation evolve as
\be
\zeta_k(\tau) \sim {\rm const}. + c_3 \tau^{-3},
\ee
so we see that one of the modes is indeed constant, as in slow-roll inflation. However, the other mode is {\it growing} in time, and is therefore far more important at late times ($\tau \to 0$).\footnote{This growth of scalar fluctuations at late times indicates that the background around which we are perturbing is not an attractor of the dynamics, in contrast to the the usual slow-roll solution.} In this case, it is the time-dependent growing mode solution which obtains a scale-invariant spectrum.  The solutions to the equation~\eqref{eq:usrlineareom} corresponding to this growing mode are
\be
\zeta_{q}(\tau) = \frac{i}{ \sqrt{4\epsilon_0q^3} M_{\rm Pl}H^2\tau^3}(1+iq\tau)e^{-iq\tau},
\ee
which we see acquires a scale-invariant spectrum, 
\be
\langle\zeta_{-\vec q}\zeta_{\vec q}\rangle' = \frac{1}{4M_{\rm Pl}^2H^4\tau^6\epsilon_0 q^3},
\ee
allowing ultra-slow roll to be compatible with observation. Notice that since these solutions evolve as
\be
\zeta_{q}(\tau) \sim \tau^{-3}.
\ee
in the long-wavelength limit, the dilation symmetry~\eqref{eq:infdil} does not satisfy the physical mode condition in this background. This is despite the fact that the dilation symmetry continues to generate a profile consistent with the equations of motion of the theory, and therefore is still an adiabatic mode. (This is also reflected in the fact that $\zeta\sim{\rm const.}$ continues to be a solution to the zero-momentum equation of motion.) Ultra-slow roll therefore provides a clear example where the requirements of being an adiabatic {\it vs.} physical mode differ.

It is well-known that ultra-slow-roll models can violate Maldacena's consistency relation for late-time correlation functions~\cite{Namjoo:2012aa,Martin:2012pe}. However, we expect that our unequal-time soft theorems are satisfied in this case as well. 
Unfortunately, we are unable to verify this expectation directly, as computing correlation functions involving early-time fields in this model is quite difficult. The reason for this is the time dependence of $\epsilon$ (or equivalently $\dot\phi$). 
Since the kinetic energy in the rolling field is redshifting very quickly, at very early times it is no longer consistent to neglect the back-reaction  on the geometry.\footnote{Additionally, unlike standard slow-roll inflation, the curvature perturbation, $\zeta$, evolves outside the horizon in ultra-slow-roll inflation. This makes the model's predictions sensitive to the details of reheating, in contrast to attractor inflation models. Also, in the late time limit, $\zeta$-gauge ceases to be well-defined in ultra-slow roll.} Though we are not able to check the unequal-time relation directly in the ultra-slow-roll case, in the next Section we introduce a useful family of toy model cases which capture all the essence of this case, with the added benefit of being simple enough to explicitly calculate correlation functions.

\section{An illustrative toy model}
\label{sec:toymodel1}
Although our focus is on cosmology, many of the puzzling features of cosmological Ward identities can be illustrated by a simpler toy model. In this model, many of the same puzzles that arise, for example, in ultra-slow-roll inflation can be formulated and explicitly resolved. In particular, the model we consider is simple enough that all correlation functions of interest can be explicitly computed. The model is a flat space quantum field theory with time-dependent interactions, described by the action\footnote{In this Section, everything is happening in flat space, so the time coordinate $\tau$ is normal coordinate time, {\it not} conformal time, but we use the same notation because it plays the same role that conformal time does in cosmology. In this section, time derivatives will be denoted by $'\equiv \frac{\rd}{\rd\tau}$.}
\be
S = M^2 \int\rd^3x\,\rd \tau\,f(\tau)\left(e^{3\zeta}\zeta'^2 - e^\zeta(\nabla\zeta)^2\right),
\label{eq:toymodelaction}
\ee
where we will specialize to particular functional forms for $f(\tau)$ to mimic features of certain cosmological models. One of the virtues of this model is that it divorces concerns about gauge symmetries and adiabatic modes from other puzzles involving the Ward identities of the theory---as this model is devoid of any gauge invariance---all the symmetries involved are global.

Aside from spatial translations and rotations, the action~\eqref{eq:toymodelaction} has two nonlinearly realized symmetries for any $f(\tau)$, which act on the field as\footnote{Note that these are of precisely the same form as the nonlinearly-realized dilation and special conformal transformation symmetries in the action for inflationary fluctuations, with $f(\tau)$ playing the role of $a^2\epsilon$. Indeed, the action has been constructed so as to manifest the symmetries~\eqref{eq:toydilation} and~\eqref{eq:toysct}. Though these symmetries arise essentially from diffeomorphisms in the gravitational context, this action can be thought of completely independently of its gravitational origins; its symmetries are traditional global symmetries.}
\begin{align}
\label{eq:toydilation}
\delta_D{\zeta}&= 1+ \vec{x} \cdot \vec\nabla \zeta,\\
\delta_{K^i}\zeta&=2x^i +\left(2x^i \vec{x}\cdot\vec\nabla -x^2 \nabla^i \right)\zeta.
\label{eq:toysct}
\end{align}
In the following, we will consider three distinct functional forms for $f(\tau)$, each of which illustrates some interesting features of cosmological consistency relations. Specifically, we will consider
\begin{itemize}

\item {\it Flat space}: $f(\tau) = {\rm const}$. In this example, the background is flat space. This provides a simplified test case to investigate the consequences of the symmetries~\eqref{eq:toydilation} and~\eqref{eq:toysct}, similar to the ghost condensate example of~\cite{Goldberger:2013rsa}. Interestingly, in this case we will find that the Ward identity corresponding to the symmetry~\eqref{eq:toysct} is not satisfied. As we will explain, this is due to the appearance of an ${\cal O}(q\tau)$ piece in the long-wavelength limit of $\zeta$, which causes a failure of Condition 3 of Section~\ref{sec:physicalmodes} for this symmetry.

\item {\it Slow roll like}: $f(\tau) = (H\tau)^{-2}$, where $\tau\in(-\infty,0)$. This situation is representative of normal slow-roll inflation, where the scale factor evolves similarly in conformal time. For our simplified model, this time dependence actually has an extra emergent symmetry, which acts on $\zeta$ as
\be
\delta_T\zeta= 1- \tau\partial_\tau\zeta,
\label{eq:flatslowrollsymm}
\ee
and leaves the action invariant up to a temporal boundary term. This symmetry does not have an analogue in actual cosmological inflation. 

\item {\it Ultra-slow roll like}: $f(\tau) = (H\tau)^{4}$, with $\tau\in(-\infty,0)$. The last situation we consider is analogous to ultra-slow-roll inflation. In this case, there is another emergent symmetry, which acts as
\be
\delta_T\zeta = \tau^{-3}\left(1 +\frac{1}{2}\tau\partial_\tau  \zeta \right).
\label{eq:flaturssymm}
\ee
Under this symmetry, the lagrangian again shifts by a time boundary term. This is the analogue in our toy model of the shift adiabatic mode identified in~\cite{Finelli:2017fml,Finelli:2018upr}.
\end{itemize}
\subsection{Testing the standard late-time soft theorems}
The various symmetries are global symmetries of the interacting theory~\eqref{eq:toymodelaction} and therefore have corresponding Ward identities in the quantum theory. The soft theorems in the toy model (\ref{eq:toymodelaction}) corresponding to the symmetries (\ref{eq:toydilation}) and (\ref{eq:toysct}) are the same as those in cosmology; the leading-order soft behavior is governed by the dilation consistency relation:
\be
\lim_{\vec{q}\to0} \frac{1}{P(q)}\vev{\zeta_{\vec q}\zeta_{\vec k_1}\cdots\zeta_{\vec k_N}}'_c = -\left(3(N-1) + \sum_{a=1}^N \vec k_a \cdot \vec\nabla_{k_a}\right)\vev{\zeta_{\vec k_1}\cdots\zeta_{\vec k_N}}'_c\,,
\label{eq:dilationid}
\ee
which is the Ward identity for the symmetry~\eqref{eq:toydilation}. Here $P(q) \equiv \langle\zeta_{\vec q}\zeta_{-\vec q}\rangle'$ is the 2-point function for the field $\zeta$. The ${\cal O}(q)$ soft behavior is also constrained by an identity; it satisfies the special conformal (SCT) consistency relation
\be
\lim_{\vec{q}\to0} \frac{\partial}{\partial q^i}\left(\frac{1}{P(q)}\vev{\zeta_{\vec q}\zeta_{\vec k_1}\cdots\zeta_{\vec k_N} }'_c\right) = \frac{1}{2}\sum_{a=1}^N\left(-6\nabla_{k_a}^i+ k_a^i\nabla_{k_a}^2-2\vec k_a\cdot\vec\nabla_{k_a}\nabla_{k_a}^i\right)\langle\zeta_{\vec k_1}\cdots\zeta_{\vec k_N}\rangle'_c\,,
\label{eq:SCTid}
\ee
which follows from~\eqref{eq:toysct}.

As we have argued in Section~\ref{sec:physicalmodes}, the late-time soft theorems should only hold under some set of restrictions. In particular, we require that the nonlinear symmetry transformation generates the physical growing mode solution to the equation of motion. The toy model we have introduced provides a number of examples where we can check these late-time identities explicitly, and see that they are indeed only satisfied when the three conditions laid out in~\ref{sec:physicalmodes} hold.

In the following we will consider each of the enumerated time dependences for $f(\tau)$, display the late-time Ward identities following from the symmetries~\eqref{eq:toydilation}--\eqref{eq:flaturssymm}, and check whether they are satisfied in perturbation theory. In order to streamline the presentation, we just describe the results of checks of the Ward identities associated to the various symmetries, the details of the computations can be found in Appendix~\ref{app:toymodel}.

\subsubsection{Flat space}
\label{sec:flatspacetoymodel1}
We first consider the case where $f(\tau)= {\rm constant}$, so that the action takes the form
\be
S = M^2\int\rd^3x\,\rd \tau\left(e^{3\zeta}\zeta'^2 - e^\zeta(\nabla\zeta)^2\right).
\label{eq:flatspacetoylagrangian}
\ee
Since this action has the non-linearly realized symmetries~\eqref{eq:toydilation} and~\eqref{eq:toysct}, we want to perform the most basic check of the identities (\ref{eq:dilationid}) and (\ref{eq:SCTid})---that is, we want to verify the relation between the 3-point function and the power spectrum. Using standard techniques, it is possible to compute these correlation functions in this toy model, the results are tabulated in Appendix~\ref{sec:toyflatspacecorrs}. Using these formulae, we can check that indeed the Ward identity associated to the dilation symmetry~\eqref{eq:toydilation} is satisfied for $N=2$.

We expect that the sub-leading ${\cal O}(q)$ behavior of the soft limit should obey the identity~\eqref{eq:SCTid}. However, it turns out that this identity is actually not satisfied---that is, we have
\be
\lim_{\vec{q}\to0} \frac{\partial}{\partial q^i}\left(\frac{1}{P(q)}\langle\zeta_{\vec q}\zeta_{\vec k_2}\zeta_{\vec k_3}\rangle'\right) \neq \frac{1}{2}\left(-6\nabla_{k_2}^i+ k_2^i\nabla_{k_2}^2-2\vec k_2\cdot\vec\nabla_{k_2}\nabla_{k_2}^i\right)\langle\zeta_{\vec k_2}\zeta_{-\vec k_2}\rangle',
\ee
as can be verified explicitly from the expressions in Appendix~\ref{sec:toyflatspacecorrs}. This violation of an equal time Ward identity associated to a global symmetry is a signal that one of the conditions in Section~\ref{sec:physicalmodes} is violated. In this case it is the somewhat technical Condition 3: we see that in the soft limit the mode functions have a piece at ${\cal O}(q)$, which does not vanish fast enough at small momentum.

\subsubsection{Slow roll}
\label{eq:toymodelslowroll1}
We next consider the case where $f(\tau) = (H\tau)^{-2}$,  which is an analogue model for slow-roll inflation. Here, in addition to the dilation~\eqref{eq:toydilation} and SCT~\eqref{eq:toysct} symmetries, we see an example of an emergent symmetry,~\eqref{eq:flatslowrollsymm}, which is a symmetry of the action
\be
S = M^2\int\rd^3x\,\rd \tau\frac{1}{H^2\tau^2}\left(e^{3\zeta}\zeta'^2 - e^\zeta(\nabla\zeta)^2\right).
\ee
The corresponding Ward identities for dilation and SCT are identical to the $f(\tau) = {\rm constant}$ case, and are given by~\eqref{eq:dilationid} and~\eqref{eq:SCTid}. The late-time Ward identity corresponding to the emergent symmetry~\eqref{eq:flatslowrollsymm} reads
\be
\lim_{\vec{q}\to0} \frac{1}{P(q)}\langle\zeta_{\vec q}\zeta_{\vec k_1}\cdots\zeta_{\vec k_N}\rangle' = -\tau\partial_\tau\langle\zeta_{\vec k_1}\cdots\zeta_{\vec k_N}\rangle'\,.
\label{eq:srolltimedid}
\ee 
Note that this symmetry does not have an analogue in slow-roll inflation, but it is nevertheless a symmetry of this toy model.
We cannot derive this late-time identity using the path integral techniques we have used to this point, because it involves a shift by the time-derivative of $\zeta$. We can, however, employ other methods which directly give late-time identities. Though one of our main points is that these identities are less reliable because they make implicit assumptions, we can nevertheless check if the identity so-derived is satisfied in our toy model.
We explicitly compute the power spectrum and 3-point function for this model in Appendix~\ref{app:srcorrelator}. From the explicit expressions there, we can check that the Ward identities associated to dilations and special conformal transformations are satisfied, along with the identity~\eqref{eq:srolltimedid}.

To the extent that this model is representative of slow-roll inflation, it is not surprising that the Ward identities are satisfied, but it is worth trying to understand why they are valid. 
In this model, the linearized classical solutions~\eqref{eq:srollmodef} go to a constant as $\vec q\to 0$, which matches the time-dependence of the symmetry transformations $\delta_D$ and $\delta_T$, so the physical mode conditions are satisfied. Additionally, the leading momentum dependence in the soft limit starts at ${\cal O}(q^2)$, which is necessary for the $\delta_K$ Ward identity to be satisfied, as we will show in Section~\ref{sec:SCT}.

\begin{table}
\centering
\begin{tabular}{| l | c | c | c | }
	\hline
  Conventional identity	& {\bf Dilation} & ~~{\bf SCT}~~ & {\bf Accidental shift}  \\ \hline
	Flat space: $f(\tau)={\rm constant}$ &   {\color{darkgreen}\cmark} &  {\color{darkred}\xmark}& \shade  \\		\hline
	 Slow roll: $f(\tau)= (H\tau)^{-2}$ &  {\color{darkgreen}\cmark} &  {\color{darkgreen}\cmark} &  {\color{darkgreen}\cmark}  \\\hline
	   Ultra-slow roll: $f(\tau)= (H\tau)^{4}$ &  {\color{darkred}\xmark} &  {\color{darkred}\xmark} &  {\color{darkgreen}\cmark} \\\hline
\end{tabular}
\caption{\small A summary of the which late-time Ward identities are satisfied in the model~\eqref{eq:toymodelaction} for different time dependences.}
\label{tab:toymodelsummary}
\end{table}

\subsubsection{Ultra-slow roll}
\label{sec:usrtoy1}
We now turn to what is possibly the most interesting case, where
$f(\tau)=(H\tau)^4$. This model in many respects behaves like ultra-slow-roll inflation. In addition to the dilation and SCT symmetries, the action
\be
S = M^2\int\rd^3x\,\rd \tau\,(H\tau)^4\left(e^{3\zeta}\zeta'^2 - e^\zeta(\nabla\zeta)^2\right),
\ee
also has an additional symmetry~\eqref{eq:flaturssymm}: $\delta_T\zeta = \tau^{-3}\left(1 +\frac{1}{2}\tau\partial_\tau  \zeta \right)$. This last symmetry is the analogue of the shift adiabatic mode of ultra-slow roll identified in~\cite{Pajer:2017hmb}. This symmetry gives rise to an equal time identity
\begin{equation}
\lim_{\vec{q}\to0} \frac{1}{P(q)} \vev{\zeta_{\vec q}\zeta_{\vec k_1}\cdots\zeta_{\vec k_N}}' = \frac{1}{2} \tau \partial_\tau \vev{\zeta_{\vec k_1}\cdots\zeta_{\vec k_N}}',
\label{eq:accidentalshiftwardusr}
\end{equation}
where we have cancelled off a factor of $\tau^{-3}$ common to both sides.

Similarly to the previous cases, we would like to check the Ward identities for this time dependence. Using the formulae from Appendix~\ref{sec:usrcorrelator} it is apparent that the dilation and SCT Ward identities are {\it not} satisfied, while the identity~\eqref{eq:accidentalshiftwardusr} {\it is} satisfied. This is similar to the ultra-slow-roll case where the shift identity is the only one satisfied at late times.

In order to understand this result, note that the mode functions evolve as $\zeta \sim \tau^{-3}$ at zero momentum. This is precisely the same time dependence that is generated by the shift symmetry~\eqref{eq:flaturssymm}, while dilations generate a long-wavelength mode that is constant. For this reason, we should expect that only the shift Ward identity will be satisfied.

We collect together the results from this toy model in Table~\ref{tab:toymodelsummary}, cataloging which Ward identities are satisfied and violated for each of the functional forms for $f(\tau)$ we have considered.

\subsubsection{Comment on symmetries involving $\partial_\tau \zeta$ and physical mode condition}

We have seen that in the slow-roll like and ultra-slow-roll like cases there are emergent shift symmetries in which the linear part is proportional to time derivative of $\zeta$: $\delta_{\rm L}\zeta \sim \partial_\tau\zeta$. They are both physical modes and the equal time identities for these symmetries are satisfied.
Although our identity \eqref{eq:mainID} does not apply to symmetries involving $\partial_\tau \zeta$, because the path integral techniques we use cannot handle in a straightforward way symmetries that mix $\zeta$ and $\dot\zeta$, one can still perturbatively prove that the identities are satisfied in both cases because of the physical mode condition (details are given in Appendix \ref{sec:pertproof}). 
\begin{figure}
\centering
  \includegraphics[scale=0.6]{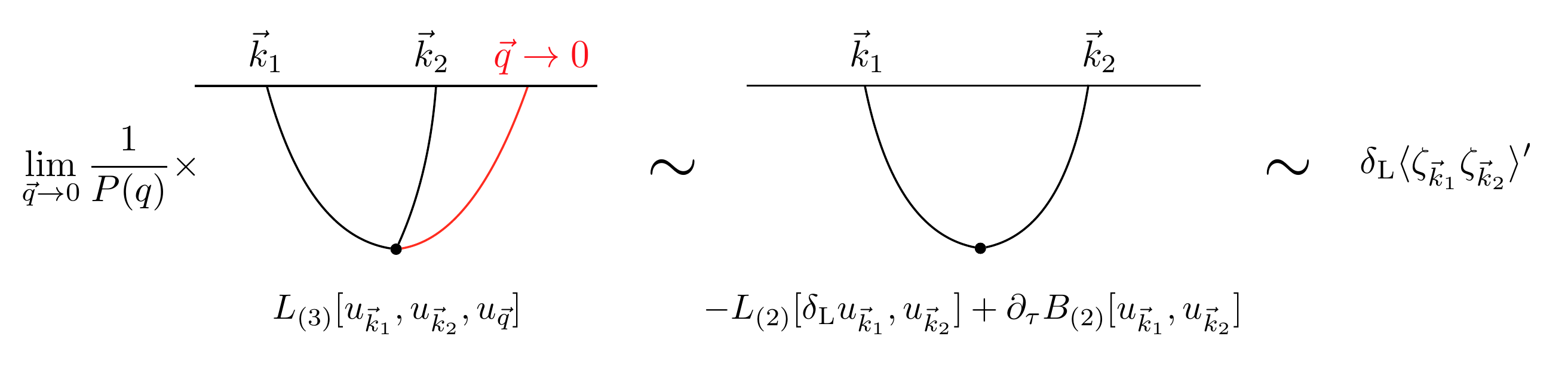}
  \caption{\small Perturbative proof of the use of physical mode. $L_{(3)}[ \delta_{\rm NL}u_q, u_{k_1},  u_{k_2} ] $ is the interaction vertex inside the in-in time integral  $\vev{\zeta_{\vec{q}}(\tau) \zeta_{\vec{k}_1}(\tau)\zeta_{\vec{k}_2}(\tau)}'  = 2 {\rm Im} \bigg( u^*_q(\tau) u^*_{k_1}(\tau)  u^*_{k_2}(\tau) \int^{\tau} \rd \tau' L^{(3)}[ u_q, u_{k_1},  u_{k_2} ](\tau')  \bigg) +{\rm perms.}$ }
   \label{fig:pertproof}
\end{figure}
Schematically we prove it in the following way. Given that the physical mode condition is satisfied, $ \lim_{\vec q \to 0} u_{\vec q}(\tau) \propto \delta_{\rm NL}(\tau) $, 
 when one perturbatively computes the three point function in the squeezed limit, the three point vertex in the time integral can be replaced by the linear transformation of the quadratic action and a boundary term,
\begin{equation}
 L_{(3)}[ \delta_{\rm NL}u_q, u_{k_1},  u_{k_2} ] +{\rm perms}.   =  -   L_{(2)}[ \delta_{\rm L} u_{k_1},  u_{k_2} ]+ \partial_{\tau}B_{(2)}[u_{k_1},u_{k_2}]+{\rm perms},
\end{equation}
as illustrated in Figure \ref{fig:pertproof}. One can then verify that it is equal to the linear transformation of the two point function.
Note that such replacement is not legitimate when the physical mode condition is not satisfied.

\subsection{Testing the early-late time soft theorems}
\label{sec:toymodel2}
We now apply the formalism of section~\ref{sec:generalderivation} to the toy model~\eqref{eq:toymodelaction}. Our goals are twofold: firstly, we would like to check the general Ward identity \eqref{eq:mainID} in a variety of situations to verify that {\it it is satisfied}. A summary of the checks is given in Table \ref{tab:toymodelsummaryIFF}. We will see additionally that our improved understanding of the regime of validity of the traditional Ward identities will allow is to resolve many of the puzzles encountered in the previous subsections.

Let us write down the identity for dilation and special conformal transformation for the toy model as they are symmetries for any function $f(\tau)$.
In the case of dilation, the operator ${\cal D}_q(\tau_i)  =1$ and $\delta_{\rm L}$ is the operator appearing on the right hand side of~\eqref{eq:dilationid}. Putting this together, the identity~\eqref{eq:mainID} for dilations reads
\be
\lim_{\vec{q}\to0}\, {\cal E}_i(q)\langle\zeta^f_{\vec k_1}\cdots\zeta^f_{\vec k_N}\zeta^i_{\vec q} \rangle'+{\rm c.c.} = -\left(3(N-1) + \sum_{a=1}^N \vec k_a \cdot \vec\nabla_{k_a}\right)\langle\zeta^f_{\vec k_1}\cdots\zeta^f_{\vec k_N}\rangle'\,.
\label{eq:flatuneqD}
\ee
Similarly, we can write down the identity for the special conformal symmetry; in this case, we have $D_{\vec q}(\tau_i) =2 i\partial_{\vec q}$ so the identity takes the form
\be
\partial_{\vec q}\left({\cal E}_i(q)\langle\zeta^f_{\vec k_1}\cdots\zeta^f_{\vec k_N}\zeta^i_{\vec q}\rangle'+{\rm c.c.}\right)\bigg\rvert_{\vec q=0}  = \frac{1}{2}\sum_{a=1}^N\left(-6\nabla_{k_a}^i+ k_a^i\nabla_{k_a}^2-2\vec k_a\cdot\vec\nabla_{k_a}\nabla_{k_a}^i\right)\langle\zeta^f_{\vec k_1}\cdots\zeta^f_{\vec k_N}\rangle'\,.
\label{eq:flatuneqSCT}
\ee

\subsubsection{Flat space}
As before, we begin our discussion with the flat space case, where the relevant symmetries are dilations and special conformal transformations.  As a simple check of the identities \eqref{eq:flatuneqD} and \eqref{eq:flatuneqSCT}, 
it is possible to compute the unequal-time 3-point function involved, the result is tabulated in Appendix~\ref{app:flatunequal}. Examining its squeezed limit, it is straightforward to check that the identity~\eqref{eq:flatuneqD} is satisfied.
It is also possible to check the special case of this identity relating the 3-point and 2-point correlation functions, and see that it is indeed satisfied. So, we see that both of the early-late time identities are satisfied in this case, in contrast to the late-time identities considered in Section \ref{sec:flatspacetoymodel1}, where the identity associated to SCT was violated.

\subsubsection{Slow roll}
We next consider the toy model which mimics slow-roll inflation, $f(\tau)\sim \tau^{-2}$. We saw that each of the equal-time Ward identities associated to dilation, SCT, and the emergent shift symmetry were satisfied in this particular model, and we would like to check that our early-time identities are satisfied as well. 

The identity associated to the time-shift symmetry~\eqref{eq:flatslowrollsymm} is more subtle. This symmetry mixes $\zeta$ with $\dot\zeta$. It is somewhat unclear how to implement this transformation in the fixed-time Schr\"odinger field theory formalism we are employing. For this reason, we do not derive unequal-time Ward identities associated to the time shift symmetries in this case and in the ultra-slow-roll case.\footnote{It should be possible to derive identities for these symmetries by considering the hamiltonian path integral, where presumably these symmetries would mix $\zeta$ and $\Pi_\zeta$.} This is not much of a loss, as it is precisely for these symmetries that there are already reliable late-time identities, discussed in Sec.~\ref{sec:toymodel1}.

By computing the vacuum wavefunctional and unequal time 3-point function, we can check that the identities associated to dilations \eqref{eq:flatuneqD} and SCTs \eqref{eq:flatuneqSCT} are satisfied in this case. These checks are described in Appendix~\ref{app:uneqtoysr}.

\begin{table}
\centering
\begin{tabular}{| l | c | c |  }
	\hline
New unequal-time identity	& {\bf Dilation} & ~~{\bf SCT}~~ \\ \hline
	Flat space: $f(\tau)={\rm constant}$ &   {\color{darkgreen}\cmark} &   {\color{darkgreen}\cmark} \\		\hline
	 Slow roll: $f(\tau)= (H\tau)^{-2}$ &  {\color{darkgreen}\cmark} &  {\color{darkgreen}\cmark}  \\\hline
	   Ultra-slow roll: $f(\tau)= (H\tau)^{4}$ &   {\color{darkgreen}\cmark} &   {\color{darkgreen}\cmark} \\\hline
\end{tabular}
\caption{\small A summary of the which unequal-time Ward identities are satisfied in the model~\eqref{eq:toymodelaction} for different time dependences. Note that in this case---in contrast to the late-time identities---all of the identities are satisfied. We do not consider symmetries which involve time derivatives of $\zeta$, because these are difficult to implement in the lagrangian path integral formalism we have employed.}
\label{tab:toymodelsummaryIFF}
\end{table}

\subsubsection{Ultra-slow roll}
Finally, we consider the case which is analogous to ultra-slow-roll inflation, $f(\tau)\sim \tau^4$. As discussed in Section~\ref{sec:usrtoy1}, the model has three global symmetries for this choice of time dependence. Two of them are dilation and SCT, which are universal for any $f(\tau)$. 

As in the slow-roll case, there is an additional emergent symmetry in this case, which acts as $\delta_T\zeta = \tau^{-3}\left(1 +\frac{1}{2}\tau\partial_\tau  \zeta \right)$. The early-time Ward identities associated to dilations and SCTs are the same as in the previous sections, eqs.~\eqref{eq:flatuneqD} and~\eqref{eq:flatuneqSCT}. Using the formulae in Appendix~\ref{app:usruneqcorr} we can check that both of these identities are satisfied. For example, for dilations the left-hand side of the consistency relation reads:
\be
\lim_{\vec q\to0}{\cal E}_i(q)\vev{\zeta^f_{\vec{k}_2} \zeta^f_{\vec{k}_3} \zeta^i_{\vec{q}} }'+{\rm c.c.} = -\frac{1}{2H^4M^2\tau_f^4 k_2}+\cdots
\ee
while the right hand side is
\be
-\left(3 +  \vec k \cdot \vec\nabla_{k}\right) \langle\zeta_{\vec k}\zeta_{-\vec k}\rangle' = -\frac{1}{2H^4M^2 k \tau_f^4},  
\ee
so we see that the early-late time identity holds in this case. The same is true for the SCT early-late identity.
Note that this is in stark contrast to the equal-time identities considered in Section~\ref{sec:usrtoy1}. There, we found that {\it neither} of these equal-time identities were satisfied. We therefore see that the Ward identities derived in Section~\ref{sec:generalderivation} are substantially more robust as they are free of further assumptions (physical mode condition). Again, we do not derive an identity for the time-shift symmetry because it involves a shift by $\zeta'$.

\section{More on ultra-slow roll and physical modes}
\label{sec:USRphysicalmode}
When we discussed ultra-slow-roll inflation in Section~\ref{sec:usruneqtime}, we only considered the unequal-time identities. However, as we have seen from the toy model case, there can be situations where physical modes involve shifts of the field involving time derivatives. In these cases, it is possible to write down late-time identities, but these cannot be obtained in a straightforward way from early-time identities, because handling them in the path integral is subtle.

In the ultra-slow-roll background---similar to our toy ultra-slow-roll case---the theory has an emergent symmetry, which essentially arises from the underlying shift symmetry of the scalar theory driving the background~\cite{Finelli:2017fml,Finelli:2018upr}. In this case, there is an additional diffeomorphism which is an adiabatic mode (it solves the lapse and shift constraints). For the simplest ultra-slow-roll model, this shift symmetry acts on $\zeta$ as 
\begin{equation}
  \delta \zeta_k = (2\pi)^3 \delta (\vec{k}) \left( \frac{H}{\dot{\Phi}} + \frac{1}{2}\int^t \rd t' \dot{\Phi} \right) + \frac{1}{a \dot{\Phi}} \zeta_k'-\frac{1}{2}\left(\int^t \rd t'\dot{\Phi}\right)(3+\vec{k}\cdot \partial_{\vec k}) \zeta_k,
\end{equation}
where the lower integration limit is chosen such that no time independent part appears. 
For the model~\eqref{eq:simpleUSR}, the nonlinear part of the symmetry acts on $\zeta$ as
\be
\delta_T \zeta = \tau^{-3}+\cdots,
\ee
so the physical mode condition is satisfied and we should expect that it is possible to write down a late-time identity corresponding to this symmetry. Indeed, this is possible, and can be generalized to other models of shift-symmetric inflation which share similarities with ultra-slow roll~\cite{Finelli:2017fml,Finelli:2018upr}. The identity for the above physical mode symmetry takes the form
\begin{align}
  &\lim_{\vec{q} \to 0} \left( \frac{H}{\dot{\Phi}} + \frac{1}{2}\int^t \rd t' \dot{\Phi}\right) \frac{1}{P_{\zeta}(q)}\vev{\zeta_f(\vec{q}) \zeta_f(\vec{k}_1)\zeta_f(\vec{k}_2)}' \nonumber \\
   &~~~~~~~~~~~~= \frac{1}{a(t) \dot{\Phi}} \frac{\partial}{\partial \tau} \vev{\zeta_f(\vec{k}_1) \zeta_f(-\vec{k}_1) }' - \frac{1}{2}\left(\int^t \rd t'\dot{\Phi}\right)\left(3+\vec{k}_1\cdot \partial_{\vec k_1}\right)  \vev{\zeta_f(\vec{k}_1) \zeta_f(-\vec{k}_1) }'.
\label{eq:USRid}
\end{align}
The first term on the right hand side generates $\simeq6\vev{\zeta_{\vec k} \zeta_{-\vec k}}'$. The second term on the right hand side is sub-leading in the slow-roll expansion (in $\epsilon_0$), and so we have verified that the identity is satisfied at leading order.\footnote{Note that the identity~\eqref{eq:USRid} is superficially different from eq.\,(26) in~\cite{Finelli:2017fml}. However, we expect that they are actually equivalent. This is highly non-obvious, but if one follows the OPE procedure advocated in~\cite{Finelli:2017fml} for the symmetry~\eqref{eq:flaturssymm} in our toy model of ultra-slow-roll inflation, the resulting 3-point identity takes the form
\be
\lim_{\vec{q}\to0} \frac{1}{P(q)} \vev{\zeta_{\vec q}\zeta_{\vec k_1}\zeta_{\vec k_2}}' = -\left[\frac{\tau}{6}\left(3+k\partial_k\right)\langle\zeta_{\vec k}\zeta_{-\vec k}\rangle'+\frac{\tau^2}{12}\partial_t\langle\zeta_{\vec k}\zeta_{-\vec k}\rangle'\right]\partial_{\tau}\log\langle\zeta_{\vec q}\zeta_{-\vec q}\rangle'\Big\rvert_{\vec q\to 0}-\left(3+k\partial_k\right)\langle\zeta_{\vec k}\zeta_{-\vec k}\rangle'
\ee
which is of a strikingly different form from~\eqref{eq:accidentalshiftwardusr}. However, an explicit computation reveals that they are actually equivalent. We therefore suspect that the same is true in the ultra-slow-roll inflationary case, though verifying this expectation is substantially harder because it requires computing to next order in slow-roll parameters.
}

This instance (and the related time-shift examples in the toy mode case) of the physical mode condition is somewhat different from the dilation case in that it is not possible to write down an early-time identity which becomes the late-time identity in a suitable limit.
Nevertheless, it is clear that that the physical mode condition plays a key role: in all cases we know of, symmetries which satisfy the physical mode condition have corresponding late-time identities. In addition to the examples considered in this paper, the nonlinear symmetries in~\cite{Creminelli:2012qr} satisfy the physical mode condition, and have corresponding Ward identities, which are satisfied.

It is worth emphasizing that, while the presence of this shift adiabatic mode and its associate late-time soft theorem is completely consistent with our picture of inflationary symmetries, our main focus in this paper is somewhat different. Our central goal was to ascertain the fate of the dilation adiabatic mode, which is still present in ultra-slow-roll inflation, but which is not the dominant growing mode at late times. We have seen that this adiabatic mode still has an associated soft theorem, but it requires considering unequal-time correlation functions, in contrast to the more familiar late-time consistency relations.

\section{Conclusions}
Cosmological soft theorems are powerful statements which apply to large classes of models. An observed violation would have profound implications for our understanding of the early universe. Consequently, it is extremely important to understand in detail the conditions under which these relations can be violated, what such a violation would imply, and more generally to understand all aspects of the soft theorems as well as possible.

We have revisited the derivation of inflationary soft theorems, motivated by a number of puzzles involving the well-known relations. In doing so, we have uncovered a generalized consistency relation~\eqref{eq:mainID}, which involves a soft mode insertion in the initial state. This soft theorem has wider applicability than the standard consistency relations, where both the soft and hard modes are evaluated at late times. We have also elucidated the requirements for this unequal-time statement to reduce to the known equal-time version---in particular the so-called {\it physical mode} condition must be satisfied: the field profile generated by a symmetry must match the time dependence of the growing mode solution of the equation of motion. We have shown explicitly that this unequal time relation is satisfied in slow-roll/attractor models of inflation, and have argued that it should also hold in ultra-slow roll. To support this assertion, we have also introduced a toy model which captures many aspects of cosmological field theory. In this model there are precise analogues of many of the puzzles in the inflationary context, and we have shown explicitly how they are resolved.

Given that our investigation was motivated by a number of puzzling
observations about the standard soft theorems, it is worthwhile to
revisit these and summarize their resolution. First, we have clarified
the precise role of the {\it adiabatic mode condition} in the derivation of
inflationary soft theorems. The adiabatic mode condition is: that
the large gauge transformation of interest generates a mode that
is the long wavelength limit of a physical mode (but there is no
requirement that this physical mode is the growing mode {\it i.e.}, one that actually dominates at late times). 
Under this condition, the large gauge transformation effectively acts as a
nonlinearly realized global symmetry. Its implication is
the existence of an unequal-time soft theorem where the soft
mode is at an early time. 
The exact form is given in \eqref{eq:mainID}, which can be
schematically represented as:
\begin{eqnarray}
\label{eq:mainIDschematic}
{\rm adiabatic \,\, mode\,\, condition} \quad \rightarrow \quad \lim_{\vec q \to 0} {\cal E}_i (q) \langle {\cal
  O}_{\phi_f}(\vec k_1...\vec k_N)\phi_i(\vec{q}) \rangle_c {}' +
  {\,\rm c. c.} 
\sim \langle {\cal O}_{\phi_f}(\vec k_1...\vec k_N) \rangle_c {}'\, ,
\end{eqnarray}
where ${\cal E}_i (q)$ is a complex quantity with its real part
related to the early time power spectrum at momentum $q$.
The subscripts $i$ and $f$ refer to initial and final times.
This early-late-time soft theorem does {\it not}
guarantee that there is a late-time version for correlation
functions corresponding to this symmetry. 
The existence of such a late-time soft theorem requires something
more. We have termed this additional requirement {\it the physical mode
condition}---that the nonlinearly-inserted soft mode evolves in time in
the same way that a physical {\it growing-mode} long-wavelength
solution to the equations of motion does. 
By growing-mode we mean the dominant mode {\it i.e.}, the one that
would dominate at late times.
When this is satisfied, early and late time correlation functions with
a soft insertion are related in a simple way, allowing our early time
identity to be promoted to the known late-time statement, given in
\eqref{finaltimeWard} and schematically represented as:
\be
\label{finaltimeWardschematic}
 {\rm physical \,\, mode\,\, condition} \quad \rightarrow \quad \lim_{\vec q \to 0}{1\over
   P_{\phi_f} (q)}  
\langle {\cal O}_{\phi_f}(\vec k_1...\vec k_N)\phi_f(\vec{q})
   \rangle_c {}'  \sim \langle {\cal O}_{\phi_f}(\vec k_1...\vec k_N) \rangle_c {}' \, .
\ee
This is the standard consistency relation where all modes, including
the soft one, are at late times.

We were also motivated by the following observation: in the
ultra-slow-roll inflation model, it is possible to violate Maldacena's
consistency relation (and its SCT generalization)
\cite{Namjoo:2012aa,Martin:2012pe}, despite the fact that spatial dilation and SCT
generate good adiabatic modes and are therefore effectively good global 
symmetries.
What then, is the fate of the dilation and SCT symmetries in these
models? Surely it should have some physical consequences. We have
shown that the consequence of this symmetry is the unequal-time
identity derived in Section~\ref{sec:generalderivation} and shown
schematically above \eqref{eq:mainIDschematic}, which we
expect to be satisfied in all cases where dilation (or SCT) is a
symmetry of the action. 
In the case of ultra-slow roll, it is precisely the violation of the
physical mode condition that invalidates the late-time (equal-time)
statement \eqref{finaltimeWardschematic} corresponding to the dilation
or SCT symmetry.
However, as noted by~\cite{Finelli:2017fml,Finelli:2018upr}, in these
models there can be a second adiabatic mode, which does
satisfy the physical mode condition (see discussion in
Section~\ref{sec:USRphysicalmode}), thus giving rise to a modified
late-time consistency relation. This interesting relation pointed out by
\cite{Finelli:2017fml,Finelli:2018upr} requires additional
transformations beyond spatial dilation and SCT.
Our main focus in this paper is slightly different: what are the physical consequences of spatial
dilation and SCT on their own, given that they still generate good
adiabatic modes in the case of ultra-slow-roll? The answer is
\eqref{eq:mainIDschematic}.

In this paper, we have focused on single field models, but there is a parallel question in the multi-field context; in this case there are also symmetries of the action whose corresponding late-time soft theorems can be violated.
We expect that the techniques introduced in this note will be illuminating in this case as well---one goal would be to gain more insight into the violation of cosmological consistency relations by heavy particles during inflation, possibly opening up new avenues to search for them observationally.

A third motivating puzzle does not involve inflation per se, but rather the
analogous (classical) soft theorems in the large scale structure
context. Understanding these relations is no less important, as they
allow us analytic access to some non-perturbative aspects of LSS
physics. There are two consistency relations in that
context. We are not going into detailed justification of them here,
and refer the readers to \cite{Kehagias:2013yd,Peloso:2013zw} for
the original discussion.\footnote{See also
  \cite{Horn:2014rta} for a discussion of the conceptual separation of
  these two consistency relations.}
For simplicity, we phrase it in terms of the three-point function:
\begin{eqnarray}
\label{potentialshift}
\lim_{\vec q \rightarrow 0} \,\, q^2 {\langle \delta (\vec q, \tau) \delta
  (\vec k_1, \tau_1) \delta (\vec k_2, \tau_2) \rangle {}' \over 
\langle \delta (\vec q, \tau) \delta (-\vec q, \tau) \rangle {}' } = 0
  \, ,
\end{eqnarray}
where $\delta(\vec q, \tau)$ is the density fluctuation at momentum
$\vec q$ and time $\tau$. This is the consistency relation that
follows from shift symmetry in the Newtonian gravitational potential.
There is an additional consistency relation associated with shifting
the gravitational potential by a linear gradient:
\begin{eqnarray}
\label{potentialLinear}
\lim_{\vec q \rightarrow 0} \,\, \partial_{q^j}  \left[ q^2 {\langle \delta (\vec q, \tau) \delta
  (\vec k_1, \tau_1) \delta (\vec k_2, \tau_2) \rangle {}' \over 
\langle \delta (\vec q, \tau) \delta (-\vec q, \tau) \rangle {}' }
  \right] = 
- \sum_{a=1}^2 {D(\tau_a) \over D(\tau)}
{k_a {}^j }
\langle \delta(\vec k_1 , \tau_1) \delta(\vec k_2, \tau_2) \rangle {}'
  \, ,
\end{eqnarray}
where $D(\tau)$ is the linear growth factor at time $\tau$ for the
density fluctuation $\delta$ in a matter + cosmological constant
universe {\it i.e.}, $\delta(\tau) \propto D(\tau)$ in linear perturbation
theory, regardless of momentum.
In both consistency relations, the times $\tau, \tau_1, \tau_2$ are arbitrary.
An explicit perturbative computation of the three-point function
gives:
\begin{eqnarray}
\label{LSS3point}
&& \lim_{\vec q \rightarrow 0} \,\, q^2 {\langle \delta (\vec q,
  \tau) \delta 
  (\vec k_1, \tau_1) \delta (\vec k_2, \tau_2) \rangle {}' \over 
\langle \delta (\vec q, \tau) \delta (-\vec q, \tau) \rangle {}' } 
= - \sum_{a=1}^2 {D(\tau_a) \over D(\tau)}
\left( {\vec q \cdot \vec k_a + O(q^2)} \right)
\langle \delta(\vec k_1 , \tau_1) \delta(-\vec k_1, \tau_2) \rangle {}'
\nonumber \\
&& \quad \quad + O(q^4) {D(\tau)^2 \over D(\tau_1) D(\tau_2)}
{\langle \delta(\vec k_1, \tau_1) \delta(-\vec k_1, \tau_2) \rangle'
\langle \delta(\vec k_2, \tau_1) \delta(-\vec k_2, \tau_2) \rangle'
\over \langle \delta(\vec q, \tau) \delta(-\vec q, \tau) \rangle'}
\end{eqnarray}
The first line is consistent with the soft theorems expressed in
\eqref{potentialshift} and \eqref{potentialLinear}. 
The question is under what condition the second line can be ignored.
Suppose the mass density power spectrum in the soft limit scales as
$\langle \delta(\vec q) \delta (-\vec q) \rangle' \sim q^m$, the term
on the second line scales as $q^{4-m}$. 
For this not to cause a break down of \eqref{potentialshift}, we need
$m < 4$. To respect \eqref{potentialLinear}, we need $m < 3$.
In other words, the soft power spectrum cannot be too blue.
From our derivation in Section \ref{sec:generalderivation},
it is not immediately apparent why such a constraint should exist---the assumed
initial condition there is gaussian random, consistent with the way
the (classical) perturbative three-point function above is obtained.
What is interesting is to analyze this perturbative result in a limit
that corresponds to the formulation of the early-late-time soft
theorem \eqref{eq:mainIDschematic} {\it i.e.}, let's take the early
time limit for $\tau$, time of the soft mode.
We see that the terms on the first line of \eqref{LSS3point} scale as
$D(\tau)^{-1}$, whereas the potentially problematic term (on the second
line) scales as $D(\tau)^0$.\footnote{Note that the power spectrum
$\langle \delta(\vec q, \tau) \delta(-\vec q, \tau) \rangle'$ has time dependence
$\sim D(\tau)^2$.} This means that in the early time limit for the
soft mode, the potentially problematic term can be ignored.
This supports the idea that the early-late-time version of the
consistency relations is more robust.\footnote{The reader might wonder what causes the break-down
of the later-time version in the case of $m < 4$ or $m < 3$. 
There seems to be a technical reason: the break-down of perturbation
theory itself when the soft power spectrum is very blue, i.e. the
perturbative expressions for the $2-$ and $3-$point functions cannot be trusted.
}
More discussions of classical soft theorems, like the ones in large
scale structure, can be found in Appendix \ref{app:classicalcheck}.

The results of this paper provide a firmer conceptual understanding of inflationary soft theorems, and provide a more general soft theorem, which should hold in a larger class of models. There are a number of natural future directions which will be interesting to consider. As we have discussed, we are optimistic that the techniques introduced in this note will be useful to gain more insight into violations of the consistency relation in cases with additional fields or non-vacuum initial conditions. Additionally, considering internal squeezed limits from the perspective of the path integral would be quite interesting---this may shed some light on the fate of the disconnected contribution whose presence we cannot rule out (see Appendix~\ref{app:4point}). It would also be instructive to connect our results to similar ones in the context of the flat space $S$-matrix~\cite{Avery:2015rga}. In particular, there should be some unified origin of the cosmological consistency relations and flat space gauge theory soft theorems. Finally, we have focused on the translation of the unequal-time consistency relation into a purely late time relation, but it is worth considering whether the unequal time relation can be tested on its own---it is interesting to speculate that since $\dot\zeta\neq0$ in ultra-slow-roll models that some information about the initial state could be captured by quantum correlations of perturbations. We hope to return to all of these considerations in the near future.

\vspace{-.2cm}
\paragraph{Acknowledgements:} We would like to thank Miguel Campiglia,
Andy Cohen, Paolo Creminelli, Angelo Esposito, Garrett Goon, Xinyu Li, Alberto Nicolis, Toshifumi Noumi, Bob Penna, Rachel Rosen, Martin Sloth and Yi Wang for helpful discussions.
This work is supported in part by 
NASA grant NXX16AB27G and DOE grant DE-SC011941. The research of SW is supported in part by the Croucher Foundation.

\appendix

\section{Normalization of correlation functions and and the $(N+2)$-point term}
\label{app:4point}

Our two goals in this Appendix are: (1) show how to enforce the
normalization of the wavefunctional; (2) argue that the $(N+2)$-point term
(the term ${\cal V}$ defined in \eqref{Vdef}) vanishes in the $\tau_i \rightarrow
-\infty$ (far past) limit. 

A simple way to deal with the normalization issue is to include
a denominator when thinking about an expectation value.
In other words, replace \eqref{eq:genfunctional} by
\be
\langle{\cal O}_{\vp_f}(\vec k_1,\cdots, \vec k_N)\rangle = {\cal N}  \int{\cal D}\vp_f\, {\cal O}_{\vp_f}(\vec k_1,\cdots, \vec
k_N)\,\left\lvert\Psi[\vp_f]\right\rvert^2 , \quad  {\cal N} = \frac{1}{ \int {\cal D}\vp_f |\Psi[\vp_f]|^2 }
\label{eq:genfunctionalNorm}
\ee
Defined this way, there is no need to explicitly keep track of how the
integration measure might change by an overall normalization under
the symmetry transformation of interest.
Recall that we obtain \eqref{eq:wald00} by considering the first order variation of
the numerator $\int {\cal D}\vp_f {\cal O}_{\vp_f}
\left\lvert\Psi[\vp_f]\right\rvert^2$. 
This should be augmented by the variation in the
normalization factor ${\cal N}$, which can be read off from \eqref{eq:wald00} by setting
${\cal O}_{\vp_f} = 1$ and $\delta\vev{1}= 0$. We see that
under the symmetry transformation of interest:
\begin{align}
\label{vary1}
0 &=\delta{\cal N} -\left[ \int{\cal D}\vp_f  \Psi^*[\vp_f]\left(\int{\cal D}\vp_i
  \langle\vp_f\rvert\vp_i\rangle\left(\int\frac{\rd^3p}{(2\pi)^3}
{\cal E}_i (p) \vp_i(\vec p)
\delta_{\rm L}\left[\vp_i(-\vec p)\right]\right)\Psi_0[\vp_i]\right)
   + {\rm c. c.} \right] \nonumber \\
   &=\delta{\cal N} - \left[ \int\frac{\rd^3p}{(2\pi)^3} {\cal E}_i (p)\langle 0_{\rm in} | \phi_i(\vec p) \delta_{\rm L} \phi_i(-\vec p) |0_{\rm in} \rangle +  ``{\rm c. c.}"\right] .
\end{align}
This variation thus involves a two-point function of $\vp_i$ (there's
also a one-point function contribution which has been set to zero). Both terms are formally divergent, however the above relation is helpful in canceling divergent terms in the identity. The correct consistency relation follows from the variation of the
expectation value defined in \eqref{eq:genfunctionalNorm}.
There is thus a correction to \eqref{eq:wald00} coming from $\delta{\cal N}$ of the form  $\delta{\cal N}\langle {\cal O}_{\vp_f}
\rangle$ where we can replace $\delta {\cal N}$ using \eqref{vary1}. The result is \eqref{eq:almostfinalward} after
removing complete sets of states, with ${\cal
  V}$ defined in \eqref{Vdef}, reproduced here:
\begin{align}
\label{eq:almostfinalwardApp}
\tiny
& {\cal V} + 
D_q(t_i)\left({\cal E}_i (q) \langle 0_{\rm in} | {\cal
  O}_{\phi_f}(\vec k_1...\vec k_N)\phi_i(\vec{q}) | 0_{\rm in}
   \rangle\right)\bigg\rvert_{\vec q =0} + {\rm c. c.} 
= \langle 0_{\rm in} | \delta{\cal O}_{\phi_f}(\vec k_1...\vec k_N)
  | 0_{\rm in} \rangle \, ,
\end{align}
and 
\begin{eqnarray}
\label{VdefApp}
\tiny
&& {\cal V} \equiv \int\frac{\rd^3p}{(2\pi)^3} {\cal E}_i (p) 
\Big[ \langle 0_{\rm in} | {\cal O}_{\phi_f}(\vec k_1...\vec k_N) \phi_i(\vec p) \delta_{\rm L} \phi_i(-\vec p) |
  0_{\rm in} \rangle 
\nonumber \\
&&
\qquad \qquad  - \langle 0_{\rm in} | {\cal O}_{\phi_f}(\vec
   k_1...\vec k_N) | 0_{\rm in} \rangle 
\langle 0_{\rm in} | \phi_i(\vec p) \delta_{\rm L} \phi_i(-\vec p) |
  0_{\rm in} \rangle \Big] +  ``{\rm c. c.}" \, .
\end{eqnarray}
The subtracted piece in ${\cal V}$ comes from
the variation of the denominator as explained above.
Written as such, we see that 
in eq. (\ref{eq:almostfinalwardApp}), ${\cal V}$
is an $N+2$-point function (with a certain disconnected piece
subtracted off), the second term is an $N+1$-point function,
and the right hand side is an $N$-point function.

We would like to investigate under what conditions ${\cal V}$ 
vanishes.
When one calculates the $(N+2)$-point functions inside of ${\cal V}$ in the interaction picture, each $\phi_i (\vec p)$ comes with rapid oscillation due to
presence of $e^{\pm i p \tau_i}|_{\tau_i =-\infty}$. We will make use of this fact.
In ${\cal V}$ the divergent contraction $\vev{{\cal O}_{\phi_f}} \vev{\phi_i(\vec p) \delta_{\rm L} \phi_i(-\vec p)}'(2\pi)^3 \delta^3({\vec{p} -\vec p})$ in the first term is canceled by the second term. We are left with a connected term and many disconnected contractions. 
\begin{align}
\mbox{I.}& \; \vev{{\cal O}_{\phi_f}(\vec k_1...\vec k_N) \phi_i(\vec p )\delta_{\rm L}\phi_i(-\vec p)}'_c(2\pi)^3\delta(\vec k_1 +\vec k_2),  \\
\mbox{II.}& \; \vev{{\cal O}_{\phi_f}(\vec k_1...\vec k_I)} \vev{ {\cal O}_{\phi_f}(\vec k_I...\vec k_J)\phi_i(\vec p)}'_c\vev{{\cal O}_{\phi_f}(\vec k_J...\vec k_N)\delta_{\rm L}\phi_i(-\vec p)}'_c \nonumber \\
& \; \times(2\pi)^6 \delta^3 \left(\sum_{n=I}^J \vec k_n+\vec p\right)\delta^3\left(\sum_{n=J}^N\vec k_n-\vec p\right) +``\mbox{partitions of } \{\vec k_1,\dots,\vec k_N\}",  
\end{align}
We first focus on the connected term (I). It gives an integral of the form 
\begin{equation}
\int\frac{\rd^3p}{(2\pi)^3} (\dots) e^{2 ip \tau_i} 
\end{equation}
in the cases of interest. Since the integrand is infinitely
oscillatory as $\tau_i \rightarrow -\infty$, the integral vanishes as
long as the function inside the parenthesis is smooth enough in $p$.
For disconnected contractions (II), after performing the integral over
$p$ using one of the
delta functions, we have
\begin{equation}
 {\cal E}_{|\vec k_J + \dots + \vec k_N|}\vev{{\cal O}_{\phi_f}(\vec k_1...\vec k_I)} \vev{ {\cal O}_{\phi_f}(\vec k_I...\vec k_J)\phi_i(\vec p)}'_c\vev{{\cal O}_{\phi_f}(\vec k_J...\vec k_N)\delta_{\rm L}\phi_i(-\vec p)}'_c (2\pi)^3 \delta^3 \left(\vec k_I+\dots +\vec k_N \right)
\end{equation}
which now contains an oscillatory function $e^{2i |\vec k_J + \dots +
  \vec k_N| \tau_i}$.  As long as $|\vec k_J + \dots + \vec k_N|  \neq
0$, this is also infinitely oscillatory so does not contribute to the
identity. There is the possibility that this subset of momenta $\vec
k_J \dots \vec k_N$ sum up to zero (a configuration of momenta which
is squeezed internally). The momenta configuration we are interested in is the hierarchical region that the soft mode at initial time is already outside the horizon $q \tau_i \to 0$ while the momenta of other modes (or any combination of them) are not so soft such that $ |\vec k_J + \dots + \vec k_N| \tau_i \to -\infty$.  However, it could happen that  $|\vec k_J + \dots + \vec k_N| \tau_i \to 0$ such that $|\vec k_J + \dots + \vec k_N|$ approaches $0$ at the same rate as $q$. For example, in the case of $N=3$ we have the disconnected contribution of the form 
\begin{equation}
{\cal E}_{k_3}\vev{ {\cal O}_{\phi_f}(\vec k_1,\vec k_2)\phi_i(\vec k_3)}'_c\vev{{\cal O}_{\phi_f}(\vec k_3)\delta_{\rm L}\phi_i(-\vec k_3)}'_c (2\pi)^3 \delta^3 \left(\vec k_1+\vec k_2+\vec k_3 \right).
\end{equation}
When $k_3\to0$ as the same rate of $q$ such that $k_3 \tau_i \to 0$, this is a double soft limit and it has a non-trivial structure of the form presented in \cite{Joyce:2014aqa}.
It is possible that
this $(N+2)$-point term has a non-negligible contribution in some double soft or internal collapsed limits.\footnote{We thank Xinyu Li for discussions on related issues.}

The subtle possibility that $|\vec k_J + \dots + \vec k_N|  \to 0$
does not happen for $N=2$ which relates three point function and two
point function that we checked in this paper so we are safe to remove
the four point terms. In addition, we can see from the classical check
in Appendix \ref{app:classicalcheck} that indeed the four point
function is negligible in the large separation limit between
final and initial times.

\section{Transformation of the wavefunction under large gauge transformations}
\label{app:wavefgaugetrans}
One of the more possibly surprising elements of our derivation of an unequal-time Ward identity in Section~\ref{sec:pathintderiv} is that we do not demand that the wavefunctional is invariant under the relevant symmetry transformations.
On one hand this seems peculiar because typically we imagine that the physical meaning of a symmetry is that the probabilities of physical configurations are not changed by symmetry transformations---particularly in the inflationary case where the relevant symmetry transformations are {\it gauge} transformations. On the other hand, it is common in spontaneous symmetry breaking for the dynamics to possess symmetries which are violated by particular choices of states. It is in this sense that we can think of large gauge transformations as spontaneously broken. Under these transformations, states transform in a nontrivial way, in particular these (broken) symmetries insert soft gauge fields.

In this Appendix we wish to make these results more intuitive. We will consider the cases of electromagnetism and gravity: in both cases, the wavefunction is not strictly gauge invariant, but rather it transforms under gauge transformations that do not go to zero at spatial infinity. 
\subsection{Electromagnetism}
We first consider the case of electromagnetism, to see that the wavefunctional is not necessarily invariant under all gauge transformations, but rather transforms nontrivially under large gauge transformations.\footnote{Our discussion follows that of Preskill, which can be found in~\href{http://www.theory.caltech.edu/\%7Epreskill/ph230/notes2000/230Lectures7-10-Page53-109.pdf}{\tt these QFT notes}.}

We begin by briefly describing Schr\"odinger field theory for Maxwell electromagnetism. More details can be found in~\cite{Hatfield:1992rz}. The action for a massless spin-1 field is
\be
S = -\frac{1}{4}\int\rd^4xF_{\mu\nu}F^{\mu\nu},
\ee
and is invariant under U$(1)$ gauge transformations that act as $\delta A_\mu = \partial_\mu\Lambda(x)$.
Performing a $3+1$ split, the action can be written as
\be
S=\int\rd^4x\left(\frac{1}{2}\dot A_i^2-\dot A_i \partial_i A_0+\frac{1}{2}(\partial_iA_0)^2-\frac{1}{2}\partial_i A_j\partial_i A_j+\frac{1}{2}\partial_iA_j\partial_j A_i\right),
\ee
where $\dot A_i \equiv \partial_0 A_i$.
If we define the electric and magnetic fields by
\begin{align}
E^i &\equiv F^{0i}  = -\dot A_i+\partial^iA_0\\
B_i &\equiv \epsilon_{ijk}F_{jk}=\epsilon_{ijk}\partial_j A_k,
\end{align}
then the lagrangian is simply (up to a boundary term) ${\cal L} = \frac{1}{2}\left(E^2 - B^2\right)/2$.
In order to construct the hamiltonian, we need the momenta canonically conjugate to the components of $A$:
\begin{align}
\Pi_0 &\equiv \frac{\delta{\cal L}}{\delta\dot A_0} = 0\\
\Pi_i &\equiv \frac{\delta{\cal L}}{\delta\dot A_i}  = \dot A_i-\partial_iA_0 = -E_i.
\end{align}
Notice that the electric field is the canonically conjugate variable to the vector potential, while $A_0$ enforces a constraint and has no conjugate momentum. Using this, we can write the hamiltonian:
\be
H = \int\rd^3 x\,\Pi_i \dot A_i - {\cal L} 
= \int\rd^3 x\left(\frac{1}{2}(E^2+B^2) -\partial_iE_i A_0\right).
\label{eq:EMhamiltonian}
\ee
Maxwell's equations then follow from Hamilton's equations. The equation corresponding to $A_0$ is
\be
\vec\nabla\cdot \vec E = 0.
\ee
Because of  its vanishing momentum, $A_0$ is {\it not} dynamical, but rather enforces the constraint that $\vec\nabla\cdot\vec E = 0$ (Gauss' law).

\subsubsection{Quantization}
The presence of the Gauss' law constraint makes quantization of this system somewhat subtle.
We want to impose the canonical commutation relations on the $A_i, E_i$ pair:
\be
\label{eq:canonicalcommutator}
\left[A_i(\vec x, t), E_j(\vec y,t)\right] =- i\delta_{ij}\,\delta (\vec x-\vec y).
\ee
The simplest thing we could do would be to impose $\vec\nabla\cdot \vec E = 0$ as an operator statement ({\it i.e.} that it annihilates all states in the Hilbert space). However, this is not compatible with the commutator~\eqref{eq:canonicalcommutator}, because the right-hand side is not divergence-less. The best we can do is to work in enlarged Hilbert space which includes states for which $\vec\nabla\cdot\vec E \left\lvert\Psi\right\rangle \neq0$, but then demand that all physical states are annihilated by the constraint $\vec\nabla\cdot \vec E\,\lvert\Psi\rangle_{\rm phys.} =0$.\footnote{An alternative option, which is less convenient for us, is to modify the canonical commutation relations so that Gauss' law does hold as an operator statement.}

We are interested in the wavefunctional representation of states, so a useful gauge to work in is temporal gauge, $A_0 = 0$.
This gauge does not completely fix the gauge symmetry, it is preserved by any time-independent gauge transformation.\footnote{There is actually a further restriction on residual large gauge transformations, which is the E\&M analogue of the adiabatic mode condition. Residual gauge parameters must be both time-independent and harmonic~\cite{Mirbabayi:2016xvc}.}
Substitution of the temporal gauge condition into the hamiltonian~\eqref{eq:EMhamiltonian} causes us to lose Gauss' law as an equation of motion, but this is ok because we are going to impose it by hand by only calling things which satisfy it physical states.
Time evolution in the Hilbert space of states is generated by the hamiltonian:
\be
i\partial_t \lvert\Psi\rangle = H\lvert\Psi\rangle,
\ee
where physical states are the ones annihilated by Gauss' law.

We want to work in the representation of the Hilbert space in terms of $A_i(x)$ field eigenstates:
\be
A_i(x)\lvert a\rangle = a_i(x)\lvert a\rangle.
\ee
Any state then has a wavefunctional representation, $\Psi[a_i]\equiv \langle a\rvert \Psi\rangle$.
Correlation functions are defined as in~\eqref{eq:schrodingercorrs}. In this basis, the momentum operator acts as a functional derivative with respect to $a_i$
\be
E_i(\vec x)  = -i\frac{\delta}{\delta{a_i(\vec x)}},
\label{eq:elecfieldmomentumrel}
\ee
so that the Gauss constraint takes the form
\be
\label{eq:gausslaw}
\partial_i\frac{\delta}{\delta a_i(\vec x)}\Psi[a]=0.
\ee

With this information, we are now able to determine how the wavefunctional transforms under a large gauge transformation. Under a general gauge transformation, $a_i\mapsto a_i+\partial_i\Lambda$, the wavefunctional shifts as
\be
\delta\Psi[a] = \int\rd^3x \frac{\delta\Psi[a]}{\delta a_i(\vec x)}\partial_i \Lambda(\vec x).
\ee
Notice that if we integrate this expression by parts (keeping the boundary term), we obtain
\be
\delta\Psi[a] = -\int\rd^3x \Lambda(\vec x)\partial_i\frac{\delta\Psi[a]}{\delta a_i(\vec x)}+\oint_{S^2} n^i \Lambda(\vec x) \frac{\delta\Psi[a]}{\delta a_i(\vec x)}.
\ee
For a physical configuration, Gauss' law tells us that the first term vanishes, and we are left with the boundary term
\be
\label{eq:largegauge}
\delta\Psi[a] = \oint_{S^2} n^i \Lambda(\vec x) \frac{\delta\Psi[a]}{\delta a_i(\vec x)} = i\oint_{S^2} \vec n\cdot \vec E\,\Lambda(\vec x)\,\Psi[a].
\ee
where $n^i$ is the unit normal to the 2-sphere at infinity and in the second equality we have used~\eqref{eq:elecfieldmomentumrel}. For small gauge transformations, where $\Lambda\to0$ as $x\to\infty$, this is zero and the wavefunction does not transform. However, under large gauge transformations, this can be nonzero. 
In particular, for $\Lambda = {\rm const}.$, the state transforms by a phase which is the total electric charge. We can write this expression more covariantly as
\be
\delta\Psi[a] = i\oint_{S^2} \Lambda(\vec x)\star F\,\Psi[a].
\ee
The large gauge charges appearing here are the same as the ones of interest in flat space scattering~\cite{Strominger:2017zoo}.

\subsection{Gravity}
We have just seen that under a large gauge transformation, the wavefunction of electromagnetism transforms by the corresponding charge supported on the sphere at infinity. We now outline the same construction in the gravitational case. For simplicity, we consider pure gravity, but the addition of matter should be straightforward. We will see that in this case as well, the wavefunctional is not strictly invariant under gauge transformations, but rather can transform nontrivially under large gauge transformations. See also~\cite{Pimentel:2013gza} for a recent discussion of the wavefunctional approach to gravity.

The action for gravity is given by the Einstein--Hilbert term (plus the GHY boundary term)\footnote{For simplicity, we set $M_{\rm Pl} = 1$ in this Section.}
\be
S = \frac{1}{2}\int\rd^4x\sqrt{-g}R+\oint \sqrt{h}K.
\ee
In the canonical formalism, we perform a $3+1$ split into ADM variables, in terms of which the metric is parameterized as~\cite{Arnowitt:1962hi,DeWitt:1967yk,Wald:1984rg}
\be
\rd s^2 = -N^2 \rd t^2 +h_{ij}(\rd x^i+N^i\rd t)(\rd x^j+N^j\rd t).
\ee
Inserting this decomposition into the Einstein--Hilbert action yields
\be
S = \frac{1}{2}\int \rd^4x \sqrt{h}N\left(R^{(3)}+K_{ij}K^{ij} - K^2\right),
\ee
where $K_{ij}$ is the extrinsic curvature of a spatial slice
\be
K_{ij} = N^{-1}\left(\frac{1}{2}\dot h_{ij}-\nabla_{(i} N_{j)}\right).
\ee
From this, we can extract the momenta conjugate to the coordinates, $h_{ij}, N_i, N$:
\begin{align}
\Pi^{ij} &\equiv \frac{\delta{\cal L}}{\delta\dot h_{ij}}= \sqrt{h}\left(K^{ij}-Kh^{ij}\right), \\
\Pi^i &\equiv \frac{\delta{\cal L}}{\delta\dot N_i} = 0,\\
\Pi &\equiv  \frac{\delta{\cal L}}{\delta\dot N} = 0.
\end{align}
As in electromagnetism, the lapse, $N$, and shift, $N_i$, impose constraints, so their conjugate momenta vanish. Again, we can write the hamiltonian by legendre transforming the lagrangian~\cite{Wald:1984rg}
\be
H = \int\rd^3x\sqrt{h} \left(-NR^{(3)}+h^{-1}N\left( \Pi^{ij}\Pi_{ij}-\frac{1}{2}( \Pi_i^i)^2\right)-2N_i\nabla_j\left(h^{-\frac{1}{2}}\Pi^{ij}\right)
\right),
\ee
where all indices are raised and lowered with $h_{ij}$.
The Einstein equations then follow from Hamilton's equations. The ones that we are interested in are the analogues of the Gauss' law constraint---the equations of motion for $N$ and $N_i$. The lapse enforces the hamiltonian constraint
\be
-R^{(3)}+h^{-1}\left( \Pi^{ij}\Pi_{ij}-\frac{1}{2}( \Pi_i^i)^2\right) = 0,
\ee
while the shift imposes the momentum constraint
\be
\nabla_j\left(h^{-\frac{1}{2}}\Pi^{ij}\right) = 0.
\ee

\subsubsection{Quantization}
We now consider the canonical quantization of the system. We pick a gauge for $h_{ij}$ where it is transverse-traceless and express states in the eigenbasis of this operator at time $t$:
\be
\Psi[ h_{ij}] \equiv \langle h\rvert \Psi\rangle,
\ee
where $\hat h_{ij}(\vec x)\lvert h\rangle = h_{ij}(\vec x)\lvert h\rangle$. As in the electromagnetism case, we cannot impose the constraints as operator equations, but we can demand that physical states satisfy
\begin{align}
H \Psi[h_{ij}] &= 0\\
\nabla_j\left(h^{-\frac{1}{2}}\frac{\delta \Psi[h]}{\delta h_{ij}(\vec x)}\right) &= 0,
\label{eq:mom1const}
\end{align}
where we have used the Schr\"odinger representation of the momentum
\be
\Pi_{ij}(\vec x) = -i\frac{\delta}{\delta h_{ij}(\vec x)}.
\label{eq:gravmomdefn}
\ee
The hamiltonian constraint will not be of much importance to us. Instead, we want to investigate how the wavefunctional transforms under residual spatial diffeomorphisms: $h_{ij}\mapsto h_{\ij}+\nabla_{(i}\xi_{j)}$. The wavefunction shifts as
\be
\delta\Psi[h] = \int\rd^3x \sqrt{h}\frac{\delta \Psi[h]}{\delta h_{ij}(x)}\nabla_{(i}\xi_{j)} = -\int\rd^3x\sqrt{h} \xi_j\nabla_i\frac{\delta \Psi[h]}{\delta h_{ij}(x)}+\oint_{S^2} n_i \xi_j \frac{\delta \Psi[h]}{\delta h_{ij}(x)}.
\ee
Where in the second equality we have integrated by parts. The first term vanishes due to the momentum constraint~\eqref{eq:mom1const} (the spatial derivative is spatial metric-compatible). We can use~\eqref{eq:gravmomdefn} to express the boundary term in terms of the momentum:
\be
\delta\Psi[h]  = i\oint_{S^2}\sqrt{\sigma} n_i \xi_j\left(K^{ij}-K h^{ij}\right)\, \Psi[h],
\ee 
where $\sigma_{ab}$ is the induced metric on the sphere at infinity. This can be recognized as the integral of the Brown--York stress tensor~\cite{Brown:1992br}, contracted with the diffeomorphism parameter, $\xi$. We therefore see that the gravitational wavefunctional transforms by the integral of the Brown--York stress tensor at infinity:
\be
\delta\Psi[h]  = i\oint_{S^2}\sqrt{\sigma} n_i \xi_jT^{ij}_{\rm BY}\, \Psi[h],
\label{eq:bycharge}
\ee
which plays an analogous role to the electric charge (or other large gauge charges) in electromagnetism.

\subsubsection{Discussion}
We now would like to comment on the relevance of these results to the Ward identities we derived in Section~\ref{sec:pathintderiv}. The main lesson we want to abstract from this discussion is that in general the wavefunction can (and will) transform under diffeomorphisms that do not go to zero at spatial infinity. How then, are refs.~\cite{Pimentel:2013gza,Goldberger:2013rsa} able to derive the consistency relations by assuming that the wavefunction is gauge-invariant? The resolution of this apparent paradox lies in the fact that for slow-roll models, the late-time wavefunctional has simple transformation properties under dilations and SCT: essentially only the gaussian part shifts, corresponding to a shift of the 1-point function, while all the higher coefficient functions conspire to cancel against each other, essentially because of Maldacena's consistency relation~\cite{Kundu:2015xta}. Therefore, if one only ever asks questions about correlation functions with more than one external leg, the overall shift of the wavefunctional is irrelevant. 

Another way of phrasing this is that for attractor models, the 1PI effective action is invariant under large gauge transformations, up to a shift of the quadratic piece, producing a tadpole term. This term is, however, irrelevant if one is interested in higher-point correlation functions, which correspond to taking multiple functional derivatives of the 1PI action.\footnote{We thank Angelo Esposito for pointing this out to us.}

It is also worth stressing that in this paper, we have assumed that the {\it initial} wavefunctional is a gaussian, and therefore we know its transformation properties. We then infer the transformation properties of the late-time wavefunction by time-evolving from the shifted initial state. An alternative would be to make some assumption about the transformation properties of the late-time wavefunction. This is the tack taken in~\cite{Kundu:2015xta} to derive the consistency relations. However, the assumption that only the gaussian part of the late-time wavefunction shifts is apparently incorrect in ultra-slow-roll models, as they violate the consistency relation.\footnote{In principle, it should be possible to {\it derive} the late-time transformation behavior of gravity coupled to matter by generalizing~\eqref{eq:bycharge} to this case.}

\section{Perturbative verification of the physical mode condition}  
\label{sec:pertproof}
%
%
%

The discussion in Section~\ref{sec:physicalmodes} is somewhat heuristic, so in this Appendix we will give an explicit proof in perturbation theory that
only symmetries satisfying the physical mode condition possess an equal time identity relating three-point functions with two-point functions. The argument may look similar to arguments employed in~\cite{Ganc:2010ff, RenauxPetel:2010ty,Creminelli:2011rh} in the context of slow-roll inflation, but for us the time dependence of long-wavelength modes will be crucial. Most of the explicit formulae we write down are in the context of the toy model introduced in Section~\ref{sec:toymodel1}, but the technique is fully general, it is just for notational convenience that we specialize to the algebraically-simpler case.

\subsection{Soft limit of the 3-point function in perturbation theory}
\label{sec:pertproof1}

In perturbation theory, the tree-level equal-time 3-point function is given by the expression 
\be
 \vev{\zeta_{\vec{q}}(\tau) \zeta_{\vec{k}_1}(\tau)\zeta_{\vec{k}_2}(\tau)}'  = -i\vev{[\zeta_{\vec{q}}(t) \zeta_{\vec{k}_1}(\tau)\zeta_{\vec{k}_2}(\tau) ,  \int^{\tau} \rd \tau' H^{(3)}_{\rm int}(\tau')  ]}',
\ee
where $H^{(3)}_{\rm int}$ is the cubic interaction hamiltonian. In the following, we are going to take advantage of the fact that, at this order, the interaction hamiltonian is related to the lagrangian in a very simple way: $H^{(3)}_{\rm int} = -L^{(3)}$, where $L^{(3)}$ denotes the cubic lagrangian. We therefore have
\be
 \vev{\zeta_{\vec{q}}(\tau) \zeta_{\vec{k}_1}(\tau)\zeta_{\vec{k}_2}(\tau)}'  = i\vev{[\zeta_{\vec{q}}(\tau) \zeta_{\vec{k}_1}(\tau)\zeta_{\vec{k}_2}(\tau) ,  \int^{\tau} \rd \tau' L^{(3)}(\tau')  ]}'.
\ee
In order to actually compute this commutator, we expand the fields in modes as
\be
\hat\zeta(x) = \int\frac{\rd^3k}{(2\pi)^3} e^{-i\vec k\cdot\vec x}\left( u^*_{k} \hat a_{\vec k}^\dagger+u_{k} \hat a_{-\vec k}\right),
\ee
so that the above takes the schematic form 
\be
 \vev{\zeta_{\vec{q}}(\tau) \zeta_{\vec{k}_1}(\tau)\zeta_{\vec{k}_2}(\tau)}'  = 2 {\rm Im} \bigg( u^*_q(\tau) u^*_{k_1}(\tau)  u^*_{k_2}(\tau) \int^{\tau} \rd \tau' L^{(3)}[ u_q, u_{k_1},  u_{k_2} ](\tau')  \bigg) +{\rm perms.},
\label{eq:pertproofeq1}
\ee
where we think of the lagrangian as a function of the mode functions themselves and perms. denotes the additional momentum orderings coming from Wick contractions. We now want to introduce some symmetry information: assume that the underlying action has a symmetry of the form $\delta\zeta = \delta_{\rm NL}\zeta+\delta_{\rm L}\zeta$, where $\delta_{\rm NL}$ and $\delta_{\rm L}$ are parts of the field variation independent of $\zeta$ and first order in $\zeta$, respectively. In order for this to be a symmetry up to ${\cal O}(\zeta^2)$, there must be a cancellation between the linear variation of the quadratic action and the nonlinear variation of the cubic action, up to a boundary term
\be
\int\rd^4x\left(\delta_{\rm L}{\cal L}^{(2)}+\delta_{\rm NL}{\cal L}^{(3)} \right) = \int\rd^4x\,\partial_\mu B^\mu.
\ee
We want to use this fact to relate $L^{(3)}$ to $L^{(2)}$ written in terms of the mode functions, $u_k$, as\footnote{The operators $\delta_{\rm NL}$ and $\delta_{\rm L}$ acting on the mode functions should just be thought of as notation capturing the effects of substituting mode functions into the symmetry transformed lagrangian. For example, for the symmetry
\be
\delta \zeta = \tau^{-3} + \frac{1}{2}\tau^{-2} \partial_\tau\zeta ,
\ee
the shifted mode function is
\be
\delta u_q = (2\pi)^3\delta^3(\vec q)\tau^{-3} + \frac{1}{2} \tau^{-2}u'_q  \equiv   (2\pi)^3 \delta^3(\vec q) \delta_{\rm NL}u_q +\delta_{\rm L} u_q.
\ee
}
\be
L^{(3)}[ \delta_{\rm NL}u_q, u_{k_1},  u_{k_2} ] +{\rm perms}.   =  -  L^{(2)}[ \delta_{\rm L} u_{k_1},  u_{k_2} ]+ \partial_\tau B[u_{k_1},u_{k_2}]+{\rm perms},
\label{eq:perturproofl3l2}
\ee
where $\partial_\tau B$ is the contribution from possible (temporal) boundary terms. In order to utilize this symmetry information, we want to match the long-wavelength limit of the mode functions to the time dependence of the nonlinear symmetry. We note that the mode function itself has the following expansion as $\vec q\to 0$
\be
 u_q(t) \overset{q\rightarrow 0}{\longrightarrow} g(\tau)q^\alpha(1+q^2\tau^2+\cdots).
\ee
Consider taking the $\vec q\to 0$ limit of~\eqref{eq:pertproofeq1}; if $g(\tau)$ matches the time dependence of $\delta_{\rm NL}u_q =  g(\tau)$ we can think of $u_{q=0}(\tau)$ as $\delta_{\rm NL}u_q$ and use~\eqref{eq:perturproofl3l2} inside the integral, along with replacing $u_q^*$ with its long-wavelength expression to obtain
\begin{align}
\lim_{q\to0} \vev{\zeta_{\vec{q}} \zeta_{\vec{k}_1}\zeta_{\vec{k}_2}}' 
= \lim_{q\to0}-2 q^{2\alpha} g(\tau)  \mbox{Im} \bigg( u^*_{k_1}(\tau)u^*_{k_2}(\tau)\Big(  \int^{\tau}\rd t'L^{(2)}&[\delta_{\rm L} u_{k_1}, u_{k_2}](\tau') -B_{k_1}\big\rvert_{\tau} \Big)+{\rm perms}. \bigg).
\label{eq:pertproofeq3}
\end{align}  
Here we have taken advantage of the momentum-conserving delta function on the right-hand-side to write the boundary term contributions as functions purely of one of the moment, $\vec k_1$.
Note that the boundary terms, $B_{k_1}$, only contribute when they contain time derivatives, $u'_k$. 
It is worth emphasizing that this trick can only be done when the time dependence of $u_{q=0}$ and $\delta_{\rm NL}u_q$ match. This is the perturbative manifestation of the physical mode condition. 

We can, in fact, simplify things further. We know the explicit form of the quadratic lagrangian---it will generically be of the form of the quadratic part of~\eqref{eq:toymodelaction}, even in the inflationary case---its variation under a linear symmetry takes the form:
\be
\delta{\cal L}^{(2)} = \frac{\delta{\cal L}^{(2)}}{\delta \zeta}\delta_{\rm L}\zeta+2\partial_\tau \left(M^2f(\tau)\zeta'\delta_{\rm L}\zeta\right) -2 \partial_i\left(M^2f(\tau)\partial^i\zeta\delta_{\rm L}\zeta\right),
\ee
When we substitute in mode functions for the $\zeta$ fields in~\eqref{eq:pertproofeq3}, this implies that when evaluated with mode functions $u_k$ we can replace
\begin{align}
L^{(2)}[\delta_{\rm L} u_{k_1}, u_{k_2}] \to  \partial_\tau\left(M^2f(\tau)\delta_{\rm L}u_{k_1}  u'_{k_2}\right),
\end{align}
where we have dropped both the $\delta {\cal L}^{(2)}/\delta\zeta$ and spatial total derivative terms because they do not contribute to the answer. The fact that this is a boundary term lets us do the integral in~\eqref{eq:pertproofeq3} simply.  It is convenient to then divide both sides by the power spectrum and then use the fact that
\be
\lim_{\vec q\to 0}\langle\zeta_{\vec q}\zeta_{-\vec q}\rangle'  \simeq g(\tau)^2 q^{2\alpha}+\cdots
\ee
to obtain the expression
\begin{align} 
\lim_{q\to0} \frac{1}{P_\zeta(q)} \vev{\zeta_{\vec{q}} \zeta_{\vec{k}_1}\zeta_{\vec{k}_2}}'   = -\frac{4}{g(\tau)} \mbox{Im} \left( u^*_{k_1}({\tau})u^*_{k_2}({\tau}) \Big[M^2f(\tau) u'_{k_1}({\tau}) \delta_{\rm L}u_{k_2}({\tau}) -  B_{k_1}(\tau)\Big]  \right) _{k_2=k_1}.
\end{align}
A final simplification is to interpret the factor $1/g(\tau)$ on the right hand side as the nonlinear transformation of the field $\zeta$ and move it to the left hand side. We get the final expression for the squeezed limit
\begin{align}  
\label{eq:sqzed}
\lim_{q\to0} \delta_{\rm NL}\zeta\frac{1}{P_\zeta(q)} \vev{\zeta_{\vec{q}} \zeta_{\vec{k}_1}\zeta_{\vec{k}_2}}'   = -4 \mbox{Im} \left( u^*_{k_1}({\tau})u^*_{k_2}({\tau}) \Big[M^2f(\tau) u'_{k_1}({\tau}) \delta_{\rm L}u_{k_2}({\tau}) -B_{k_1}(\tau) \Big]  \right)_{k_2=k_1}.
\end{align}

This formula provides a new way of computing the squeezed limit in the case that there is a non-linear symmetry generating the adiabatic mode. Since we derived this formula using the symmetry relation between the second and third order action, it certainly contains symmetry information. If we attempt to re-run the argument with a symmetry which is not a physical mode, we will be unable to relate the cubic action to the quadratic one to obtain an identity.

\subsection{Late-time dilation in slow roll}
In order to complete the previous argument, we would like to show explicitly that the right hand side of~\eqref{eq:sqzed} is related to a symmetry variation of the 2-point function.

To orient ourselves, let's first see how we can extract Maldacena's consistency relation from this argument. For concreteness, we continue to consider the toy model of Section~\ref{eq:toymodelslowroll1}, but an identical argument applies to inflation, though the details are slightly more complicated.

We specialize to $f(\tau) = 1/(H\tau)^2$, and consider the dilation symmetry. The first thing to check is whether it is a physical mode. In position space, the nonlinear part of the dilation shifts $\zeta$ by a constant, $\delta_{\rm NL}\zeta = 1$. By examining the slow-roll mode function~\eqref{eq:srollmodef}, we see that it is time-independent as $\vec q\to 0$, so dilations are indeed physical modes and the derivation of the previous section applies.

In Fourier space, dilations act on the field as
\be
\delta\zeta_{\vec{k}} = (2\pi)^3 \delta^3(k) - \left(3+\vec{k}\cdot\vec{\partial}_k\right)\zeta_{\vec{k}},
\ee
from this, the mode functions inherit the ``transformation" rules
\be
\delta_{\rm NL}u_k = 1~~~~~~~~~~~~~~~~~~~\delta_{\rm L}u_k = - \left(3+\vec{k}\cdot \vec{\partial}_k\right)u_k.
\ee
In order to compute~\eqref{eq:sqzed}, we require the boundary term by which the action shifts in momentum space. This is most straightforwardly computed by Fourier transforming the quadratic and cubic terms in~\eqref{eq:toymodelaction}, the result of which is\footnote{This is really the only place where the full inflationary derivation is more complicated. It requires Fourier transforming and varying the full inflationary cubic action.}
\begin{align}
S_2 &= M^2\int\frac{\rd k}{(2\pi)^3}\rd \tau f(\tau)\left( \zeta'_{\vec k}\zeta'_{-\vec k}-k^2 \zeta_{\vec k}\zeta_{-\vec k}\right)\\
S_3 &= M^2 \int\frac{\rd k_1}{(2\pi)^3}\frac{\rd k_2}{(2\pi)^3}\frac{\rd k_3}{(2\pi)^3}\rd \tau f(\tau) \delta(\vec k_1+\vec k_2+\vec k_3)\left(3\zeta_{\vec k_1}\zeta'_{\vec k_2}\zeta'_{\vec k_3}+\vec k_2\cdot\vec k_3\zeta_{\vec k_1}\zeta_{\vec k_2}\zeta_{\vec k_3}\right).
\end{align}
In fourier space, the dilation symmetry acts on $\zeta_{\vec k}$ as 
\be
\delta\zeta_{k} = (2\pi)^3\delta(\vec k)-\left(3+\vec k\cdot\vec\partial_k\right)\zeta_{\vec k}.
\ee
Varying the quadratic and cubic actions under this symmetry and keeping boundary terms when integrating by parts, we find that up to cubic order, the action shifts by a boundary term in momentum space under the dilation symmetry:
\be
\delta S = M^2\int\frac{\rd k}{(2\pi)^3}\rd \tau f(\tau)\vec\partial_k\cdot\left(-\vec k\zeta'_{\vec k}\zeta'_{-\vec k}+\vec k k^2\zeta_{\vec k}\zeta_{-\vec k}\right).
\ee
In order to determine the contribution to~\eqref{eq:sqzed}, we write this term as a total time derivative, plus terms proportional to the linear equations of motion
\begin{equation}
M^2f(\tau)\vec\partial_k\cdot\left(-\vec k\zeta'_{\vec k}\zeta'_{-\vec k}+\vec k k^2\zeta_{\vec k}\zeta_{-\vec k}\right)=   \partial_{\vec{k}}\cdot\left(\vec{k} \zeta_{\vec{k}} \frac{\delta{\cal L}^{(2)}}{\delta\zeta} \right) - \partial_\tau\Big[ M^2 f(\tau)(3+\vec{k}\cdot\partial_{\vec{k}})\big( \zeta_{\vec{k}} \zeta'_{-\vec{k}} \big) \Big].
\end{equation}
When we re-write this in terms of the linear mode functions, $u_k$, the term involving $\delta{\cal L}^{(2)}/\delta\zeta$ will vanish, and we are left with
\be
B_k = -M^2 f(\tau)(3+\vec{k}\cdot\partial_{\vec{k}})\big( u_{\vec{k}}  u'_{-\vec{k}} \big) 
\ee
Putting this all together, equation~\eqref{eq:sqzed} takes the form
\begin{align}  
\lim_{q\to0} \frac{1}{P_\zeta(q)} \vev{\zeta_{\vec{q}} \zeta_{\vec{k}_1}\zeta_{\vec{k}_2}}'   = 4M^2f(\tau) \mbox{Im} \left( u^*_{k_1}({\tau})u^*_{k_1}({\tau}) \Big[  u'_{k_1}({\tau}) \delta_{\rm L}u_{k_1}({\tau}) - \delta_{\rm L}\big( u_{k_1}  u'_{k_1} \big)  \Big]  \right).
\end{align}
After some algebra and repeated use of the Wronskian condition $ u_{k} u'^*_{k} - u^*_{k}u'_{k} = \frac{i}{2f(\tau)M^2}$ for the mode functions, we find that this simplifies precisely to Maldacena's consistency relation
\begin{align}  
\lim_{q\to0} \frac{1}{P_\zeta(q)} \vev{\zeta_{\vec{q}} \zeta_{\vec{k}_1}\zeta_{\vec{k}_2}}'   = \delta_{\rm L}\lvert u_{k_1}\rvert^2 =  -\left(3+\vec{k}_1\cdot\partial_{k_1}\right)\langle\zeta_{\vec k_1}\zeta_{-\vec k_1}\rangle'.
\end{align}
Similar manipulations in the other situations identified in Sec.~\ref{sec:toymodel1} which possess physical modes allow us to derive the late-time identities associated to these symmetries in perturbation theory.

\subsection{Special conformal transformations}
\label{sec:SCT}
So far we have only discussed late-time identities at leading order in $\vec q$. In Sections~\ref{sec:flatspacetoymodel1} and~\ref{sec:usrtoy1} we saw that there are cases where the sub-leading momentum identities, associated with special conformal transformations, fail to hold as well. In the ultra-slow-roll case this is perhaps unsurprising, as we implicitly use the dilation identity in writing down the SCT identity, so if the dilation Ward identity is not satisfied, there is no reason for the SCT one to be. The flat space case discussed in Section~\ref{sec:flatspacetoymodel1} is more subtle. We would like to understand the underlying reason for the failure of the SCT relation in this case.

It turns out that the necessary and sufficient conditions for the SCT identity to hold (perturbatively) in a single field model are that:
\begin{enumerate}
 \item Three point correlation functions are rotationally-invariant and satisfy the dilation identity.\footnote{This requirement is necessary because in writing the identity~\eqref{eq:SCTid} in terms of primed correlators, we have used the both the dilation identity~\eqref{eq:dilationid} and rotational invariance of the correlation functions~\cite{Maldacena:2011nz,Goldberger:2013rsa,Hinterbichler:2012nm}.}
 \item The mode function $u_q(\tau)$ in the soft $q$ limit, apart from overall factor $q^\alpha$, has no order $q^1$ term, that is
 \begin{equation}
   u_q(\tau) \overset{q\rightarrow 0}{\longrightarrow} g(\tau)q^\alpha\left( 1+{\cal O}(q^2\tau^2) \right).
 \end{equation}
\end{enumerate}

This can be seen relatively straightforwardly by adapting the argument of Section~\ref{sec:pertproof1}. The only major change is that when we take the soft limit of the $u_q^*$ mode that is outside the integral, we now keep sub-leading terms in the $\vec q \to 0$ limit. The left hand side of the identity is now a function of all three momentum magnitudes, which we will denote ${\cal G}(q,k_2,k_2)$ the SCT identity involves taking $\partial_{q^i}$ of this. We obtain an expression of the form
\begin{equation}
 \frac{\partial}{\partial q^i} \Big( \left. {\cal G}(q,k_1,k_2) \right|_{\vec{k}_2=-\vec{q}-\vec{k}_1} \Big) = \left.\left( \frac{q^i}{q}\frac{\partial}{\partial q}{\cal G}(q,k_1,k_2)   + \frac{q^i+k^i_1}{k_2} \frac{\partial}{\partial k_2}{\cal G}(q,k_1,k_2)  \right)\right|_{\vec{k}_2=-\vec{q}-\vec{k}_1}.
 \label{eq:sctexpan}
\end{equation}
The first term on the right-hand side depends on the direction of  $\vec{q}$, it therefore violates the SCT identity unless $\lim_{q\to 0}\partial_q {\cal G} =0 $. Thus we require that the mode function contains no $q\tau$ term (apart from the overall $q^\alpha$ factor).\footnote{One may also worry that the interaction terms somehow induce pieces that scale like $q$. However, inspection of the form of the types of interactions that appear in inflation shows that this does not happen.} 

When this condition is satisfied, we can use the dilation identity on the right hand side of~\eqref{eq:sctexpan} to derive the SCT identity. We take the $\vec q \to 0$ limit of both side. This commutes with $\partial_{k_2}$ so the second term on the right-hand side of~\eqref{eq:sctexpan} gives
\begin{align}
 \lim_{q \to 0}\frac{k_1^a}{k_1} \left.\left(\frac{\partial}{\partial k_2}\frac{ \vev{\zeta_{\vec{q}} \zeta_{\vec{k}_1}\zeta_{\vec{k}_2}}'}{\vev{\zeta_{\vec{q}}\zeta_{-\vec{q}}}'} \right)\right|_{\vec{k}_2=-\vec{q}-\vec{k}_1} &  = \frac{k_1^a}{k_1}  \left.\left(\frac{\partial}{\partial k_2}\left(-3 - k_1\frac{\partial}{\partial k_1} - k_2\frac{\partial}{\partial k_2}\right) \vev{\zeta_{\vec{k}_1}\zeta_{-\vec{k}_1}}'\right)\right|_{\vec{k}_2=-\vec{k}_1} \nonumber \\
 & = - \frac{1}{2}k^i_1 \left( \frac{4}{k_1}\frac{\partial}{\partial k_1 } + \frac{\partial ^2}{\partial k_1^2}   \right)\vev{\zeta_{\vec{k}_1}\zeta_{-\vec{k}_1}}',
\end{align}
which is precisely how the SCT operator acts on the 2-point function.

We are now able to understand that the flat space model violates the SCT identity because its mode function, $u_k(\tau) =  \frac{1}{\sqrt{4q }M} e^{-iq\tau}$ has a piece which is ${\cal O}(q)$ in the soft limit.

\subsection{The time shift identity in the ultra-slow-roll toy model}
We can apply similar arguments to the emergent shift symmetry in the ultra-slow-roll toy model to derive a late-time identity. 
Again, we have the symmetry statement in momentum space as,
\begin{align}
 \delta_{\rm NL} {\cal L}_{(3)} &= -\delta_{\rm L} {\cal L}_{(2)} + M^2\partial_{\tau} \left[ \frac{f(\tau)}{2 \tau^2} \zeta'_{\vec k}  \zeta'_{-\vec k} - \frac{f(\tau)}{2 \tau^2} k^2\zeta_{\vec k}  \zeta_{-\vec k}   - 9 \frac{f(\tau)}{\tau^4}\zeta_{\vec k} \zeta_{-\vec k}  \right] ,\nonumber \\
    & = -{\rm eom}_{(2)}\delta_{\rm L} \zeta- M^2\partial_{\tau} \left[f(\tau)\left( \delta_{\rm L} \zeta_{\vec k} \zeta'_{-\vec k}+ \zeta'_{\vec k} \delta_{\rm L}\zeta_{-\vec k}   \right) \right]  \nonumber\\
 &\quad   + M^2\partial_{\tau} \left[ \frac{f(\tau)}{2 \tau^2} \zeta'_{\vec k}  \zeta'_{-\vec k} - \frac{f(\tau)}{2 \tau^2} k^2\zeta_{\vec k}  \zeta_{-\vec k}  - 9 \frac{f(\tau)}{\tau^4}\zeta_{\vec k} \zeta_{-\vec k}  \right]
\end{align}
Terms inside $\partial_{\tau}[\dots]$ without $\zeta'_k$ do not contribute when using \eqref{eq:sqzed}. Therefore the squeezed limit is 
\begin{align}
  \lim_{\vec{q} \to 0} \frac{1}{\tau^3} \frac{1}{P_{\zeta}(q) } \vev{\zeta_{\vec q} \zeta_{\vec k_1} \zeta_{\vec k_2}}' &= -4M^2 f(\tau) \mbox{Im} \left[ u^{*}_{k_1}u^{*}_{k_1} 2u'_{k_1} \delta_{\rm L}u_{k_1} - \frac{1}{2\tau^2} u'_{k_1} u'_{k_1} \right]  \nonumber \\
   &= u^*_{k_1} \delta_{\rm L} u_{k_1} + \delta_{\rm L}u^*_{k_1}  u_{k_1}  \nonumber \\
   & = \frac{1}{2 \tau^2} \partial_{\tau} \vev{\zeta_{\vec k_1} \zeta_{-\vec k_1}}',
\end{align}
which is the identity \eqref{eq:accidentalshiftwardusr}.

\section{Correlation functions in the toy model}
\label{app:toymodel}
In order to facilitate our checks of the Ward identities of the toy model in Section~\ref{sec:toymodel1} we collect some computations of correlation functions in the various cases of interest. In order to check the consistency relations, we will only need the action expanded up to cubic order:
\be
S = M^2\int\rd^3x\rd \tau\,f(\tau)\left(e^{3\zeta}\zeta'^2 - e^\zeta(\nabla\zeta)^2\right) = M^2\int\rd^3x\rd \tau f(\tau)\left(\zeta'^2 - (\nabla\zeta)^2+3\zeta\zeta'^2 - \zeta(\nabla\zeta)^2+\cdots\right).
\label{eq:perttoymodelaction}
\ee
In order to compute correlation functions in perturbation theory, we need the mode function solutions to the linearized equations of motion:
\be
\zeta''_k+\frac{f'}{f}\zeta'_k+k^2\zeta_k = 0.
\label{eq:zetaeomtoymodel}
\ee
Using these solutions---which we denote by $u_k(\tau)$---the equal-time power spectrum is given by
\be
\langle\zeta_{\vec k}\zeta_{-\vec k}\rangle' = u_k^*(\tau)u_k(\tau).
\ee
We can additionally compute equal-time correlation functions in interaction picture using the interactions in~\eqref{eq:perttoymodelaction} and the in-in master formula~\cite{Weinberg:2005vy}
\be
\langle{\cal O}(\tau)\rangle = \langle 0\rvert \bar Te^{i\int_{\tau_i}^\tau\rd \tau' H_{\rm int}(\tau')}{\cal O}(\tau) T e^{-i\int_{\tau_i}^\tau\rd \tau' H_{\rm int}(\tau')}\lvert 0\rangle~,
\label{eq:ininmastereq}
\ee
with $H_{\rm int}$  the interaction hamiltonian, $T$ denotes time-ordering, $\bar T$ denotes anti-time-ordering, and $\tau_i$ is an early time. We will only be interested in tree-level three-point calculations, for which the previous formula reduces to
\be
\langle{\cal O}(\tau)\rangle = -i\int_{-\infty}^{\tau}\rd \tau'\left\langle0\left\rvert\left[{\cal O}(\tau), H_{\rm int}(\tau')\right]\right\lvert0\right\rangle~,
\label{eq:inintreelevel}
\ee
and the interaction Hamilton is just minus the interaction lagrangian
\be
H_{\rm int} = -M^2\int\rd^3 x\left(3\zeta\zeta'^2 - \zeta(\nabla\zeta)^2\right).
\ee

It will also be necessary for us to compute some unequal-time correlation functions. This is done in essentially the same way as for equal time correlators, with the additional input that the initial field insertion is already in interaction picture $\zeta_{\vec{k}}(\tau_i) = \zeta^I_{\vec{k}}(\tau_i) $ in perturbation theory, so that we only time-evolve the late-time insertions as in~\eqref{eq:ininmastereq}.

\subsection{Vacuum wavefunctional} \label{app:vacuumfunctional}
The vacuum wavefunctional, $\Psi_0[\zeta_i] = \langle\zeta_i\rvert0_{\rm in}\rangle$ plays an important role in our identity. We therefore will need to compute it in a variety of examples. Fortunately, we only consider gaussian initial states, so in all cases the theories we consider will be free. Linearizing~\eqref{eq:perttoymodelaction}, we can expand the field $\zeta(x)$ and its conjugate momentum, $\Pi_\zeta(x)$, in the basis of solutions to the equation of motion~\eqref{eq:zetaeomtoymodel} 
\begin{align}
\label{eq:zetamodeexp}
\hat\zeta(x) &= \int\frac{\rd^3k}{(2\pi)^3} e^{-i\vec k\cdot\vec x}\left( u^*_{k} \hat a_{\vec k}^\dagger+u_{k} \hat a_{-\vec k}\right)\\
\hat\Pi_\zeta(x) &=\frac{\partial{\cal L}}{\partial\zeta'}= 2M^2 f(\tau)\zeta'(x)=  2M^2 f(\tau)\int\frac{\rd^3k}{(2\pi)^3} e^{-i\vec k\cdot\vec x}\left(  u'^*_{k} \hat a_{\vec k}^\dagger+u'_{k}\hat a_{-\vec k}\right).
\label{eq:pimodeexp}
\end{align}
The variables $\zeta(x)$ and $\Pi_\zeta(y)$ satisfy canonical commutation relations $[\zeta(\tau,\vec x), \Pi_\zeta(\tau,\vec y)] = i\delta(\vec x-\vec y)$, which implies that the normalization of the modes $u_k$ should be chosen to satisfy the Wronskian condition
\be
 u_{k} u'^*_{k} - u^*_{k}u'_{k} = \frac{i}{2f(\tau)M^2},
 \label{eq:wronskian}
\ee
so that the creation and annihilation operators satisfy the usual algebra $\big[\hat a_{\vec k},\hat a^\dagger_{\vec q}\big] = (2\pi)^3 \delta(\vec k - \vec q)$. This also guarantees that the modes $u_{\vec k}(\tau)\equiv u_{k}(\tau) e^{i\vec k\cdot x}$ with different $\vec k$ are orthonormal with respect to the Klein--Gordon inner product.

From the linearized action~\eqref{eq:perttoymodelaction} we can construct the hamiltonian
\be
\hat {\cal H} = \int\rd^3x\left(\frac{1}{4M^2 f(\tau)}\hat \Pi_\zeta^2 + M^2 f(\tau) (\nabla\hat\zeta)^2\right).
\ee
Recalling that in the field basis the momentum is represented as $\Pi_\zeta = -i\frac{\delta}{\delta \zeta}$, 
in principle we can use this hamiltonian to directly solve the Schr\"odinger equation
\be
i\partial_\tau\Psi[\zeta] = \hat{\cal H}\Psi[\zeta],
\ee
for the lowest-energy state. However, we will take a somewhat simpler route. We are searching for the vacuum state of the theory, which by definition satisfies $\hat a_{\vec k}\lvert 0\rangle = 0$. So, by casting the annihilation operator in Schr\"odinger picture we can solve a first-order differential equation instead.

We can use the orthonormality of the mode functions to invert~\eqref{eq:zetamodeexp}~\&~\eqref{eq:pimodeexp}
\begin{align}
\hat a_{\vec k} &= i\int\rd^3 x e^{-i\vec k\cdot x}\left(2M^2f(\tau) u'^*_{k} \zeta(x)-u^*_{k} \Pi_\zeta(x)\right)\\
\hat a_{\vec k}^\dagger &= -i\int\rd^3 x e^{i\vec k\cdot x}\left(2M^2f(\tau)u'_{k} \zeta(x)-u_{k} \Pi_\zeta(x)\right)
\end{align}
The equation $\hat a_{\vec k}\lvert 0_{\rm in}\rangle = 0$ in the basis of field eigenstates is then
\be
\left(\frac{\delta}{\delta \zeta_{\vec k}}-2iM^2f(\tau_i) \,\frac{u'^*_k(\tau_i)}{u^*_k(\tau_i)} \zeta_{\vec k}\right)\Psi_0[\zeta_{\vec k}] = 0
\ee
This equation is solved by a gaussian wavefunctional:\footnote{This expression for the vacuum wavefunctional can also be obtained from the on-shell action using the Hartle--Hawking prescription~\cite{Anninos:2014lwa}.}
\be
\Psi_0[\zeta_{\vec k}] \simeq \exp\left( \frac{i}{2} \int \frac{\rd^3k}{(2\pi)^3}2M^2 f(\tau_i)  \frac{u'^*_k(\tau_i)}{u^*_k(\tau_i)} \zeta_{\vec{k}}\zeta_{-\vec{k}} \right),
\label{eq:generalwavefunctionform}
\ee
where we have neglected the normalization factor, which is unimportant for our purposes. This wavefunctional does not quite solve the Schr\"odinger equation because its phase is not correct. It is straightforward to solve for the correction to the phase $\sim -ik\tau_i/2$, but since it is field-independent, it will not be important for our purposes, and~\eqref{eq:generalwavefunctionform} contains all the information we need.

\subsection{Flat space}
\label{sec:toyflatspacecorrs}
We first consider the case
 $f(\tau) = {\rm constant}$, so that the mode functions are the usual plane waves:
\be
u_k(\tau) =  \frac{1}{\sqrt{4k }M} e^{-ik\tau}.
\label{eq:flatmodes}
\ee

\subsubsection{Late time correlation functions}
From~\eqref{eq:flatmodes}, it is straightforward to extract the power spectrum:
\be
 \langle\zeta_{\vec k}\zeta_{-\vec k}\rangle'= P(k) = \frac{1}{4M^2k}.
\ee
Using the in-in formula~\eqref{eq:inintreelevel} we can compute the late-time 3-point function:
\be
 \vev{\zeta_{\vec{k}_1}\zeta_{\vec{k}_2}\zeta_{\vec{k}_3}}'= -\frac{ k_1^2+k_2^2+k_3^2+6 k_1 k_2+6 k_1k_3+6 k_2 k_3}{32M^4 k_1 k_2 k_3  (k_1+k_2+k_3)}.
 \ee
We will be interested in the soft limit of the correlation function, which up to ${\cal O}(q)$ is given by
 \be
 \lim_{q\to 0} \frac{1}{P(q)}\vev{\zeta_{\vec{q}} \zeta_{\vec{k}_1}\zeta_{\vec{k}_2}}' = -\frac{1}{2M^2k_1} + \frac{\vec q\cdot\vec k_1-2q\,k_1}{4M^2k_1^3}+{\cal O}(q^2)
\ee
To facilitate checks of the Ward identities, we also would like to catalogue the action of the dilation and SCT operators on the 2-point function of $\zeta$:\footnote{A useful formula for the SCT operator acting on a scalar function of momenta is
\be
\frac{1}{2}\left(-6\nabla_{k}^i+ k^i\nabla_{k}^2-2\vec k_a\cdot\vec\nabla_{k}\nabla_{k}^i\right)f(k) = -\frac{1}{2}\left(\frac{4}{k}\partial_k+\partial_k^2\right)f(k),
\ee
which simplifies checking the consistency relations.
}
\begin{align}
\label{eq:toyflatdil2pt}
\delta_D\langle\zeta_{\vec k}\zeta_{-\vec k}\rangle'=-\left(3 +  \vec k \cdot \vec\nabla_{k}\right) \langle\zeta_{\vec k}\zeta_{-\vec k}\rangle'& = -\frac{1}{2M^2k} = -2\langle\zeta_{\vec k}\zeta_{-\vec k}\rangle' \\
\delta_{K^i}\langle\zeta_{\vec k}\zeta_{-\vec k}\rangle'=\frac{1}{2}\left(-6\nabla_{k}^i+ k^i\nabla_{k}^2-2\vec k_a\cdot\vec\nabla_{k}\nabla_{k}^i\right) \langle\zeta_{\vec k}\zeta_{-\vec k}\rangle'& =  \frac{k^i}{4M^2k^3} .
\label{eq:toyflatsct2pt}
\end{align}
Using these formula, it is straightforward to check that 
\be
\lim_{\vec{q}\to0} \frac{1}{P(q)}\langle\zeta_{\vec q}\zeta_{\vec k_2}\zeta_{\vec k_3}\rangle' = -\left(3 +  \vec k_2 \cdot \vec\nabla_{k_2}\right)\langle\zeta_{\vec k_2}\zeta_{-\vec k_2}\rangle'\,,
\ee
but that
\be
\lim_{\vec{q}\to0} \frac{\partial}{\partial q^i}\left(\frac{1}{P(q)}\langle\zeta_{\vec q}\zeta_{\vec k_2}\zeta_{\vec k_3}\rangle'\right) \neq \frac{1}{2}\left(-6\nabla_{k_2}^i+ k_2^i\nabla_{k_2}^2-2\vec k_2\cdot\vec\nabla_{k_2}\nabla_{k_2}^i\right)\langle\zeta_{\vec k_2}\zeta_{-\vec k_2}\rangle'.
\ee
So, we see that the late-time dilation identity is satisfied, but the SCT identity is {\it not}.

\subsubsection{Unequal time correlation functions}
\label{app:flatunequal}
In order to check the identity~\eqref{eq:mainID} in in the simplest case we require two components: the wavefunctional coefficient, ${\cal E}_i(q)$, and the 3-point function where one insertion is in the initial state, while the other insertions are at some late time. 

The vacuum wavefunctional can be obtained from the general formula~\eqref{eq:generalwavefunctionform}
\be
\Psi_0[\zeta] \simeq\exp\left(-\frac{1}{2}\int\frac{\rd^3q}{(2\pi)^3} 2M^2q\,\zeta_{\vec q}\zeta_{-\vec q}\right).
\ee
We therefore see that in this case
\be
{\cal E}_i(q) = 2M^2q.
\ee
We can also compute the unequal time correlation function involving one initial-time insertion and two late-time insertions:
\begin{equation}
\vev{\zeta^f_{\vec{k}_2} \zeta^f_{\vec{k}_3} \zeta^i_{\vec{q}} }' =\frac{ e^{iq(\tau-\tau_i)}(k_2+k_3) \left(-5 q^2+k_2^2+6 k_2 k_3+k_3^2\right)}{32 q\,k_2 k_3  (q^2-(k_2+k_3)^2)M^4 }.
\end{equation}
Notice that this correlation function is in general complex-valued because it involves fields at different times. Combining these two quantities and taking the soft limit, we obtain
\be
\lim_{\vec q\to0}{\cal E}_i(q)\vev{\zeta^f_{\vec{k}_2} \zeta^f_{\vec{k}_3} \zeta^i_{\vec{q}} }'+{\rm c.c.} = -\frac{1}{2k_2 M^2}+\frac{\vec q\cdot\vec k_2}{4k_2^3 M^2} +{\cal O}(q^2)
\ee
In this case, the power spectrum is just $\langle\zeta_{\vec k}\zeta_{-\vec k}\rangle = 1/(2{\rm Re}{\cal E}_i(q)) = 1/(4M^2 k)$, as above. We can therefore use the formulae~\eqref{eq:toyflatdil2pt} and~\eqref{eq:toyflatsct2pt} for the variations of the 2-point functions. Using these expressions we can check that~\eqref{eq:flatuneqD} and~\eqref{eq:flatuneqSCT} are satisfied.

\subsection{Slow roll} 
\label{app:srcorrelator}
We next consider the case which mimics slow-roll inflation, where
$f(\tau) = (H\tau)^{-2}$. In this case, the solutions to~\eqref{eq:zetaeomtoymodel} take the form
\be
u_k(t) =  \frac{H}{\sqrt{4k^3}M}(1+ik\tau) e^{-ik\tau}.
\label{eq:srollmodef}
\ee
\subsubsection{Late time correlation functions}
The power spectrum is then given by
\be
 \langle\zeta_{\vec k}\zeta_{-\vec k}\rangle' = \frac{H^2}{4M^2k^3}\left(1+k^2\tau^2\right).
 \label{eq:toysroll2pt}
\ee
The late-time 3-point correlation function in this model is given by
\begin{align}
\vev{\zeta_{\vec{k}_1}\zeta_{\vec{k}_2}\zeta_{\vec{k}_3}}' 
=\frac{H^4}{32 k_1^3 k_2^3 k_3^3 k_t^2M^4} \Bigg[  &-\sum_{i}k_i^5  +\sum_{i\neq j} (-2k_i^4k_j+ 3k_i^3k_j^2) +\sum_{\rm cyclic}(8k_1^2k_2^2k_3-2k_1^3k_2k_3 )\nonumber \\
&  + \tau^2 \Big( \sum_{i\neq j}( -k_i^5k_j^2-7k_i^4k_j^3) + \sum_{i\neq j\neq k}( -2k_i^4k_j^2k_k-6k_i^3k_j^3k_k +k_i^3k_j^2k_k^2)  \Big) \nonumber \\
& + \tau^4k_t k_1^2k_2^2k_3^2\big(  k_1^2+k_2^2+k_3^2+6 k_1 k_2+6 k_1k_3+6 k_2 k_3 \big) \Bigg].
\end{align}
In the squeezed limit, this becomes
\be
 \lim_{q\to 0} \frac{1}{P(q)}\vev{\zeta_{\vec{q}} \zeta_{\vec{k}_2}\zeta_{\vec{k}_3}}' =  -\frac{H^2\tau^2}{2M^2k_2}+\frac{H^2\tau^2}{4M^2}\frac{\vec q\cdot \vec k_2}{k_2^3}+{\cal O}(q^2).
\ee
We can also compute the right hand sides of the consistency relations:
\begin{align}
\label{eq:toymodelsrdilation2pt}
\delta_D\langle\zeta_{\vec k}\zeta_{-\vec k}\rangle'=-\left(3 +  \vec k \cdot \vec\nabla_{k}\right) \langle\zeta_{\vec k}\zeta_{-\vec k}\rangle'& = -\frac{H^2 \tau^2}{2M^2k},  \\
\delta_{K^i}\langle\zeta_{\vec k}\zeta_{-\vec k}\rangle'=\frac{1}{2}\left(-6\nabla_{k}^i+ k^i\nabla_{k}^2-2\vec k_a\cdot\vec\nabla_{k}\nabla_{k}^i\right) \langle\zeta_{\vec k}\zeta_{-\vec k}\rangle'& =  \frac{H^2\tau^2k^i}{4M^2k^3},\\
\delta_T\langle\zeta_{\vec k}\zeta_{-\vec k}\rangle'=-\tau\partial_\tau \langle\zeta_{\vec k}\zeta_{-\vec k}\rangle' &= -\frac{H^2\tau^2}{2M^2k}.
\label{eq:toymodelsrtime2pt}
\end{align}
Using these expressions, it is straightforward to check that all the Ward identities~\eqref{eq:dilationid},~\eqref{eq:SCTid} and~\eqref{eq:srolltimedid} are satisfied.

\subsubsection{Unequal time correlation functions}
\label{app:uneqtoysr}
We now turn to the computation of unequal-time correlation functions. The vacuum wavefunctional is given by
\be
\Psi_0[\zeta] \simeq\exp\left(-\frac{1}{2}\int\frac{\rd^3q}{(2\pi)^3} \frac{2q^2M^2}{iH^2\tau_i(1-iq\tau_i)}\,\zeta_{\vec q}\zeta_{-\vec q}\right),
\ee
which implies the wavefunctional coefficient is
\be
{\cal E}_i(q) =  \frac{2q^2M^2}{iH^2\tau_i(1-iq\tau_i)}.
\ee
As expected, this is related to the power spectrum~\eqref{eq:toysroll2pt} by $ \langle\zeta_{\vec q}\zeta_{-\vec q}\rangle'  =1/(2{\rm Re}\,{\cal E}_i(q))$.

The unequal time correlation function between two late-time insertions and one initial insertion is given by
\begin{align}
\vev{\zeta^f_{\vec{k}_2} \zeta^f_{\vec{k}_3} \zeta^i_{\vec{q}} }' =&\frac{iH^4e^{iq(\tau_i-\tau_f)}(k_2+k_3)(iq\tau_i-1)}{32M^2  q^3k_2^3k_3^3 (k_2+k_3-q)^2(k_2+k_3+q)^2} \Bigg( i(1+iq\tau_f)\Big[(k_2^2-k_3^2)^2(k_2^2+k_3^2 +3k_2k_3)     \nonumber \\
&\quad -2 q^2 \left(3 k_2^4+3 k_3^4+7 k_2^3 k_3+7 k_2 k_3^3+2 k_2^2 k_3^2\right)  +5 q^4 \left(k_2^2+k_3^2-k_2 k_3\right)\Big]  \nonumber\\
&\quad +i \tau_f^2k_2^2k_3^2\Big[9 q^4-2 q^2 \left(k_2^2+k_3^2+8 k_2 k_3\right)+(k_2+k_3)^2 \left(k_2^2+k_3^2+6 k_2 k_3\right) \Big] \nonumber\\
&\quad -\tau^3_fq \,k_2^2 k_3^2  (k_2+k_3-q)(k_2+k_3+q)\left(k_2^2+k_3^2+6 k_2 k_3-5q^2\right) \Bigg).
\end{align}
Combining this with the gaussian wavefunctional coefficient, we can compute the squeezed limit of this correlator:
\be
\lim_{\vec q\to0}{\cal E}_i(q)\vev{\zeta^f_{\vec{k}_2} \zeta^f_{\vec{k}_3} \zeta^i_{\vec{q}} }'+{\rm c.c.} = -\frac{H^2 \tau_f^2}{2M^2 k_2}+\frac{H^2\tau_f^2 \vec q\cdot \vec k_2}{4M^2 k_2^3}+{\cal O}(q^2).
\ee
Then, using the action of the various differential operators on the 2-point function~\eqref{eq:toymodelsrdilation2pt}--\eqref{eq:toymodelsrtime2pt}, it is easy to check that the identities~\eqref{eq:flatuneqD} and~\eqref{eq:flatuneqSCT} are satisfied.

\subsection{Ultra-slow roll}  
\label{sec:usrcorrelator}
Finally, we consider the case that mimics ultra-slow roll, with $f(\tau) = (H\tau)^4$. In this case, the mode functions are given by
\be
u_{k}(\tau) = \frac{i}{ \sqrt{4k^3} MH^2\tau^3}(1+ik\tau)e^{-ik\tau}.
\label{eq:usrmodefunction}
\ee

\subsubsection{Late time correlation functions}
Using the mode functions, we can extract the power spectrum:
\be
\langle\zeta_{\vec k}\zeta_{-\vec k}\rangle' = \frac{1}{4H^4M^2k^3\tau^6}\left(1+k^2\tau^2\right).
\ee
We can also compute the equal time 3-point function, which takes the form
\begin{align}
\vev{\zeta_{\vec{k}_1}\zeta_{\vec{k}_2}\zeta_{\vec{k}_3}}' &= \frac{1}{256H^8 M^4  k_1^3 k_2^3 k_3^3\,\tau^{12}} \Bigg[  -24\sum_ik_i^3  +4 \tau^2 k_t^2 \bigg( \sum_ik_i^3 -2 \sum_{i\neq j}k_i^2k_j +6k_1k_2k_3 \bigg) \nonumber \\
&\quad  +\tau^4\bigg(2\sum_{i\neq j}\Big(k_i^5k_j^2-k_i^6k_j \Big)  +\sum_{i \neq j \neq k}\Big(2 k_i^4k_j^2k_k-8k_i^3k_j^2k_k^2 \Big) \bigg) \nonumber \\
&\quad + \tau^3 k_t \big(\sum_i k_i^2\big) \big(\prod_{\rm cyclic}(k_1+k_2-k_3) \big)\Big(e^{ik_t t}\mbox{Ei}(-ik_t \tau) \prod_i(i+k_i \tau) +{\rm c.c.} \Big) \Bigg].
\end{align}
where ${\rm Ei}$ is the exponential integral function: $\mbox{Ei}(-ik_t \tau) = \int^\tau \rd \tau' \frac{e^{-i k_t\tau'}}{\tau'}$.

The soft limit of the 3-point function is given by
\be
\lim_{q\to 0} \frac{1}{P(q)}\vev{\zeta_{\vec{q}} \zeta_{\vec{k}_2}\zeta_{\vec{k}_3}}' =-\frac{3+2k_{2}^2\tau^2}{4H^4M^2k_2^3\tau^6}+ \frac{9+2k_2^2\tau^2}{8H^4M^2k_2^5\tau^6} \vec q\cdot\vec k_2+{\cal O}(q^2).
\label{eq:usrtoyequalsoftlimit}
\ee
We can also act on the 2-point function with the following differential operators:
\begin{align}
\label{eq:dilontoyusr2pt}
\delta_D\langle\zeta_{\vec k}\zeta_{-\vec k}\rangle'=-\left(3 +  \vec k \cdot \vec\nabla_{k}\right) \langle\zeta_{\vec k}\zeta_{-\vec k}\rangle'& = -\frac{1}{2H^4M^2 k \tau^4},  \\\label{eq:SCTontoyusr2pt}
\delta_{K^i}\langle\zeta_{\vec k}\zeta_{-\vec k}\rangle'=\frac{1}{2}\left(-6\nabla_{k}^i+ k^i\nabla_{k}^2-2\vec k_a\cdot\vec\nabla_{k}\nabla_{k}^i\right) \langle\zeta_{\vec k}\zeta_{-\vec k}\rangle'& =  \frac{k^i}{4H^4M^2 k^3\tau^4},\\
\delta_T\langle\zeta_{\vec k}\zeta_{-\vec k}\rangle'=\frac{1}{2}\tau\partial_\tau \langle\zeta_{\vec k}\zeta_{-\vec k}\rangle' &=-\frac{3+2k^2\tau^2}{4H^4M^2k^3\tau^6}.
\end{align}
From these expressions, it is straightforward to check that the identities~\eqref{eq:dilationid} and~\eqref{eq:SCTid} are {\it not} satisfied, while the identity~\eqref{eq:accidentalshiftwardusr} {\it is} satisfied.

\subsubsection{Unequal time correlation functions}
\label{app:usruneqcorr}
Finally, let's consider unequal time correlation functions in this model. 
Inserting the mode function~\eqref{eq:usrmodefunction} into~\eqref{eq:generalwavefunctionform} we can derive the vacuum wavefunctional
\be
\Psi_0[\zeta] \simeq\exp\left(-\frac{1}{2}\int\frac{\rd^3q}{(2\pi)^3} i H^4 M^2 \left(6\tau_i^3-\frac{2\tau_i^5  q^2}{1-i q \tau_i  } \right)\zeta_{\vec{q}} \zeta_{-\vec{q}} \right).
\ee
From this we can extract the gaussian wavefunctional coefficient
\be
{\cal E}_i(q)=i H^4 M^2 \left(6\tau_i^3-\frac{2\tau_i^5  q^2}{1-i q \tau_i  } \right).
\ee
The unequal time correlation function with one initial insertion is given by
\begin{align}
\vev{\zeta^f_{\vec{k}_2} \zeta^f_{\vec{k}_3} \zeta^i_{\vec{q}} }'  &=\frac{e^{iq(\tau_i-\tau_f)}(1-iq\tau_i)}{512 H^8 M^4 \tau_f^9\tau_i^3  q^3 k_2^3 k_3^3  } \Bigg(  4(k_2+k_3)\bigg[-12(1+iq\tau_f)(k_2^2+k_3^2-k_2k_3)   \nonumber \\
 &\quad + 2 \tau_f^2 \big(q^2 (k_2^2+k_3^2-k_2 k_3)+(k_2+k_3)^2 (k_2^2+k_3^2-3 k_2 k_3)\big) \nonumber \\
 &\quad-iq\tau_f^3\big(k_2^4+k_3^4-q^4 -10k_2^2k_3^2\big)  -k_2 k_3 \tau_f^4 \big((k_2-k_3)^2-q^2) \left(q^2+k_2^2+k_3^2\right)  \bigg] \\
 &\quad+2\tau_f^3\big((k_2-k_3)^2-q^2 \big)\big((k_2+k_3)^2-q^2 \big)\big(k_2^2+k_3^2+q^2 \big)  \nonumber \\
 &\quad~~ \times \Big[i(1-ik_2\tau_f)(1-ik_3\tau_f)e^{i(q+k_2+k_3)\tau_f}\mbox{Ei}(-i(q+k_2+k_3)\tau_f) \nonumber  - \Big\{(k_2,k_3)\leftrightarrow(-k_2,-k_3) \Big\}\Big]\Bigg) .
\end{align}
Although this expression is fairly complicated, it simplifies after taking the soft limit and adding in its complex conjugate:
\be
\lim_{\vec q\to0}{\cal E}_i(q)\vev{\zeta^f_{\vec{k}_2} \zeta^f_{\vec{k}_3} \zeta^i_{\vec{q}} }'+{\rm c.c.} = -\frac{1}{2H^4M^2\tau_f^4 k_2}+\frac{ \vec q\cdot \vec k_2}{4H^4M^2\tau_f^4 k_2^3}+{\cal O}(q^2).
\label{eq:usrtoyuneqsoftlim}
\ee
Using the formulae~\eqref{eq:dilontoyusr2pt} and~\eqref{eq:SCTontoyusr2pt} it is easy to check that~\eqref{eq:flatuneqD},~\eqref{eq:flatuneqSCT} are satisfied.

\section{Classical soft theorems}
\label{app:classicalcheck}
In order to gain some insight into the quantum-mechanical Ward identities we consider in the main text, it is helpful to consider their classical analogues. In particular, the nature of perturbative quantum-mechanical calculations require us to take the initial time, $t_i$ to be infinitely far in the past relative to the time at which the late-time modes are evaluated. This makes some technical points rather opaque. In the classical case, however, it is possible to compute many quantities explicitly for arbitrary separations between $t_i$ and $t_f$, which grants a large amount of conceptual clarity, particularly for understanding the fate of the 4-point contributions discussed in Appendix~\ref{app:4point}. Additionally, this classical setting provides an arena to understand the constraint on the power spectrum of large-scale structure obtained in~\cite{Horn:2014rta}.

The classical equations of motion for a general interacting field theory can be solved perturbatively in essentially the same manner as in QFT and correlation functions between the classical field variables are constrained by classical versions of the Ward identities.

The model we consider is the model~\eqref{eq:toymodelaction}, with $f(\tau) = 1$, considered as a classical field theory in flat space:\footnote{In this Appendix, because everything is in flat space---with no time-dependent couplings---we again use $t$ as our time coordinate.}
\be
S =  \int\rd^3x\,\rd t\left(e^{3\phi}\dot\phi^2 - e^\phi(\nabla\phi)^2\right).
\ee
We will be interested in the dilation and SCT symmetries of this action, which act as
\begin{align}
\delta_D\phi&= 1+ \vec{x} \cdot \vec\nabla \phi,\\
\delta_{K^i}\phi&=2x^i +\left(2x^i \vec{x}\cdot\vec\nabla -x^2 \nabla^i \right)\phi.
\end{align}
\subsection{Identities for classical correlations}
We first consider the classical version of the Ward identities derived in Section~\ref{sec:pathintderiv}. The classical problem we want to solve is quite similar to the quantum mechanical one: instead of an initial quantum wavefunctional, we provide an initial probability distribution for classical field configurations, $\vp$ which we assume to be gaussian:
\begin{equation}
P[\vp] \simeq  \exp \left( -\frac{1}{2} \int \rd^3x \rd^3y {\cal E}(x,y)\varphi(x) \varphi(y) \right).
\end{equation}
Given a particular classical realization of the field variables, they can be evolved to late times via a path integral, which we can use to compute correlation functions of late-time field values, $\phi$
\begin{equation}
   \vev{\phi(x_1,t)\dots \phi(x_n,t)} = \int {\cal D} \phi \, \phi(x_1,t)\dots \phi(x_n,t) \delta\left(\phi-\phi_{\rm cl.}\right) P[\vp],
\end{equation}
where the delta function enforces that the fields evolve according to the classical trajectory, $\phi_{\rm cl.}$, which satisfies the equations of motion.\footnote{The quantum analogue of this delta function is $e^{iS}$ in the path integral} 

Following the strategy in Sec.~\ref{sec:pathintderiv}, we can derive an identity associated to symmetries on $\phi$ for the 2-point function
\begin{equation}
\vev{\delta\phi(x_1,t) \phi(x_2,t)}+\vev{\phi(x_1,t) \delta\phi(x_2,t)} = \frac{1}{2}\int \rd^3x \rd^3y {\cal E}(x,y)\vev{ \phi(x_1,t)\phi(x_2,t) \left[ \varphi(x) \delta \varphi(y)+\delta\varphi(x)  \varphi(y) \right] } .
\end{equation}
This identity shares many features with the quantum-mechanical identity~\eqref{eq:almostfinalward}. In particular the field variations include both terms independent of fields and terms linear in fields, so that both 3-point and 4-point functions appear on the right hand side. One of our reasons for considering the classical situation is to justify explicitly the assumption made in Section~\ref{sec:pathintderiv} that the 4-point term is irrelevant to the late-time soft theorem.

\subsubsection*{Dilation}
The identity associated with dilation in momentum space (after removing momentum conserving delta functions and only keeping leading order terms) is of the form
\begin{equation}
\lim_{\vec q \to 0} {\cal E}_{\vec q}\vev{\phi_{\vec k_1}\phi_{\vec k_2}\varphi_{\vec q}}' - (3-k_1\partial_{k_1}){\cal E}_{\vec k_1} \vev{\phi_{\vec k_1}\varphi_{-\vec k_1}}'\vev{\phi_{-\vec k_1}\varphi_{\vec k_1}}'=-(3+k_1\partial_{k_1})\vev{\phi_{\vec k_1}\phi_{-\vec k_1}}'.
\label{eq:classicaldilation}
\end{equation}

\subsubsection*{SCT}
Similarly, the identity associated to special conformal transformations reads\footnote{Deriving this identity is somewhat intricate, removing the delta functions from the identity requires using both the dilation identity~\eqref{eq:classicaldilation} and rotational invariance of correlation functions. Its derivation is presented in Appendix~\ref{app:sctfourier}.}
\begin{align}
\nonumber
\frac{\partial}{\partial {q^a}} \left( {\cal E}_{\vec q} \vev{\phi_{\vec k_1} \phi_{-\vec k_1-\vec q}\varphi_{\vec q} }' \right) \bigg|_{\vec q\to 0} &- \frac{k_1^a}{2k_1}\frac{\partial}{\partial { k_1}} \Big[(3-k_1\partial_{k_1}){\cal E}_{\vec k_1} \vev{\phi_{\vec k_1}\varphi_{-\vec k_1}}'^2 \Big] \\
& \qquad   = -\frac{k_1^a}{2} \left( \frac{4}{k_1} \partial_{k_1} + \partial_{k_1}^2 \right) \vev{\phi_{\vec k_1}\phi_{-\vec k_1}}'.
\label{eq:classicalSCTid}
\end{align}

\subsection{Perturbative solution to the equation of motion}
In the previous section we have derived the classical analogues of the soft theorems considered in the main text. We would now like to explicitly check that these identities are satisfied. Essentially, this requires solving the classical equations of motion for the late time field, $\phi$, given some initial spatial profile, $\vp$. Once we have this function, we can compute correlation functions of late time fields using the gaussianity of the initial field profiles via Wick contraction. This is essentially the same procedure employed in studies of cosmological large scale structure~\cite{Bernardeau:2001qr}. Along the way, we will see that transforming the initial data by a symmetry transformation consistently acts as the same symmetry transformation on the late time field profile, as expected.

We will only be interested in correlation functions to leading order in the nonlinearities, so we only require the action up to cubic order
\begin{equation}
S= \int \rd^3x \rd t\left( \frac{1}{2}\dot{\phi}^2 -\frac{1}{2} (\nabla \phi)^2 + \frac{3}{2}\phi \dot{\phi}^2 -\frac{1}{2}\phi (\nabla \phi)^2 +\cdots \right).
\end{equation}
The corresponding equation of motion is
\begin{equation}
-\ddot{\phi} + \nabla^2 \phi -\frac{3}{2}\dot{\phi}^2 - 3 \phi \ddot{\phi} + \frac{1}{2}(\nabla \phi)^2 + \phi \nabla^2\phi =0.
\end{equation}
Since the problem is translation invariant, it is convenient to go to momentum space, so that the equation of motion reads
\begin{equation}
\ddot{\phi}_{\vec k} +k^2\phi_{\vec k} =\int \frac{\rd^3q}{(2\pi)^3}\left[ -\frac{3}{2} \dot{\phi}_{\vec{q}}\dot{\phi}_{\vec{k}-\vec{q}} -3\ddot{\phi}_{\vec{q}} \phi_{\vec{k}-\vec{q}} -\frac{1}{2} \vec{q}\cdot(\vec{k}-\vec{q}) \phi_{\vec{q}} \phi_{\vec{k}-\vec{q}}  - (\vec{k}-\vec{q})^2\phi_{\vec{q}} \phi_{\vec{k}-\vec{q}}  \right].
\label{eq:classical3ptEOM}
\end{equation}

We solve this equation perturbatively. Schematically, it is of the form
\be
\ddot{\phi}_{\vec k}(t) +k^2\phi_{\vec k}(t) =  J_{\vec k}(t),
\label{eq:sourcedwaveeq}
\ee
where $J_{\vec k}$ is the source on the right hand side of~\eqref{eq:classical3ptEOM}. We introduce a small parameter $\alpha$ and expand the late time field in powers of $\alpha$:
\be
 \phi_k(t) = \alpha \phi^{(1)}_{\vec k}(t) +  \alpha^2 \phi^{(2)}_{\vec k}(t) + \cdots,
 \label{eq:latetimefield}
\ee
and solve order-by-order in $\alpha$, subject to the boundary condition that $\phi_{\vec k}(t_i) = \vp_{\vec k}$. The second boundary condition we impose is so-called growing mode initial conditions, which implies that the time-derivative of $\phi$ vanishes at the initial time, $\dot\phi_k(t_i) = 0$. At lowest order, we just have
\be
\ddot\phi^{(1)}_{\vec k}(t)+k^2 \phi^{(1)}_{\vec k}(t) = 0.
\ee
With our choice of boundary conditions, this is solved by
\be
\phi^{(1)}_{\vec k} (t) = \cos(k(t-t_i))\vp_{\vec k}.
\ee
We can then plug this solution back into~\eqref{eq:sourcedwaveeq} to solve the equation of motion at first order in $\alpha$. The $1^{\rm st}$ order solution will act as a source for $\phi^{(2)}$ so we can solve this equation by Green's function methods:
\be
\phi^{(2)}_k(t) = \int_{t_i}^{t}\rd t' G_k(t,t') J_k^{(1)}(t'),
\ee
where $G_k(t,t')$ is the causal Green's function
\be
G_k(t,t')= \frac{1}{k}\sin(k(t-t')) \theta(t-t'),
\ee
which satisfies the equation $\ddot{G_k}(t,t') +k^2 G_k(t,t') = \delta(t-t')$.\footnote{The boundary conditions we impose on this Green's function are that it vanishes at $t = t_i$ and that its first derivative jumps by 1 at $t = t_i$.} Explicitly, the ${\cal O}(\alpha^2)$ solution is given by
{\small
\begin{align}
\phi^{(2)}_{\vec k}(t) &=\int_{t_i}^t \rd t' \frac{\sin(k(t-t'))}{k} \theta(t-t')\int \frac{\rd^3q}{(2\pi)^3}\bigg[ \bigg(3q^2 -\frac{\vec{q}\cdot(\vec{k}-\vec{q})}{2} -\lvert\vec k-\vec q\rvert^2\bigg) \cos[q(t'-t_i)]\cos[\lvert \vec k-\vec q\rvert(t'-t_i)]  \nonumber  \\
          & \qquad \qquad  \qquad  \qquad  \qquad  \qquad  -\frac{3}{2}q\lvert\vec k-\vec q\rvert \sin[q(t'-t_i)]\sin[\lvert\vec k-\vec q\rvert(t'-t_i)] \bigg] \varphi_{\vec q} \varphi_{\vec k-\vec q}\,.
\label{eq:firstorderclassicalsoln}
\end{align}
}
For later use we record the results of doing the time integrals involved in this solution:
{\small
\begin{align}
\label{eq:Fintegral}
 {\cal F}_{\vec k,\vec q,|\vec k-\vec q|}(t)&\equiv\int_{t_i}^t \rd t'  \frac{1}{k}\sin(k(t-t'))  \cos(q(t'-t_i))\cos(|\vec k-\vec q|(t'-t_i))  \\\nonumber
 &= \frac{(q^2 + |\vec k-\vec q|^2 -k^2)\cos[k(t-t_i)] }{(k^2 - (q-|\vec k-\vec q|)^2)(k^2 -(q+|\vec k-\vec q|)^2)} + \frac{1}{2} \frac{\cos[(q-|\vec k-\vec q|)(t-t_i)]}{k^2 - (q-|\vec k-\vec q|)^2}+\frac{1}{2} \frac{\cos[(q+|\vec k-\vec q|)(t-t_i)]}{k^2 - (q+|\vec k-\vec q|)^2}, \\
 \label{eq:Gintegral}
 {\cal G}_{\vec k,\vec q,|\vec k-\vec q|}(t)&\equiv\int_{t_i}^t \rd t'  \frac{1}{k}\sin(k(t-t'))  \sin(q(t'-t_i))\sin(|\vec k-\vec q|(t'-t_i))\\\nonumber
   &= \frac{2q|\vec k-\vec q|\cos[k(t-t_i)] }{(k^2 - (q-|\vec k-\vec q|)^2)(k^2 -(q+|\vec k-\vec q|)^2)} + \frac{1}{2} \frac{\cos[(q-|\vec k-\vec q|)(t-t_i)]}{k^2 - (q-|\vec k-\vec q|)^2}-\frac{1}{2} \frac{\cos[(q+|\vec k-\vec q|)(t-t_i)]}{k^2 - (q+|\vec k-\vec q|)^2}.
\end{align}
}
In terms of these functions, the solution $\phi^{(2)}_{\vec k}(t)$ is given by
\be
\phi^{(2)}_{\vec k}(t) = \int \frac{\rd^3q}{(2\pi)^3}\bigg[ \bigg(3q^2 -\frac{\vec{q}\cdot(\vec{k}-\vec{q})}{2} -\lvert\vec k-\vec q\rvert^2\bigg)  {\cal F}_{\vec k,\vec q,|\vec k-\vec q|}(t)  -\frac{3}{2}q\lvert\vec k-\vec q\rvert  {\cal G}_{\vec k,\vec q,|\vec k-\vec q|}(t) \bigg] \varphi_{\vec q} \varphi_{\vec k-\vec q}\,.
\ee

\subsection{Symmetry transformations of the nonlinear solution}
We would like to see how symmetry transformations of the late-time field, $\phi$, and initial field, $\vp$ are related to each other. In particular, we want to check that varying the initial conditions under a symmetry is equivalent to varying the late time field configuration under the same symmetry. This is an implicit assumption we have made in the path integral derivation of the Ward identities (essentially that $\langle \phi_f +\delta\phi_f\rvert\phi_i+\delta\phi_i\rangle = \langle \phi_f\rvert\phi_i\rangle$).

\subsubsection*{Dilation}
We first consider the action of dilations on the fields. 
Under direct dilation the late time field~\eqref{eq:latetimefield} (at leading order in $\varphi_k$) transforms as
\begin{align}
\delta\phi_{\vec k}(t) &= (2\pi)^3 \delta(\vec k) - (3+k\partial_k)\phi_{\vec k}(t)  \\
&= (2\pi)^3 \delta(\vec k)   +k(t-t_i)\sin[k(t-t_i)]  \varphi_{\vec k}  - \cos[k(t-t_i)](3+  k\partial_k)\varphi_{\vec k}
\label{eq:dilationlatetimefield}
\end{align}
The question is whether we reproduce this
 same result if we instead transform the initial data as $\delta\varphi_{\vec k}=(2\pi)^3 \delta(\vec k)  -(3+k\partial_k)\varphi_{\vec k}$. This shifts the late time field as
\begin{equation}
\delta \phi_{\vec k}(t) =  (2\pi)^3 \delta(\vec k)   - \cos[k(t-t_i)](3+  k\partial_k)\varphi_{\vec k} +  \phi^{(2)}_{\vec k}(t) \Big\rvert_{\varphi_{\vec q}\, \mapsto (2\pi)^3\delta(\vec q) }+  \phi^{(2)}_{\vec k}(t) \Big\rvert_{\varphi_{\vec k-\vec q}\, \mapsto (2\pi)^3\delta(\vec k-\vec q) }.
\label{eq:earlytimedilation}
\end{equation}

Using the expressions~\eqref{eq:Fintegral} and~\eqref{eq:Gintegral} we find
\begin{subequations}
\label{eg:FGfuncts}
\begin{align}
 \lim_{q\to0}  {\cal F}_{\vec k,\vec q,|\vec k-\vec q|}(t) \left[3q^2 -\frac{1}{2}\vec{q}\cdot(\vec{k}-\vec{q}) -|\vec k-\vec q|^2 \right] &= \frac{3}{2}k(t-t_i) \sin[k(t-t_i)],  \\
  \lim_{q\to k}  {\cal F}_{\vec k,\vec q,|\vec k-\vec q|}(t) \left[3q^2 -\frac{1}{2}\vec{q}\cdot(\vec{k}-\vec{q}) -|\vec k-\vec q|^2 \right] &= -\frac{1}{2}k(t-t_i) \sin[k(t-t_i)],\\
   \lim_{q\to0}  {\cal G}_{\vec k,\vec q,|\vec k-\vec q|}(t) \left[ -\frac{3}{2}q|\vec k-\vec q| \right] &=0 , \\
  \lim_{q\to k}  {\cal G}_{\vec k,\vec q,|\vec k-\vec q|}(t) \left[ -\frac{3}{2}q|\vec k-\vec q|  \right] &=0.
\end{align}
\end{subequations}
Inserting this into~\eqref{eq:earlytimedilation} reproduces~\eqref{eq:dilationlatetimefield}, so we see that indeed we get the same result whether we transform the initial data or directly transform the final field profile.

\subsubsection*{SCT}
We would now like to perform the same check for special conformal transformations. 
Under an SCT the late time field (at leading order in $\varphi_k$) transforms as 
\begin{align}
\label{eq:latetimefieldSCTvar}
\delta^a \phi_{\vec k}(t) &= -(2\pi)^3 2i \frac{\partial}{\partial k^a}\delta(\vec k) +i\left(6 \frac{\partial}{\partial k^a}+2k^b \frac{\partial^2}{\partial k^b \partial k^a}- k^a \frac{\partial^2}{\partial k^b \partial k^b}\right) \phi_k(t)  \nonumber \\
& = -(2\pi)^3 2i \frac{\partial}{\partial k^a}\delta(\vec k) +i \cos(k(t-t_i))\left(6 \frac{\partial}{\partial k^a}+2k^b \frac{\partial^2}{\partial k^b \partial k^a}-k^a \frac{\partial^2}{\partial k^b \partial k^b}\right) \varphi_k  \\\nonumber
 &\quad - ik^a\Big[  (t-t_i)^2 \cos[k(t-t_i)]\varphi_k+ \frac{4(t-t_i)}{k}\sin[k(t-t_i)]\varphi_k +2(t-t_i)\sin[k(t-t_i)]\partial_k \varphi_{\vec k} \Big].
\end{align}
If we instead transform the initial data by $\delta^a \varphi_{\vec k} =  -(2\pi)^3 2i \frac{\partial}{\partial k^a}\delta(\vec k) + ik^a \left(\frac{4}{k}\partial_k +\partial_k^2 \right)\varphi_{\vec k} $, the induced change in late time field will be 
\begin{align}
\delta^a \phi_{\vec k}(t) & = -(2\pi)^3 2i \frac{\partial}{\partial k^a}\delta(\vec k) +i \cos(k(t-t_i))\left(6 \frac{\partial}{\partial k^a}+2k^b \frac{\partial^2}{\partial k^b \partial k^a}- k^a \frac{\partial^2}{\partial k^b \partial k^b}\right) \varphi_{\vec k} \nonumber \\
 &\quad + \left.\phi^{(2)}_{\vec k}(t) \right|_{\varphi_{\vec q}\,\mapsto -(2\pi)^3 2i \frac{\partial}{\partial q^a} \delta(\vec q) } + \left.\phi^{(2)}_{\vec k}(t) \right|_{\varphi_{\vec k-\vec q}\,\mapsto -(2\pi)^3 2i \frac{\partial}{\partial(k- q)^a} \delta(\vec k-\vec q) }.
\end{align}
We can evaluate this using properties of the ${\cal F}, {\cal G}$ functions:
{\small
\begin{align*}
 & \lim_{\vec q\to0} 2i\frac{\partial}{\partial q^a} \left( {\cal F}_{\vec k,\vec q,|\vec k-\vec q|}    \left[3q^2 \frac{1}{2}\vec{q}\cdot(\vec{k}-\vec{q}) -|\vec k-\vec q|^2 \right] \right)  = ik^a\left[\frac{1}{2}k(t-t_i)^2 \cos[k(t-t_i)] + (t-t_i)\sin[k(t-t_i)]  \right] ,  \\
  &\lim_{\vec q\to \vec k} 2i\frac{\partial}{\partial (k-q)^a}\left( {\cal F}_{\vec k,\vec q,|\vec k-\vec q|}  \left[3q^2 -\frac{1}{2}\vec{q}\cdot(\vec{k}-\vec{q}) -|\vec k-\vec q|^2 \right] \right) = -ik^a\left[\frac{3}{2}k(t-t_i)^2 \cos[k(t-t_i)] +5 (t-t_i)\sin[k(t-t_i)]  \right] , \\
  & \lim_{\vec q\to0} 2i\frac{\partial}{\partial q^a} \left( {\cal G}_{\vec k,\vec q,|\vec k-\vec q|} \left[ -\frac{3}{2}q|\vec k-\vec q| \right] \right) =0 , \\
  &\lim_{\vec q\to\vec  k} 2i\frac{\partial}{\partial (k-q)^a}\left( {\cal G}_{\vec k,\vec q,|\vec k-\vec q|} \left[ -\frac{3}{2}q|\vec k-\vec q|  \right]\right)  =0.
\end{align*}
}
Using these formulae along with~\eqref{eg:FGfuncts}, we see that this matches the late-time variation of the field profile by an SCT~\eqref{eq:latetimefieldSCTvar} after a little algebra.

\subsection{Correlation function checks}
In addition to checking the symmetry properties of the nonlinear solution $\phi = \phi^{(1)}+\phi^{(2)}+\cdots$, we can use this solution to compute classical late-time correlation functions of the fields. The correlation functions we are interested in are computed via Wick contraction of the initial fields, $\vp$. Though the initial fields are gaussian, the late-time fields can have non-trivial connected higher-point functions because the late time field is composite, schematically $\phi \sim \vp + \vp^2+\cdots$.

\subsubsection{Dilation}
The identity associated with dilation in momentum space (after removing momentum conserving delta functions and only keeping leading order terms) is of the form~\eqref{eq:classicaldilation},
where $1/{\cal E}_{\vec k}=  \vev{\varphi_{\vec k} \varphi_{-\vec k}}' = P(k) $. The right hand side, when explicitly computed is 
\begin{equation}
 -(3+k_1\partial_{k_1})\Big( \cos^2[k_1(t-t_i)] P(k_1) \Big) = k_1(t-t_i)\sin[2k_1(t-t_i)]P(k_1) - \cos^2[k_1(t-t_i)](3P(k_1) +k_1P'(k_1)).
\end{equation}
On the other hand, terms on the left of the identity can be evaluated as
\begin{align}
&\lim_{\vec q \to 0} {\cal E}_{\vec q} \vev{\phi_{\vec k_1}\phi_{\vec k_2}\varphi_{\vec q}}'  = \lim_{\vec q \to 0} {\cal E}_{\vec q} \Big( \vev{\phi^{(1)}_{\vec k_1}\phi^{(2)}_{\vec k_2}\varphi_{\vec q}}'+\vev{\phi^{(2)}_{k_1}\phi^{(1)}_{\vec k_2}\varphi_{\vec q}}' \Big) \nonumber = k_1(t-t_i)\sin[2k_1(t-t_i)]P(k_1),  \nonumber \\
  &(-3+k_1\partial_{k_1}){\cal E}_{\vec k_1}  \vev{\phi_{\vec k_1}\varphi_{-\vec k_1}}'\vev{\phi_{-\vec k_1}\varphi_{\vec k_1}}' = \cos^2[k_1(t-t_i)](-3P(k_1) -k_1P'(k_1)).
\end{align}
This confirms that the identity works for an arbitrary initial power spectrum, $P(k)$, and further that one must include four-point term coming from the linear variation of the initial functional distribution in order for the identity to be satisfied. 

Note that in the large time separation limit, $t-t_i\to \infty$, the ratio of the term involving the disconnected 4-point function to that involving the 3-point function goes to zero, it therefore can be neglected for sufficiently late-time correlation functions.

\subsubsection*{Power spectrum constraint}
So far we have been considering unequal-time Ward identities involving correlation functions with an initial state insertion. Here we see that if we want to write an identity purely in terms of late-time fields, there is a subtle constraint that must be satisfied. 

If we would like to write an equal time identity with the soft mode for large time separation $t-t_i\to  \infty$ of the form
\begin{equation}
 \lim_{\vec q \to 0} \frac{1}{\vev{\phi_{\vec q}\phi_{-\vec q}}'} \vev{\phi_{\vec k_1}\phi_{\vec k_2}\phi_{\vec q}}'
=-(3+k_1\partial_{k_1})\vev{\phi_{\vec k_1}\phi_{-\vec k_1}}'
 \end{equation}
there is an extra constraint on the power spectrum apart from the physical mode condition. The 3 point function consists of three parts at leading order,
\begin{equation}
 \frac{1}{\vev{\phi_{\vec q}\phi_{-\vec q}}'}  \left( \vev{\phi^{(1)}_{\vec k_1}\phi^{(2)}_{\vec k_2}\phi^{(1)}_{\vec q}}' +\vev{\phi^{(2)}_{\vec k_1}\phi^{(1)}_{\vec k_2}\phi^{(1)}_{\vec q}}' +\vev{\phi^{(1)}_{\vec k_1}\phi^{(1)}_{\vec k_2}\phi^{(2)}_{\vec q}}'  \right).
\end{equation}
In order for the first two terms in this expression to match with ${\cal E}_{\vec q}\vev{\phi_{\vec k_1}\phi_{\vec k_2}\varphi_{\vec q}}' $ in the squeezed limit, we must impose the physical mode condition, which is satisfied in this case because $\lim_{\vec q\to0} \cos[q(t-t_i)] = 1 $. 

The third term $\vev{\phi^{(1)}_{\vec k_1}\phi^{(1)}_{\vec k_2}\phi^{(2)}_{\vec q}}'$ gives an extra constraint on ${\cal E}_{\vec q}$. An explicit computation gives
\begin{equation}
  \lim_{\vec q\to 0}\frac{1}{\vev{\phi_{\vec q}\phi_{-\vec q}}'}  \vev{\phi^{(1)}_{\vec k_1}\phi^{(1)}_{\vec k_2}\phi^{(2)}_{\vec q}}' = \lim_{q\to 0} \frac{P(k_1)^2 \cos^2[k_1(t-t_i)])}{P(q) \cos^2[q(t-t_i)] }\left( \frac{1}{2}k_1^2(t-t_i)^2 -\cos[2k_1(t-t_i)]+1 \right).
\end{equation} 
In the large time separation limit, it also has a leading order term $(t-t_i)^2$. In order for this term not to spoil the identity, it has to vanish in the squeezed limit. Therefore the initial power spectrum is constrained to satisfy
\begin{equation}
 \lim_{\vec q\to 0} \frac{1}{P(q)} = \lim_{\vec q\to 0}{\cal E}_{\vec q}=0.
\end{equation}
This is reminiscent of the constraint present in large scale structure soft theorems that the initial power spectrum cannot be too blue in order for the identities to be valid~\cite{Horn:2014rta}. This constraint is not present for the early-late time identity.

\subsubsection{Special conformal transformations}
It is also possible to check that the identity~\eqref{eq:classicalSCTid} is satisfied. Indeed, computing directly the various terms in this identity one finds:
\begin{align}
&   \left.\frac{\partial}{\partial q^a} \left( {\cal E}_{\vec q}\vev{\phi_{\vec k_1} \phi_{-\vec k_1-\vec q}\varphi_{\vec q} }' \right) \right|_{\vec q\to 0} \nonumber \\
&~~~~~= \frac{k_1^a}{2}\left( 2 (t-t_i)^2 \cos[2k_1(t-t_i)] P(k_1)+ \frac{t-t_i}{k_1}\sin[2k_1(t-t_i)](P(k_1) + k_1 P'(k_1)) \right) \\
 &\frac{1}{2} \frac{k_1^a}{k_1} \frac{\partial }{\partial k_1} \Big[(-3+k_1\partial_{k_1}){\cal E}_{\vec k_1} \vev{\phi_{\vec k_1}\varphi_{\vec k_1}}'^2 \Big] \nonumber \\
& ~~~~~= \frac{k_1^a}{2}\left( \frac{t-t_i}{k_1}\sin[2k_1(t-t_i)](3P(k_1)+k_1P'(k_1) ) - \cos^2[k_1(t-t_i)] \left[ \frac{4}{k_1}P'(k_1) + P''(k_1) \right]\right) \\ 
 &-\frac{k_1^a}{2} \left( \frac{4}{k_1} \partial_{k_1} + \partial_{k_1}^2 \right) \vev{\phi_{\vec k_1}\phi_{-\vec k_1}}' \nonumber \\
 &~~~~~=  \frac{k_1^a}{2}\bigg( 2 (t-t_i)^2 \cos[2k_1(t-t_i)] P(k_1)  + \frac{t-t_i}{k_1} \sin[2k_1(t-t_i)](4P(k_1) + 2k_1P'(k_1) )  \nonumber \\
  & \qquad ~~~~~- \cos^2[k_1(t-t_i)] \left[ \frac{4}{k_1}P'(k_1) + P''(k_1) \right] \bigg)
\end{align}
Combining these terms, we see that the identity~\eqref{eq:classicalSCTid} is satisfied for arbitrary 
initial gaussian coefficient ${\cal E}_{\vec k}= \frac{1}{P(k)}$. In this case as well, the four point function term is sub-leading in the large time separation $t-t_i\to \infty$.

\subsection{SCT identity in momentum space}
\label{app:sctfourier}
Here we collect a final technical derivation; the translation of the classical identity for special conformal transformations
\begin{align}  \label{eq:SCTclassical}
  &\int \rd^3x \rd^3y {\cal E}(x,y)\left\{  2x^a \vev{\phi(x_1) \phi(x_2) \varphi(y)} +\frac{1}{2} \left(2x^a x^b \partial_x^b -x^2 \partial_x^a + x \to y \right) \vev{\phi(x_1) \phi(x_2)\varphi(x) \varphi(y)} \right\} \nonumber \\
  &~~~~~=  \sum_{i=1,2} \left(2x_i^a x_i^b \partial_{x_i}^b -x_i^2 \partial_{x_i}^a  \right)\vev{\phi(x_1) \phi(x_2)}
\end{align}
into fourier space. We first fourier transform $\phi(x_1)$ and $\phi(x_2)$. The first term on the left hand side and the right hand side can be written as
\begin{align}
  \int \rd^3x \rd^3y{\cal E}(x,y)  2x^a \vev{\phi(x_1) \phi(x_2) \varphi(y)} &\longmapsto - 2i \left.\frac{\partial}{\partial q^a} \left( {\cal E}_{\vec q} \vev{\phi_{\vec k_1} \phi_{\vec k_2}\varphi_{\vec q} } \right) \right|_{\vec q\to 0} \\\nonumber
  \sum_{i=1,2} \left(2x_i^a x_i^b \partial_{x_i}^b -x_i^2 \partial_{x_i}^a  \right)\vev{\phi(x_1) \phi(x_2)} &\longmapsto i \sum_{i=1,2} \left( 6\frac{\partial}{\partial k^a_i} + 2 k_i^b \frac{\partial^2}{\partial k_i^b\partial k_i^a } -k_i^a \frac{\partial^2}{\partial k_i^b\partial k_i^b  } \right)\vev{\phi_{\vec k_1} \phi_{\vec k_2} }.
\end{align}
The term with the four point function has to be treated carefully, as we would like to remove delta functions at the end. Our expectation is that like the case of the inflationary SCT identity, removing momentum delta functions will require the use of the dilation identity. Therefore the strategy is to combine things into the form of the dilation identity. The four point function term is
\begin{align}
 &\frac{1}{2} \int_{x,y} {\cal E}(x,y)\left(2x^a x^b \partial_x^b -x^2 \partial_x^a + x \to y \right) \vev{\phi(x_1) \phi(x_2)\varphi(x) \varphi(y)} \nonumber \\
  &~~~~~~~~~~~= \frac{1}{2}\int_{x,y,q,p_1,p_2} \frac{1}{(2\pi)^9} G_q e^{-i q\cdot(x-y)} i \left( \hat{Q}^a_{p_1}+\hat{Q}^a_{p_2}\right)e^{-ip_1\cdot x -ip_2\cdot y} \vev{\phi_{\vec k_1}\phi_{\vec k_2} \varphi_{\vec p_1} \varphi_{\vec p_2}}, \\
  &\hat{Q}^a_{p}\equiv  2 p^b \frac{\partial^2}{\partial p^b\partial p^a } - p^a \frac{\partial^2}{\partial p^b\partial p^b} \nonumber 
\end{align}
We integrate by part the operator $\hat{Q}^a$ to act on the four point function as  $ \left( \hat{O}^a_{p_1}+\hat{O}^a_{p_2} \right)\vev{\phi_{\vec k_1} \phi_{\vec k_2}  \varphi_{\vec p_1} \varphi_{\vec p_2}}$, with $\hat{O}^a_{p}\equiv 6\frac{\partial}{\partial p^a} + 2 k_i^b \frac{\partial^2}{\partial p^b\partial p^a } - p^a \frac{\partial^2}{\partial p^b\partial p^b} $. Then the integrals in $x,y$ can be performed to obtain delta functions $(2\pi)^3\delta(\vec p_1+\vec q)$ and $(2\pi)^3\delta(\vec p_2-\vec q)$. 

We then integrate over $\vec q$ using the delta function and then integrate by parts again the differential operator $\hat{O}^a$. This yields a term of the form
\begin{align}
 \frac{i}{2}\int_{p_1,p_2} \frac{1}{(2\pi)^3}\left( \hat{Q}^a_{p_1}+\hat{Q}^a_{p_2}\right) \left[ {\cal E}_{\vec p_2}\delta(\vec p_1+\vec p_2) \right]\vev{\phi_{\vec k_1}\phi_{\vec k_2} \varphi_{\vec p_1} \varphi_{\vec p_2}}.
\end{align}
There are 3 types of terms when $\hat{Q}^a$ acts on ${\cal E}_{\vec p_2} \delta^3(\vec{p}_1+\vec{p}_2)$,
\begin{align}
p \partial^2{\cal E}_{\vec p_2} \delta(P_T) : & \left( \hat{Q}^a_{p_2}{\cal E}_{\vec p_2} \right) \delta(\vec p_1+\vec p_2),      \\
p \partial {\cal E}_{\vec p_2}  \partial \delta(P_T) : &  2p_2^b \frac{\partial}{\partial p_2^b} {\cal E}_{\vec p_2} \frac{\partial }{\partial p_2^a} \delta(\vec p_1+\vec p_2)   + 4p_2^{[b} \frac{\partial}{\partial p_2^{a]}} {\cal E}_{\vec p_2}  \frac{\partial }{\partial p_2^b} \delta(\vec p_1+\vec p_2)  \nonumber  \\
 &=   2p_2\partial_{p_2} {\cal E}_{\vec p_2} \frac{\partial }{\partial p_2^a} \delta(\vec p_1+\vec p_2) , \\
p  {\cal E}_{\vec p_2} \partial ^2\delta(P_T) : &  \,{\cal E}_{\vec p_2} \left( 2P_T^b \frac{\partial^2}{\partial P_T^b\partial P_T^a } - P_T^a \frac{\partial^2}{\partial P_T^b\partial P_T^b} \right)\delta(\vec p_1+\vec p_2) \nonumber \\
 & \overset{\mbox{IBP}}{=} -6 {\cal E}_{\vec p_2} \frac{\partial}{\partial P_T^a} \delta(\vec p_1+\vec p_2).
\end{align}
We have used the fact that ${\cal E}_{\vec p_2} $ is rotationally invariant. Collecting terms and considering the two contractions of the four point function (the third divergent contraction is canceled as explained in \ref{app:4point})  we have 
\begin{align}
& \frac{i}{2} \int_{p_1,p_2} (2\pi)^3\left[ \hat{Q}^a_{p_2} {\cal E}_{\vec p_2} \delta(\vec p_1+\vec p_2)  - (6-2p_2\partial_{p_2}){\cal E}_{\vec p_2} \frac{\partial}{\partial P_T^a} \delta(\vec p_1+\vec p_2)  \right]  \nonumber \\
&~~~~\times\Big[ \vev{\phi_{\vec k_1}\varphi_{\vec p_1}}' \vev{ \phi_{\vec k_2} \varphi_{\vec p_2}}'\delta(\vec k_1+\vec p_1)\delta(\vec k_2+\vec p_2)+\vev{\phi_{\vec k_2}\varphi_{\vec p_1}}' \vev{ \phi_{\vec k_1} \varphi_{\vec p_2}}'\delta(\vec k_2+\vec p_1)\delta(\vec k_1+\vec p_2) \Big]  \nonumber \\
=\; &  \frac{i}2{(2\pi)^3} \bigg\{ \Big[  \hat{Q}^a_{-k_1} {\cal E}_{\vec k_1}+ \hat{Q}^a_{-k_2}  {\cal E}_{\vec k_2} \Big] \delta(\vec k_1+\vec k_2)\nonumber \\
 & \qquad + \Big[ (6-2k_2\partial_{k_2}) {\cal E}_{\vec k_2}  \vev{\phi_{\vec k_1}\varphi_{-\vec k_1}}' \vev{ \phi_{\vec k_2} \varphi_{-\vec k_2}}'   + k_1\leftrightarrow k_2 \Big]  \frac{\partial}{\partial k_T^a} \delta(\vec k_1+\vec k_2)   \bigg\} \nonumber \\
 =\;& (2\pi)^3 i (6-2k_2\partial_{k_1}) {\cal E}_{\vec k_1} \vev{\phi_{\vec k_1}\varphi_{-\vec k_1}}'^2 \frac{\partial}{\partial k_T^a} \delta(\vec k_1+\vec k_2) \nonumber \\
  &\qquad - \frac{1}{2}\left.\frac{\partial}{\partial k_2^a}\Big[(6-2k_2\partial_{k_2}) {\cal E}_{\vec k_2}  \vev{\phi_{\vec k_2}\varphi_{-\vec k_2}}'^2  \Big] \right|_{\vec{k}_2=-\vec{k}_1} \delta(\vec k_1+\vec k_2).
\end{align}
In our manipulations 
we have used momentum conservation and the delta function identity $f(x)\delta'(x-a) = f(a)\delta'(x-a) - f'(a) \delta(x-a)$. Collecting all the terms in the identity (\ref{eq:SCTclassical}), we find that the coefficient of $\frac{\partial}{\partial k_T^a} \delta^3(\vec{k}_1+\vec{k}_2)$ is exactly the dilation identity therefore is zero and the momentum conserving delta functions in the rest of the identity can be removed. The SCT identity after removing the delta function and imposing momentum conservation is,
\begin{align}
\nonumber
\frac{\partial}{\partial {q^a}} \left( {\cal E}_{\vec q} \vev{\phi_{\vec k_1} \phi_{-\vec k_1-\vec q}\varphi_{\vec q} }' \right) \bigg|_{\vec q\to 0} &- \frac{k_1^a}{2k_1}\frac{\partial}{\partial { k_1}} \Big[(3-k_1\partial_{k_1}){\cal E}_{\vec k_1} \vev{\phi_{\vec k_1}\varphi_{-\vec k_1}}'^2 \Big] \\
&~~~~~~= -k_1^a \left( \frac{2}{k_1} \partial_{k_1} + \frac{1}{2}\partial_{k_1}^2 \right) \vev{\phi_{\vec k_1}\phi_{-\vec k_1}}'.
\end{align}

\renewcommand{\em}{}
\bibliographystyle{utphys}
\addcontentsline{toc}{section}{References}
\bibliography{physicalmodesbib}

\end{document}